%% file: ms.tex
\documentclass[prx,twocolumn,floatfix]{revtex4-1}                                                                                                            
                                                                                                                                                             
\usepackage{graphicx}                                                                                                                                        
\usepackage{amssymb,amsmath}                                                                                                                                 
\usepackage[colorlinks,citecolor=blue,urlcolor=blue]{hyperref}                                                                                               
\usepackage{dcolumn} 
\usepackage{color}       
\newcommand{\ve}[1]{\boldsymbol{#1}}

\begin{document}                                                                                                                                             

\title{Phase diagram and dynamics  of the SU($\boldsymbol N$) symmetric Kondo lattice model}

\author{Marcin Raczkowski}
\affiliation{Institut f\"ur Theoretische Physik und Astrophysik, Universit\"at W\"urzburg, Am Hubland, D-97074 W\"urzburg, Germany}
 \author{Fakher F. Assaad}
\affiliation{Institut f\"ur Theoretische Physik und Astrophysik and W\"urzburg-Dresden Cluster of Excellence ct.qmat,
             Universit\"at W\"urzburg, Am Hubland, D-97074 W\"urzburg, Germany}

\date{\today}

\begin{abstract}
In heavy-fermion systems, the competition between the local Kondo physics and intersite magnetic fluctuations 
results in unconventional quantum critical phenomena which are frequently addressed within the Kondo lattice model (KLM). 
Here we study this interplay in the SU($N$) symmetric generalization of the two-dimensional half-filled KLM by 
quantum Monte Carlo simulations with $N$ up to 8.  
While the long-range antiferromagnetic (AF) order in SU($N$) quantum spin systems typically gives way to 
spin-singlet ground states with spontaneously broken lattice symmetry, we find that the SU($N$) KLM is unique in that 
for each finite $N$ its ground-state phase diagram  hosts only two phases -- AF order and the Kondo-screened phase. 
The absence of any intermediate phase  between the $N=2$ and large-$N$ cases establishes adiabatic correspondence 
between both  limits and  confirms that the large-$N$ theory is a correct saddle point of the KLM fermionic 
path integral and a good starting point to include  quantum  fluctuations.
In addition, we determine the evolution of the single-particle gap, quasiparticle residue of the doped hole at momentum $(\pi,\pi)$, 
and spin gap across the  magnetic order-disorder transition.  
Our results indicate that  increasing $N$  modifies the behavior of  the coherence temperature: while it  evolves smoothly across the 
magnetic transition at $N=2$ it develops an abrupt jump -- of up to an order of magnitude --  at larger but finite $N$.
We discuss the magnetic order-disorder transition from a quantum-field-theoretic perspective and comment on implications of 
our findings for the interpretation of experiments on quantum critical heavy-fermion compounds. 
\end{abstract}

\maketitle


\section{\label{sec:intro}Introduction}

Nowadays, we are witnessing  remarkable progress in experimental techniques and the emergence of promising 
platforms for exploring novel aspects of quantum many-body phenomena. One notable example of 
many-body physics is the Kondo effect which arises  from entanglement of the impurity spin with surrounding 
conduction electrons and the formation, below the Kondo temperature $T_K$, of a spin singlet ground state~\cite{Hewson_book}.   
In fact, the role of the electron spin can be replaced by any other quantum degree  of freedom with symmetry protected two-fold degeneracy, e.g., 
orbital momentum~\cite{Lichtenstein02} while the simultaneous presence of both a spin 
and an orbital degeneracy might lead to an SU(4) symmetric Kondo physics~\cite{Delft03,Choi05,Galpin05}. The SU(4) Kondo effect 
was already observed in carbon nanotubes, quantum dots with orbitally degenerate states, double quantum dot systems, 
and in a nanoscale silicon transistor~\cite{Sasaki04,Kouwenhoven05,Lim06,Finkelstein07,Tettamanzi12,Weymann14}.
Under a high crystalline symmetry such as the cubic one, there are chances for the spin-orbital-coupled Kondo entanglement to remain also in 
realistic systems, e.g., in rare-earth compounds~\cite{Martelli19}.

Furthermore, the advent of scanning tunneling microscopy has made it possible to fabricate artificial 
Kondo nanostructures with atomic precision~\cite{Li98,Madhavan98,Manoharan00,DiLullo12,Bucher18,Morr19,Lorente19,Wiebe19,Otte19} whose properties, 
in particular the onset of lattice effects, have recently been a subject of increasing theoretical 
attention~\cite{Morr17,Raczkowski19,Jiang19}. Magnetic atoms or organic molecules with orbital degeneracy 
deposited onto a metallic surface  provide the opportunity to realize an SU(4) symmetric Kondo effect~\cite{Minamitani12}. 
In addition, it is possible to study its evolution upon increasing the number of periodically arranged magnetic centers 
as in Ref.~\cite{Tsukahara11} where, starting from a single iron(II) phthalocyanine molecule deposited  
on top of Au(111) surface, in consecutive steps a two-dimensional superlattice was created followed by 
the theoretical analysis~\cite{Lobos14}.

Yet another very active field of research with promises to provide new insights into the  SU($N$) symmetric generalization 
of the Kondo effect~\cite{CS69,Read83CS,Otsuki07} are quantum simulations with alkali-earth-like atoms in optical lattices~\cite{Rey14}. 
Thus far, building on theoretical proposals~\cite{Foss10a,Foss10b,Gorshkov10}, subsequent experimental studies utilizing 
ytterbium and strontium isotopes  reported the observation of SU($N$) symmetric spin-exchange interactions between different 
orbitals with $N$ as large as 10~\cite{Zhang14,Scazza14,Cappellini14,Riegger18}. From a practical point of view, 
there are three crucial features required for the realization of the SU($N$) Kondo physics in such setups: 
(i) the existence of a metastable excited state playing together with the atomic ground state the role of orbital degrees of freedom 
 loaded into an orbital state-dependent optical lattice~\cite{Riegger18},
(ii) a large  nuclear spin $I>1/2$ of fermionic isotopes must decouple from the electronic degrees of freedom  to guarantee the 
  SU($N=2I+1$) spin-rotation symmetry of the interactions, and 
(iii) an antiferromagnetic (AF) character of spin-exchange interactions in the limit with one fully localized orbital. 
Although the currently used isotopes with $I>1/2$ realize ferromagnetic interorbital 
interactions~\cite{Zhang14,Scazza14,Cappellini14,Riegger18}, ongoing theoretical 
proposals~\cite{Nakagawa15,Kuzmenko16,Zhang16,Cheng17,Zhang19}, numerical simulations~\cite{Rey15,Zhong17,SV18}, and 
utilizing other isotopes of alkali-earth-like atoms~\cite{Ono19} allow one to envisage a controllable implementation 
of a Kondo-singlet state with SU($N$) symmetric interactions in single-impurity and lattice situations in the near future.

Given all these experimental developments, it is timely to consider an SU($N$) generalization of the conventional SU(2) Kondo lattice 
model (KLM) and to elucidate what kind of correlated phenomena occur under novel conditions with $N>2$ which is the purpose of this paper. 
The importance of the KLM stems from its capability to account for the essential aspects of $4f$-orbital-based  heavy-fermion systems  summarized in the 
seminal Doniach phase diagram~\cite{Doniach77,Ueda97}. 
On the one hand, the Kondo exchange interaction $J$ between the local moments and conduction electrons promotes a Kondo-screened 
paramagnetic phase in which the local moments are quenched by spins of the conduction electrons. 
On the other hand, the conduction electron-mediated RKKY exchange interaction between the local moments drives 
them into a magnetically ordered state thus leading to a quantum phase transition~\cite{RMP07,Gegenwart08,Si10}.
In some cases, the latter corresponds to a spin-density-wave transition  as described in the  
Hertz-Millis scenario~\cite{Hertz76,Millis93}.

However, accumulating experimental evidence suggests that a realistic description of various types of quantum criticality and 
non-Fermi-liquid effects in heavy-fermion systems requires, in addition to the "Kondo axis" $K$ tuning the ratio between the Kondo 
temperature $T_K$ and the strength of RKKY interactions, a second  "quantum axis" $Q\sim 1/S$ tuning the strength  of 
quantum zero‐point fluctuations of the impurity spin $S$~\cite{Coleman2010,Wirth16}. 
The magnitude of $Q$ can be tuned  by either increasing geometric frustration  or reduction of dimensionality; 
large $Q$ paves the way for the realization of exotic proposals such as local quantum criticality~\cite{Coleman01,Si01}, fractionalized 
Fermi liquids~\cite{Burdin02,Senthil03,Senthil04}, and partial Kondo screening~\cite{Motome10,Pixley14,Sato18}.

Alternatively, when the physical SU(2) spin symmetry of the quantum model is generalized to SU($N$), a large number of degrees of freedom 
makes the long-range magnetic order less likely. 
An exciting aspect of studying SU($N$) quantum  antiferromagnets in various geometries is that they allow one to pin down the role of Berry phases 
on the emergence of quantum-disordered ground state. As pointed out by Haldane~\cite{Haldane88}, the relevance of the Berry phase term 
implies that point defects (hedgehogs) of the N\'eel field  in spacetime acquire a net geometric phase. On the square lattice, 
large-$N$ calculations predicted that the proliferation of topological defects in the presence of nontrivial Berry phases naturally 
leads to columnar ${\pmb q}=(\pi,0)$ valence bond solid (VBS) order in the paramagnet~\cite{RS89,Read89,Read90}.  
Later on, the onset of VBS order at sufficiently large $N$  was confirmed by variational Monte Carlo study~\cite{Paramekanti07} 
and quantum Monte Carlo (QMC) simulations on the square~\cite{Kawashima03,Assaad05,Kawashima07,Beach09,Congjun14,Kawashima15} and   
honeycomb~\cite{Lang13,Congjun16,Congjun17} lattices.  
Extensive QMC simulations of extended SU($N$) Heisenberg models have also led to insight into the nature of the quantum phase 
transition separating N\'eel and VBS phases~\cite{Kawashima09,Kaul12,Kaul_cp,Harada13} lending strong support to the theory of 
continuous  "deconfined" quantum criticality~\cite{Senthil04a,Senthil04b}. Moreover, by extending QMC studies to the bilayer 
geometry~\cite{Kaul12b}, it has been confirmed that finite interlayer coupling renders Berry phases irrelevant at the quantum 
critical point~\cite{Sandvik94,Sandvik95,Zaanen97,Wang06}. As a result, the continuous N\'eel-VBS transition turns 
first order~\cite{Kaul12b}.

In contrast to SU($N$) Hamiltonians with direct effective spin-exchange interactions, very little is known about the phase diagram 
of the SU($N$) KLM with RKKY interactions  between the impurity spins mediated by conduction electrons.
On the one hand, coherent Kondo screening~\cite{Jarrell98,Eder98,Pruschke00,Costi02,Assaad04,Fisk08,Raczkowski10,Tanaskovic11,Haule19,Costa19} 
and the formation of the Kondo insulating  (KI) phase at half-filling can be accounted for within 
the large-$N$ approach~\cite{Burdin00}. Strictly speaking,  this mean-field approximation is formally justified only in a limit 
where the spin symmetry of the original model is extended from SU(2) to SU($N$) with $N\to\infty$. 
Nevertheless, the method is often applied to heavy-fermion models with only SU(2) symmetry \cite{Morr17} and is considered 
as a good starting point for studying dynamical properties of heavy-fermion metals  
using the $1/N$ expansion technique~\cite{Coleman83,Read84,Auerbach86,Millis87}. However,  there is no way of assessing \emph{a priori} 
the validity of a large-$N$ approach at any finite $N$. 

On the other hand, despite a few attempts to develop a controlled treatment of both magnetism and the Kondo effect within a single
large-$N$ expansion~\cite{Tsvelik00,Rech06}, its applicability remains restricted to quantum disordered phases
and thus the large-$N$ approach cannot be used to explore the full phase diagram of the model.
Another caveat of large-$N$ approximation is that finite hybridization order parameter breaks the local gauge symmetry and implies 
that the constraint of single occupancy on the $f$-sites is fulfilled only on average which motivated  the development of alternative 
approaches~\cite{Nilsson11,Becker13,Becker18}. This yields  a further  motivation for systematic studies of the SU($N$) KLM  
using an unbiased method which handles the constraint numerically exactly so as to assess the validity of 
large-$N$ approximate treatments~\cite{Burdin00}.  

Here, by performing auxiliary-field QMC simulations~\cite{alf} we shall map out the phase diagram of the SU($N$) KLM  as a function of the coupling $J/t$ and the number of flavors $N$. 
Given diverse  phenomena found upon loss of the AF order  in SU($N$) Hubbard and Heisenberg models of magnetic 
insulators~\cite{Kawashima03,Assaad05,Kawashima07,Beach09,Congjun14,Kawashima15,Lang13,Congjun16,Congjun17,Kawashima09,Kaul12,Kaul_cp,Harada13}, 
one could equally expect the emergence of novel phases in the SU($N$) KLM. Furthermore, previous QMC simulations of the SU(2) KLM  predict 
that below the magnetic energy scale $T_{\textrm{RKKY}}$, the single-particle gap scales as $J$ \cite{Assaad99,Capponi01}. 
This contrasts with an exponentially small gap found in the dynamical mean-field theory~\cite{Pruschke00},  large-$N$ limit~\cite{Burdin00}, and Gutzwiller approximation~\cite{Rice85}, 
all of them  omitting spatial fluctuations. Hence, we shall elucidate necessary conditions for recovering the large-$N$ limit in the single-particle dynamics thus providing a valuable 
benchmark of the large-$N$ approach.

\section{\label{sec:model}Model, QMC method,  and large-${\boldsymbol N}$ saddle point}


Our starting point is the SU(2) symmetric KLM at half-filling~\cite{Ueda97},
\begin{equation}
\hat{\mathcal{H}} = -t\sum_{\langle\pmb{i},\pmb{j}\rangle, \sigma } 
                  \hat{c}^{\dagger}_{\pmb{i},\sigma} \hat{c}^{}_{\pmb{j},\sigma} +
          J \sum_{\pmb{i}  }  \pmb{S}^{c}_{\pmb{i}} \cdot \pmb{S}^{f}_{\pmb{i}},  
\label{su2}
\end{equation}
where $\pmb{S}^{c}_{\pmb{i}} = \tfrac{1}{2} \sum_{\sigma,\sigma'}  \hat{c}^{\dagger}_{\pmb{i},\sigma}
\pmb{\sigma}^{}_{\sigma,\sigma'} \hat{c}^{}_{\pmb{i},\sigma'}$ are spin operators of conduction electrons and
$\pmb{S}^{f}_{\pmb{i}} = \tfrac{1}{2} \sum_{\sigma,\sigma'}  \hat{f}^{\dagger}_{\pmb{i},\sigma}
\pmb{\sigma}^{}_{\sigma,\sigma'} \hat{f}^{}_{\pmb{i},\sigma'}$ are localized spins with  $\pmb{\sigma} $ being the Pauli matrices.
The Hamiltonian (\ref{su2}) describes localized  spin 1/2 magnetic moments coupled via an AF  exchange interaction $J$ to conduction 
electrons moving  on a square lattice with a nearest-neighbor hopping amplitude $t$. Consider now a fermionic representation of the 
SU($N$) generators, 
\begin{equation}
\hat{S} ^{\mu}_{\pmb{i},\nu}   = \hat{f}^{\dagger}_{\pmb{i},\nu} \hat{f}^{}_{\pmb{i},\mu}   
 - \frac{\delta_{\mu,\nu}}{N}  \sum_{\sigma=1}^{N} \hat{f}^{\dagger}_{\pmb{i},\sigma} \hat{f}^{}_{\pmb{i},\sigma},
\label{generator}
\end{equation}
subject to the local  constraint,
\begin{equation}
        \sum_{\sigma=1}^N  \hat{f}^{\dag}_{{\pmb i},\sigma} \hat{f}^{}_{{\pmb i},\sigma} = \frac{N}{2},
\label{constraint}
\end{equation}
selecting the fully antisymmetric self-adjoint representation corresponding to a Young tableau with a single column and $N/2$ rows.
The corresponding SU($N$) generalization of the KLM (\ref{su2}) reads,
\begin{equation}
\hat{\mathcal{H}} =   \hat{\mathcal{H}}_t + \hat{\mathcal{H}}_{J} + \hat{\mathcal{H}}_U, 
\label{sun}
\end{equation}
with
\begin{align}
\hat{\mathcal{H}}_t  &= -t\sum_{\langle\pmb{i},\pmb{j}\rangle,\sigma=1}^N  \hat{c}^{\dag}_{{\pmb i},\sigma}\hat{c}^{}_{{\pmb j},\sigma},  \\
\hat{\mathcal{H}}_{J}&= -\frac{J}{2N} \sum_{\pmb{i}} \left( \hat{D}_{\pmb i}^{\dag} \hat{D}_{\pmb i}^{} 
                        + \hat{D}_{\pmb i}^{} \hat{D}_{\pmb i}^{\dag}\right), \\
\hat{\mathcal{H}}_U  &=   \frac{U_f}{N}\sum_{\pmb{i}} 
         \left [ \sum_{\sigma=1}^N \left( \hat{f}^{\dag}_{{\pmb i},\sigma} \hat{f}^{}_{{\pmb i},\sigma} - \frac{1}{2}\right) \right]^2.
\end{align}
Here, $\hat{D}_{\pmb i}^{\dag}=\sum_{\sigma=1}^N  \hat{c}^{\dag}_{{\pmb i},\sigma} \hat{f}^{}_{{\pmb i},\sigma}$ and 
we have added a Hubbard term $\hat{\mathcal{H}}_U$  for the $f$-electrons.  
Since $ \left[   \hat{\mathcal{H}}, \hat{\mathcal{H}}_U \right]  = 0 $, in the presence of the Hubbard term, charge fluctuations on the 
$f$-sites are suppressed  by Boltzmann factor $e^{-\Theta U_f /N \left( \sum_{\sigma=1}^N  \hat{f}^{\dagger}_{\sigma}  
\hat{f}^{\phantom\dagger}_{\sigma} - {\frac{N}{2} } \right)^2 }$ thus imposing the constraint (\ref{constraint}) provided that the 
projection parameter $\Theta$  is chosen to be sufficiently large. To obtain ground state properties of the Hamiltonian (\ref{sun}), 
we use a projective QMC technique based on the imaginary-time evolution of a trial wave function 
$| \Psi_\text{T}\rangle$, with $ \langle\Psi_\text{T}  |\Psi_0 \rangle  \neq 0 $,  to the ground state $|\Psi_0 \rangle$:                                
\begin{equation}
  \frac{ \langle  \Psi_0 | \hat{O} |  \Psi_0 \rangle  }{ \langle  \Psi_0  |  \Psi_0 \rangle  }  =  
  \lim_{\Theta \rightarrow \infty} 
  \frac{ \langle  \Psi_\text{T} | e^{-\Theta \hat{\mathcal{H}}  } \hat{O} e^{-\Theta \hat{\mathcal{H}}   } |  \Psi_\text{T}\rangle  }
  { \langle  \Psi_\text{T}  | e^{-2\Theta \hat{\mathcal{H}} }  |  \Psi_\text{T} \rangle  }.
\end{equation}
It relies on a Trotter-Suzuki decomposition so as to split  the imaginary-time propagation of  the single-body  $H_t$ 
and the interaction  $\hat{\mathcal{H}}_{int}=\hat{\mathcal{H}}_{J}+\hat{\mathcal{H}}_{U}$ terms  into $L_{\tau}$ steps of size 
$\Delta\tau=\Theta/L_{\tau}$ such that, 
\begin{equation}
e^{-\Theta \hat{\mathcal{H}} }= \prod_{i=1}^{L_\tau} 
e^{-\Delta\tau \hat{\mathcal{H}}_t/2 } e^{-\Delta\tau \hat{\mathcal{H}}_{int} }  e^{-\Delta\tau \hat{\mathcal{H}}_t/2 }  
+ \mathcal{O}(\Delta\tau^2).
\label{TS}
\end{equation}
 Since  $  \left[ \hat{D}^{\dagger}_{   \pmb{i} } , \hat{D}^{\phantom\dagger}_{   \pmb{i}  } \right]  \neq 0 $ some  care has to be taken in order 
 to ensure the hermiticity of the imaginary-time propagator in the Monte Carlo  approach.  First we rewrite: 
 \begin{eqnarray}
        \hat{\mathcal{H}}_J  &   =  &       \hat{\mathcal{H}}_J ^{+}  +   \hat{\mathcal{H}}_J ^{-}  \text{  with }   \nonumber \\
       \hat{\mathcal{H}}_J^{+}   & =  &   -\frac{J}{4 N}  \sum_{ \pmb{i} }  
        \left(  \hat{D}^{\dagger}_{   \pmb{i} } +    \hat{D}_{  \pmb{i} }  \right)^2    \text{  and }   \nonumber \\
         \hat{\mathcal{H}}_J^{-}   & =  &   -\frac{J}{4 N}  \sum_{ \pmb{i} }                  
        \left(i \hat{D}^{\dagger}_{   \pmb{i} } -  i \hat{D}_{  \pmb{i} }  \right)^2   
\end{eqnarray} 
where  $\hat{\mathcal{H}}_J^{\pm}  $   correspond to sums of commuting terms.  Second we approximate: 
\begin{eqnarray}
 e^{-\Delta\tau \hat{\mathcal{H}}_{int} }   =   e^{-\Delta\tau \hat{\mathcal{H}}_{U} } e^{-\frac{\Delta\tau}{2} \hat{\mathcal{H}}_{J}^{+} } e^{-\Delta\tau \hat{\mathcal{H}}_{J}^{-} } e^{-\frac{\Delta\tau}{2} \hat{\mathcal{H}}_{J}^{+} }  +  \mathcal{O}(\Delta\tau^3). \nonumber \\
\end{eqnarray}
At this point,  all the interaction terms are  in the form of perfect squares, and we can   implement the model in the ALF library \cite{alf}.

  Although the ALF library uses  discrete   fields for optimization and sampling issues,  
it is equivalent to the use of  continuous fields.   In fact decoupling the above perfect square terms with scalar fields yields for the finite temperature  grand-canonical 
partition function,  
\begin{equation}
	  Z \propto \int D \left\{ z(\ve{i}, \tau)  \right\}   e^{- N S  \left( \left\{ z(\ve{i}, \tau)  \right\}, \left\{\lambda(\ve{i}, \tau)  \right\} \right)  }, 
\end{equation}
with action, 
\begin{eqnarray}
	S   =& &  \int_{0}^{\beta} d \tau  \sum_{\ve{i}} \left(   J | z(\ve{i}, \tau) |^2 + U | \lambda(\ve{i}, \tau) |^2 \right)   \\ 
	        & &  - \ln \text{Tr}  {\cal T} e^{-\int_{0}^{\beta} d \tau  \hat{H}\left( \left\{ z(\ve{i}, \tau)    \right\},  \left\{\lambda(\ve{i}, \tau)  \right\}  \right) },   \nonumber 
\end{eqnarray}
and Hamiltonian,
\begin{eqnarray}
\hat{H}\left( \left\{ z \right\}, \left\{\lambda \right\}    \right)  =    & & 
    	-t\sum_{\langle\pmb{i},\pmb{j}\rangle}  \hat{c}^{\dag}_{{\pmb i}}\hat{c}^{}_{{\pmb j} } + J  \sum_{\ve{i}}  
\left(  z(\ve{i}, \tau) \hat{f}^{\dag}_{{\pmb i}}  \hat{c}^{\phantom\dagger}_{{\pmb i}}   +   h.c.  \right)   \nonumber \\ 
& &  + i U_f \lambda(\ve{i}, \tau)  \left( \hat{f}^{\dag}_{{\pmb i}}  
\hat{f}^{\phantom\dag}_{{\pmb i}}   - \frac{1}{2} \right). 
\end{eqnarray}
In the above, the fermions operators have  lost their flavor index.   Since the complex,  $z(\ve{i}, \tau)$, and scalar, $\lambda(\ve{i}, \tau)$,  fields couple to SU($N$) 
symmetric operators,  $N$ can be   pulled out in front of the action.   This is particularly useful for the simulations since $N$ merely  comes in as a parameter.     
Using a particle-hole transformation: 
\begin{equation}
     \hat{c}^{\dagger}_{\ve{i}}    \rightarrow   e^{i \ve{Q} \cdot \ve{i} } \hat{c}^{\phantom\dagger}_{\ve{i}}   \; \; \text{and}  \; \; 
     \hat{f}^{\dagger}_{\ve{i}}    \rightarrow  - e^{i \ve{Q} \cdot \ve{i} } \hat{f}^{\phantom\dagger}_{\ve{i}},     
\end{equation}
with $\ve{Q} = (\pi,\pi) $, one will show that  the  imaginary part of the action takes the value $i n \pi$  with $n$ an integer. 
Hence for even values of $N$,  statistical weights in Monte Carlo sampling are positive for all values of Hubbard-Stratonovitch configurations and the negative sign  problem is absent. 

In the large-$N$ limit, the saddle-point approximation: 
\begin{eqnarray}
	& & \left. \frac{\delta  S  \left( \left\{ z(\ve{i}, \tau)  \right\}, \left\{\lambda(\ve{i}, \tau)  \right\} \right)  }{ \delta  z(\ve{i}, \tau)  }   \right|_{z = z^{*},  \lambda =\lambda^{*}}   = 0,  \text{ and }  \nonumber \\  
	& & \left. \frac{ \delta  S  \left( \left\{ z(\ve{i}, \tau)  \right\}, \left\{\lambda(\ve{i}, \tau)  \right\} \right)  }{ \delta  \lambda(\ve{i}, \tau)  }   \right|_{z= z^{*}, \lambda=\lambda^{*} }   =   0 
\end{eqnarray}
becomes exact.  Assuming  space and time independent fields produces the large-$N$ mean-field theory   discussed in  Appendix~\ref{app:scales}. The Monte Carlo method can  
hence be seen as not only  accounting for all fluctuations around  the large-$N$ saddle point,  but also for assessing  if the saddle point is stable or not.

As mentioned above, our calculations were carried within the projective formulation.    To be able to pull out $N$ in front of the action,  we  use an SU($N$) symmetric trial 
wave function corresponding to the  large-$N$ saddle-point  Hamiltonian: 
\begin{equation}
	\hat{\mathcal{H}}_{\text{T}}  = -t\sum_{\langle\pmb{i},\pmb{j}\rangle,\sigma=1}^N  \hat{c}^{\dag}_{{\pmb i},\sigma}\hat{c}^{}_{{\pmb j},\sigma} + V \sum_{\ve{i}} 
\left(  \hat{D}^{\dagger}_{   \pmb{i} }  +   \hat{D}^{\phantom\dagger}_{  \pmb{i} }  \right).
\end{equation}
Unless stated otherwise,  our simulations  were carried out at finite  imaginary time step  $\Delta \tau t =  0.1 $ and to generate the trial wave function, we have used $V = 0.1t$.

\section{\label{overview} Overview of the results}

The hybridization of conduction electron states with the $f$-electron states in a 
lattice situation leads to the large Fermi surface of the heavy-fermion metal and to the hybridization gap in the KI 
phase at half-band filling. The factors controlling the large Fermi surface topology continue to attract considerable 
attention~\cite{Ogata07,Martin08,Fabrizio08,Vojta08,Hackl08,Otsuki09,Bercx10,Benlagra11,Becca13,Peters15,Lenz17}.

As discussed in Appendix~\ref{app:scales}, the number of flavors $N$ is a control parameter which tunes the relative importance of the RKKY interaction 
and the Kondo energy scale. Here, we are interested in the following questions: 
(i) Does the order-disorder phase transition exist for any $N>2$ at all or just the opposite --  
does one immediately reach the large-$N$ limit with only the KI phase? 
(ii) Assuming  that  the phase transition  continues to exist, is the continuous nature of the transition specific to the $N=2$ case? 
(iii) Does a larger $N$ stabilize any  intervening phase in the vicinity of the magnetic phase transition, e.g., VBS order? 
(iv) Given that at the mean-field level with a frozen $f$-spin Ansatz, magnetic ordering and Kondo screening are not 
compatible~\cite{Capponi01}, what are the single-particle spectral properties of the AF phase at finite $N>2$?   
(v) Does the quasiparticle (QP) dispersion continue to feature a flat band extending up to $\pmb{k}=(\pi,\pi)$  point signaling 
remnant Kondo screening of the impurity spins?  If so, how does the QP residue  evolve as a function of $N$ and across the 
magnetic order-disorder transition?

\begin{figure}[t!]
\begin{center}
\includegraphics[width=0.45\textwidth]{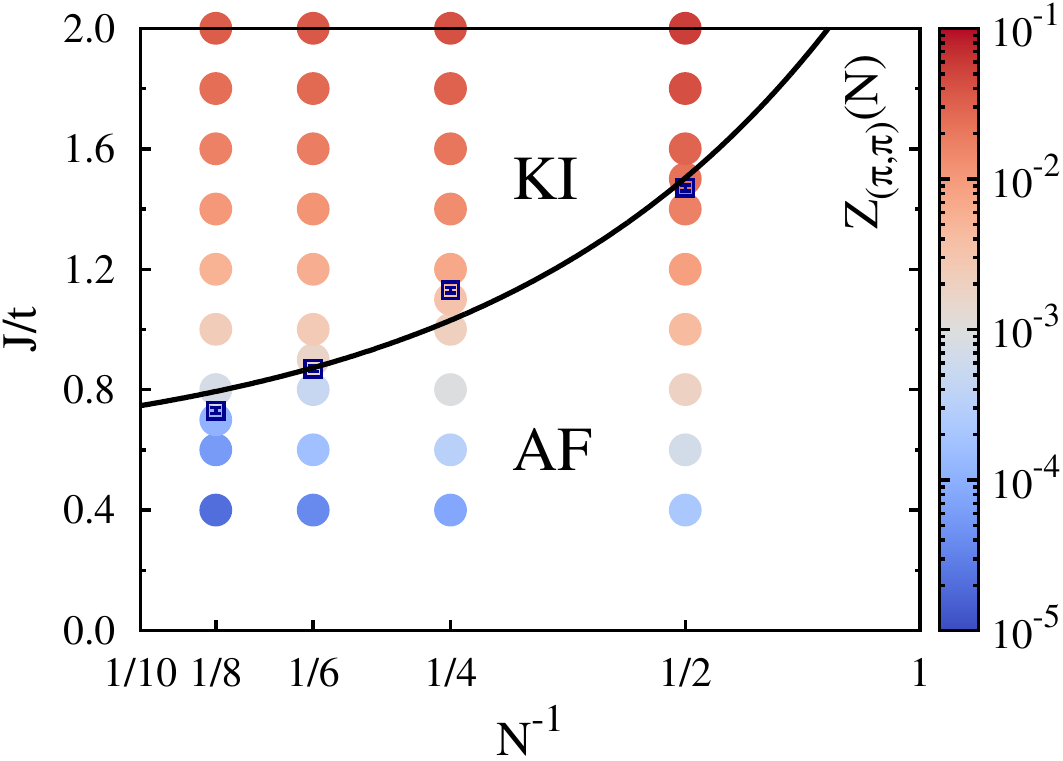}
\end{center}
\caption
{Ground-state phase diagram  of the SU($N$) KLM as a function of the inverse of the number of fermionic flavors $N$ and Kondo coupling $J$ with 
antiferromagnetic (AF) and Kondo insulating (KI) phases at half-filling;  
empty squares indicate magnetic order-disorder transition points extracted from the behavior of the staggered moment 
Eq.~(\ref{moment}) in the thermodynamic limit; solid line is a fit to a finite-$N$ functional form of the critical coupling  $J_c\propto 1/\ln N$ 
obtained by comparing the Kondo temperature $T_K$ to the magnetic energy scale $T_{\textrm{RKKY}}$, see Appendix~\ref{app:scales}. 
Color-coded circles correspond to the quasiparticle (QP) residue $Z_{\pmb k}$ of the doped hole at momentum ${\pmb k}=(\pi,\pi)$, see Sec.~\ref{charge} 
for raw data with errorbars. 
}
\label{PD}
\end{figure}

A partial answer to these questions is given in Fig.~\ref{PD} showing the ground-state phase diagram of the SU($N$) 
KLM as a function of the inverse of the number of fermion components $N$ and Kondo coupling $J$. 
Our main result is that the SU($N$) KLM in the fully antisymmetric self-adjoint representation supports magnetic  ordering for each considered value of $N$, 
and that no other phases  aside from the Kondo insulator and N\'eel state intervene, see Secs.~\ref{spin} and ~\ref{vbs}.  

Intuitively,  we expect the $J=0$ and $N \rightarrow \infty$ point to be singular.    For the ordering of limits   $\lim_{N \rightarrow \infty }  \lim_{J 
\rightarrow 0} $  we expect an AF ground state whereas for $ \lim_{J \rightarrow 0} \lim_{N \rightarrow \infty }  $ we expect a paramagnetic one. 
Figure~\ref{PD} confirms this point of view:  the magnetic order-disorder transition point (empty squares), extracted from the behavior 
of the staggered moment Eq.~(\ref{moment}) in the thermodynamic limit,  shifts upon increasing $N$ to smaller values of $J/t$ which enhances the domain of stability 
of the KI phase at the expense of the AF state. 
While the RKKY scale varies as $\frac{1}{N}$, the Kondo scale is $N$ independent  such that comparing scales yields an estimate of the critical coupling 
$J_c(N) \propto \frac{1}{\ln{N}}$ in the large-$N$ and small $J$ limit (see Appendix \ref{app:scales}).   We have used this form  to plot a guide to the eye 
for the phase boundary  in Fig.~\ref{PD}.

In addition, color-coded circles in Fig.~\ref{PD} correspond to the QP residue $Z_{\pmb k}$ of the doped hole at momentum $(\pi,\pi)$ extrapolated to the thermodynamic limit. 
We extracted this quantity directly on the imaginary-time axis by fitting the tail of the single-particle Green's function for conduction electrons to the form 
$Z_{\pmb{k}}e^{-\Delta_{qp}(\pmb{k}) \tau }$ where $\Delta_{qp}$ is the single-particle gap.
At $N=2$, and in  the magnetically ordered phase, we observe a remarkable coexistence of Kondo screening and antiferromagnetism that stands at odds 
with the mean-field result predicting only a very narrow coexistence  region~\cite{MF00,Capponi01}.  
We show in Sec.~\ref{charge} that this does not carry over over to larger values of $N$ where we observe an abrupt drop in  the QP residue $Z_{(\pi,\pi)}$ across the 
magnetic transition, see Fig.~\ref{PD}.  

Finally, in Secs.~\ref{spin_dynamic} and \ref{charge_dynamic} we investigate the impact of an enhanced Hamiltonian symmetry on 
the spin excitation spectrum and single-particle spectral function, respectively.
We proceed now to discuss numerical results which led us to  the above phase diagram.

\section{\label{results} Numerical  results}

Numerical results were obtained with an $N$-dependent projection parameter ranging from $2\Theta t=50$ for $N=2$ to $2\Theta t=400$ for $N=8$, chosen sufficiently 
large to guarantee  the convergence to the ground state $|\Psi_0\rangle$, see Appendix~\ref{app:conv}.
Physical observables have been extrapolated to the thermodynamic limit based on the QMC  data obtained on lattice sizes
ranging from $6\times 6$ to $14\times 14 $ with periodic boundary conditions. Finite-size scalings and representative raw QMC data are presented 
in Appendices~\ref{app:moment}, \ref{app:delta}, and \ref{app:spin}.

\subsection{\label{spin} Spin degrees of freedom}

\begin{figure}[t!]
\begin{center}
\includegraphics[width=0.45\textwidth]{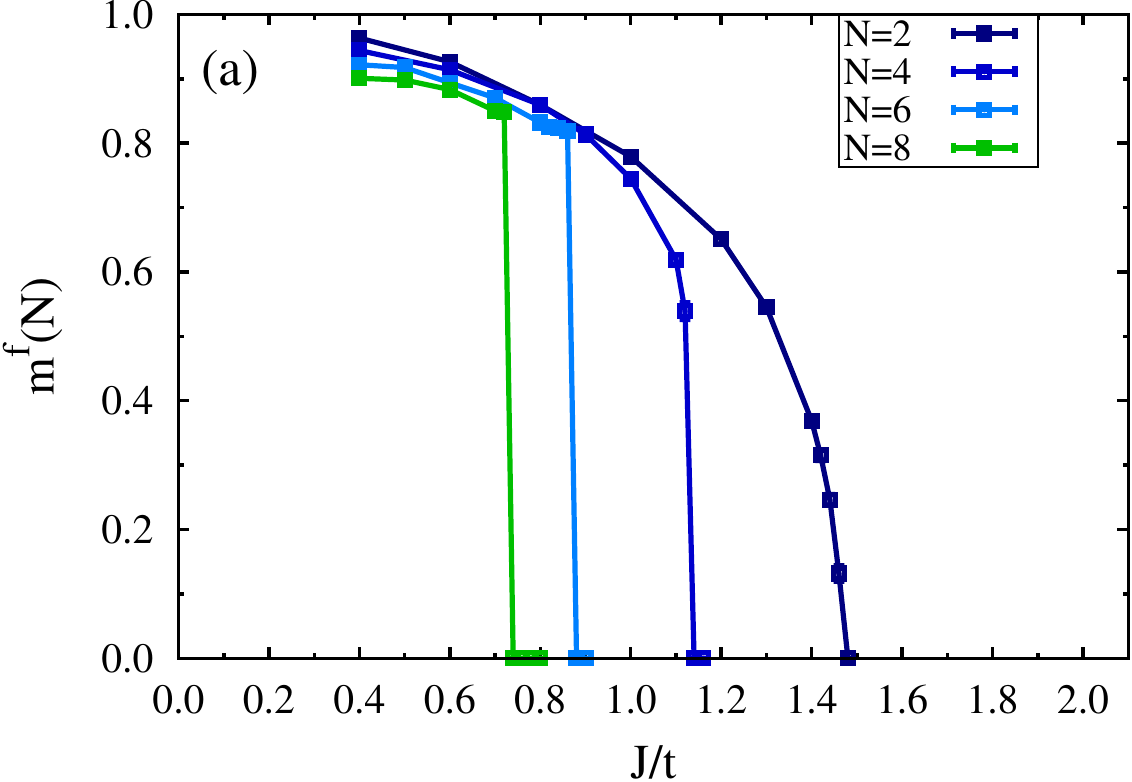}
\includegraphics[width=0.45\textwidth]{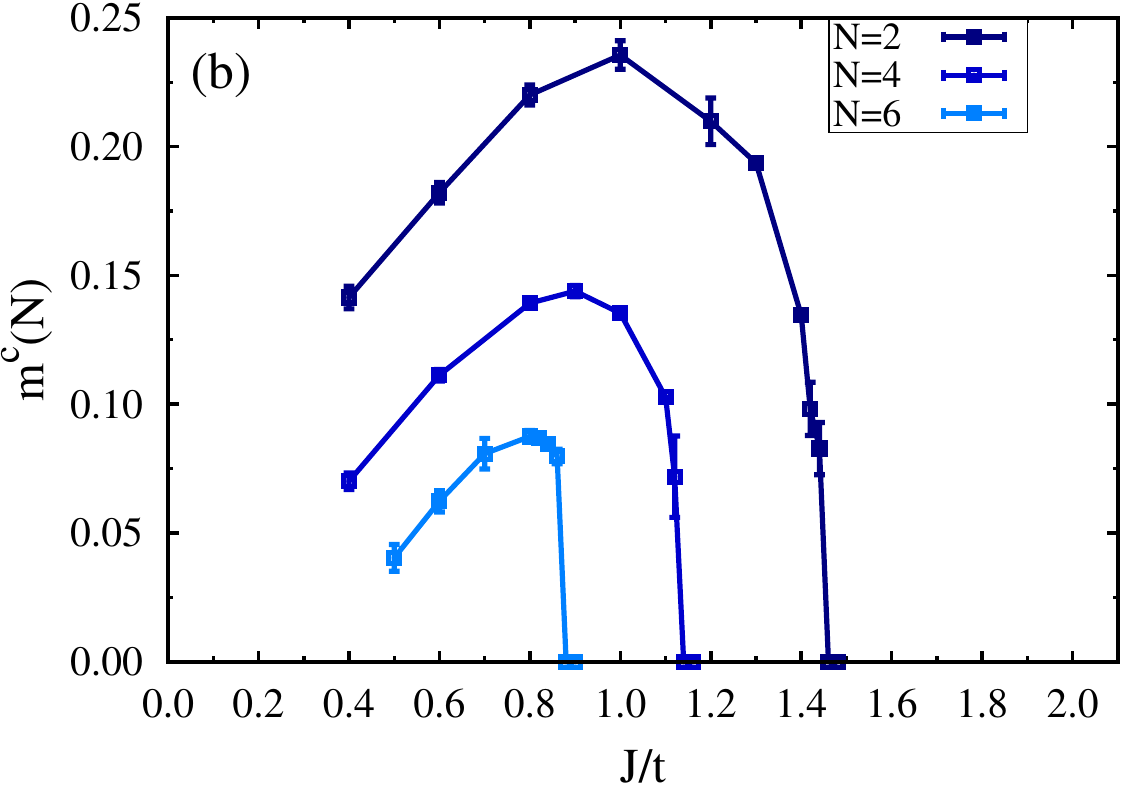}
\includegraphics[width=0.45\textwidth]{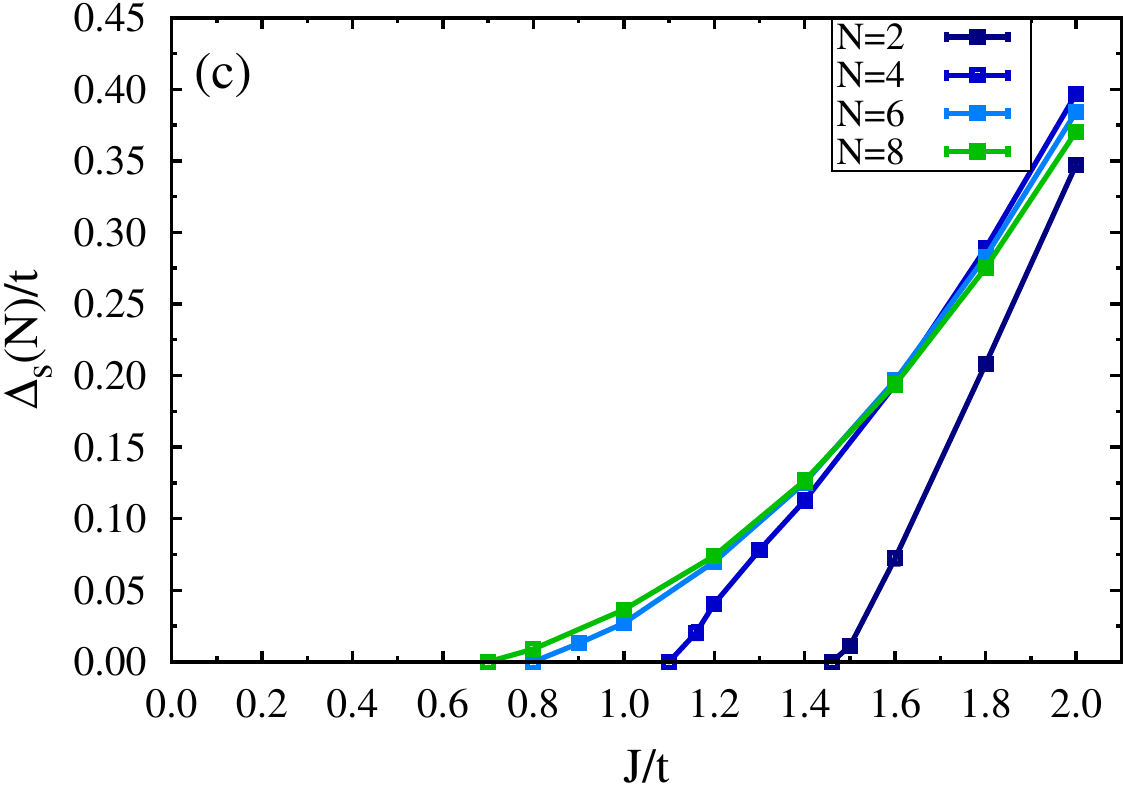}
\end{center} 
\caption
{
Staggered moment $m_{}^{\alpha}$ of the: (a) $f$-electrons and (b) $c$-electrons as well as (c) the spin gap $\Delta_s$ of the SU($N$) KLM  
after extrapolation the QMC data to the thermodynamic limit, see Appendices~\ref{app:moment} and \ref{app:spin}.
For $N=6$ with $J/t<0.5$ and for $N=8$ for all values of $J$, we were not able to distinguish  $m^c$ 
from zero. 
}
\label{Spin}
\end{figure}

We define the N\'eel state  for the SU($N$) quantum antiferromagnet  as:
\begin{equation}
\label{SUN_Neel.eq}
        | \Psi_{\text{N\'eel}} \rangle  = \prod_{\ve{i}\in A}  \hat{f}^{\dagger}_{\ve{i},1} \cdots   \hat{f}^{\dagger}_{\ve{i},N/2}  
        \prod_{\ve{i} \in B} \hat{f}^{\dagger}_{\ve{i},N/2+1} \cdots   \hat{f}^{\dagger}_{\ve{i},N}  | 0\rangle.
\end{equation}
For the square lattice, we can split the lattice  into two sublattices    A and B   such that the nearest neighbors of  one sublattice 
belong to the other.  For this N\'eel state, one will show that  for $\ve{i} \ne \ve{j} $ 
\begin{equation}
        \frac{4}{N} \sum_{\nu,\mu} \langle  \Psi_{\text{N\'eel}} | 
\hat{S}^{f,\mu}_{\pmb{j},\nu}   \hat{S}^{f,\nu}_{\pmb{i},\mu}  | \Psi_{\text{N\'eel}}  \rangle =e^{i \ve{Q}\cdot \left( \ve{i} -  \ve{j} \right)}. 
\end{equation} 
We hence adopt the following definition of the spin-spin correlation function,  
\begin{equation}
     S^{\alpha}_{}(\pmb{Q})  =   \frac{4}{N} \frac{1}{L^2}  \sum_{\ve{i}, \ve{j}, \mu,\nu} e^{i \ve{Q}\cdot \left( \ve{i} -  \ve{j} \right)} \langle   
\hat{S}^{\alpha,\mu}_{\pmb{j},\nu}   \hat{S}^{\alpha,\nu}_{\pmb{i},\mu}  \rangle, 
\end{equation}
with $\alpha=c,f$.
To  pin down the nature of ground state of the SU($N$) KLM,  we compute  the staggered moment, 
\begin{equation}
m_{}^{\alpha} =  \sqrt{  \lim_{L\to \infty} \frac{S^{\alpha}(\pmb{Q})}{L^2}}.  
\label{moment}
\end{equation}
The corresponding finite-size scaling is presented in Appendix~\ref{app:moment} and the extrapolated values for localized (conduction) electrons are 
plotted versus $J/t$ in Fig.~\ref{Spin}(a) [Fig.~\ref{Spin}(b)], respectively. On the one hand, the QMC data confirms that increasing $N$ suppresses 
magnetism by shifting the magnetic order-disorder transition point from $J_c/t \simeq 1.47(1)$ in the SU(2) symmetric case to progressively lower values of $J/t$:  
for $N=4$ we find $J_c/t \simeq 1.13(1)$; for $N=6$ and $N=8$ we find the transition points at 0.87(1) and  0.73(1), respectively. 
In this respect, the effect of finite $N$ bears a similarity to that generated by geometric frustration, e.g., by next-nearest-neighbor
hopping $t'$~\cite{Martin08,Bercx10}. 
On the other hand, the data in Fig.~\ref{Spin}(a) suggests that in the magnetically ordered phase, the $f$-local moment $m^f$ remains large since 
up to $N=8$ it exceeds 80\% of the N\'eel value. Furthermore, while $m^f$ seems to grow continuously below $J_c$ at $N=2$, 
one finds a rapid buildup of the $f$-local moment for larger $N$.

Once the magnetic order disappears at $J_c$, the ground state is expected to develop a finite spin gap $\Delta_s(\pmb{q})$, i.e., the energy difference between the 
singlet $S=0$ ground state and  the lowest excited spin triplet $S=1$ state with momentum $\pmb{q}$. To compute $\Delta_s(\pmb{q})$ without resorting to 
analytic continuation,  we consider the imaginary-time displaced spin correlation functions, 
\begin{equation}
S(\pmb{q},\tau) = \sum_{\mu,\nu}\langle \hat{S}^{\mu}_{\nu}(\pmb{q},\tau) \hat{S}^{\nu}_{\mu} (-{\pmb{q}})\rangle,  
\label{S_tau}
\end{equation}
where $\hat{S}^{\mu}_{\nu}(\pmb{q},\tau)= \hat{S}^{f,\mu}_{\nu}(\pmb{q},\tau) + \hat{S}^{c,\mu}_{\nu}(\pmb{q},\tau)$ is the total spin. 
The spin gap $\Delta_s(\pmb{q})$ can be extracted  from the asymptotic behavior of $S(\pmb{q},\tau)$ at  $\tau t \gg 1$ since 
$S(\pmb{q},\tau) \propto \exp \left( -\tau  \Delta_s(\pmb{q}) \right)$. Here, we focus on the AF wavevector $\pmb{Q}=(\pi,\pi)$, 
i.e., $\Delta_s \equiv {\rm min}_{\pmb{q}} \Delta_s(\pmb{q}) = \Delta_s(\pmb{Q})$.
A linear extrapolation of finite-size QMC estimates of $\Delta_s(N)$ to the thermodynamic limit, see Appendix~\ref{app:spin}, 
leads to the results plotted in Fig.~\ref{Spin}(c). 
For each $N$ we find that opening of the spin gap coincides with the vanishing of the magnetic moment.  
As shown in Fig.~\ref{Spin}(c), upon increasing $N$ the $J$-dependence of the spin gap approaches asymptotically 
the large-$N$ behavior $\propto e^{-t/J}$.

\begin{figure}[t!]
\begin{center}
\includegraphics[width=0.45\textwidth]{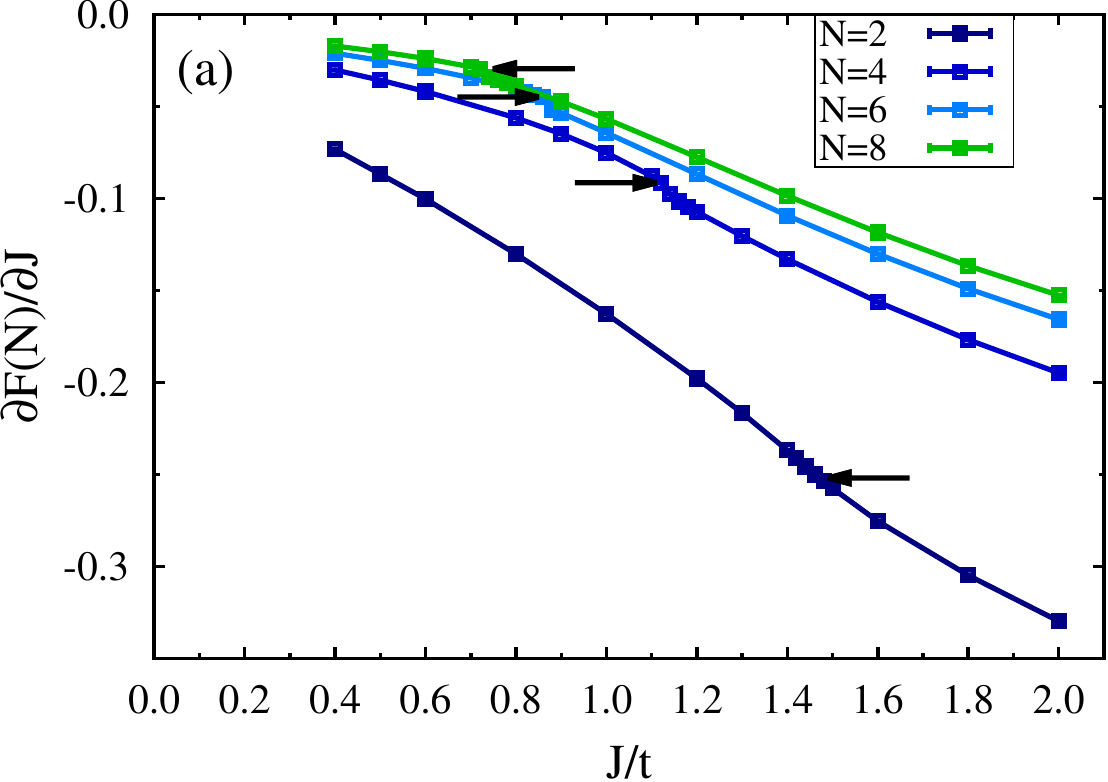}
\includegraphics[width=0.45\textwidth]{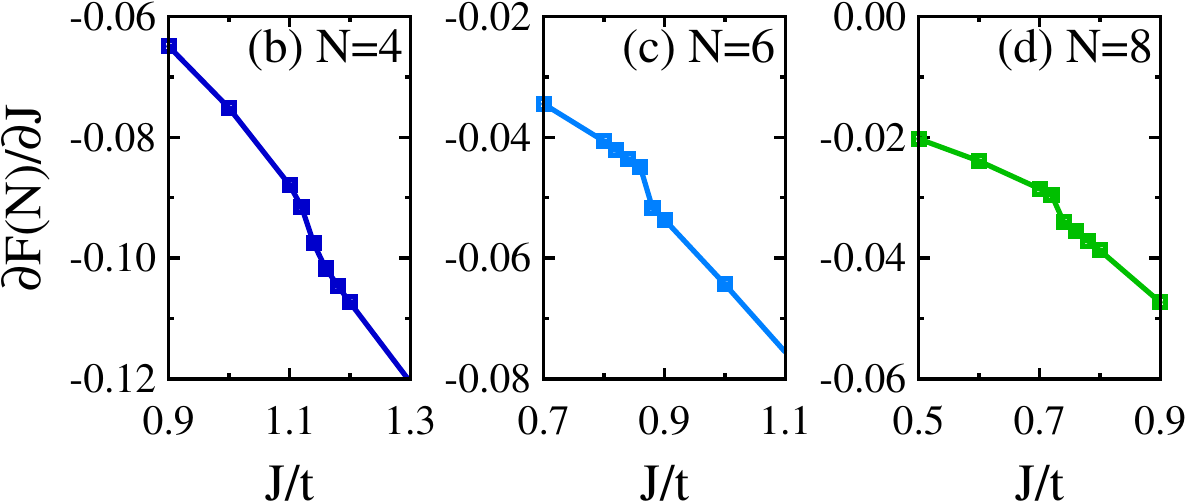}
\end{center}
\caption
{(a) Expectation value of the Kondo interaction term $\hat{\mathcal{H}}_{J}$  corresponding to the derivative of the free energy
with respect to $J$ obtained on the $14\times 14$ lattice. The arrows indicate magnetic order-disorder transition points.  
Bottom panels show a close-up around the transition point: (b) $N=4$, (c) $N=6$, and (d) $N=8$. }
\label{Scf}
\end{figure}

While the spin gap  evolves smoothly across the  transition,  the magnetization shows an abrupt change, especially at larger values of $N$. This poses the question of the 
nature of the  transition,  first order or continuous.    To provide more insight,   we plot in Fig.~\ref{Scf}  the free-energy derivative,
\begin{equation}
	  \frac{\partial F}{\partial J }    = \frac{1}{L^2} \frac{ \langle  \hat{\mathcal{H}}_{J}  \rangle }{J}, 
\end{equation}
obtained on our largest $14\times 14$ lattice. A progressively steeper evolution of this quantity across the transition point suggests that the  phase transition becomes 
first order upon increasing $N$.

\subsection{\label{charge} Charge degrees of freedom}

An important quantity of direct experimental relevance  is the QP residue $Z_{\pmb{k}}$. Indeed, 
since the effective QP mass $m^{*}\propto \frac{1}{Z_{\pmb{k}}}$, the behavior of $Z_{\pmb{k}}$ along the Fermi surface 
reveals how electron interactions modify properties of a metal. 
Given that  QMC simulations are restricted to the half-filled case, one possibility to get insight into properties of 
the metallic state at small doping is to consider the problem of a single-hole doped into the insulating phase. Then,  
assuming a rigid band scenario, one can estimate the QP residue 
$Z_{\pmb{k}} = \left|\langle \Psi_0^n | c^{\dagger}_{\pmb{k},\sigma} | \Psi_0^{n-1} \rangle  \right|^{2}$ 
of the doped hole at momentum $\pmb k$ together with the corresponding QP gap 
$\Delta_{qp}(\pmb{k})=E_0^{n}-E_0^{n-1}(\pmb{k})$, directly from the long-time behavior of the imaginary-time Green's function: 
\begin{equation} 
\label{Green}
        G(\pmb{k},\tau) = \frac{1}{N}\sum_{\sigma}
\langle \Psi_0^n | c^{\dagger}_{\pmb{k},\sigma}(\tau) c^{}_{\pmb{k}, \sigma} | \Psi_0^n \rangle  
\stackrel{\tau \to \infty}{\to}  Z_{\pmb{k}}e^{-\Delta_{qp}(\pmb{k}) \tau }. 
\end{equation}
Here,  $E_0^n$ is the ground-state energy at half-filling with $n$ particles while $E_0^{n-1}(\pmb{k})$ corresponds 
to the energy eigenstate with momentum $\pmb{k}$ in the $n-1$ particle  Hilbert space.
Typical raw data of $G(\pmb{k},\tau)$ from QMC simulations with different system sizes $L$ and the extrapolation to the thermodynamic 
limit of finite-size estimates of $\Delta_{qp}\equiv {\rm min}_{\pmb{k}} \Delta_{qp}(\pmb{k}) = \Delta_{qp}(\pmb{k}=(\pi,\pi))$  
and $Z_{(\pi,\pi)}$ are presented  in Appendix~\ref{app:delta}.

\begin{figure}[t!]
\begin{center}
\includegraphics[width=0.45\textwidth]{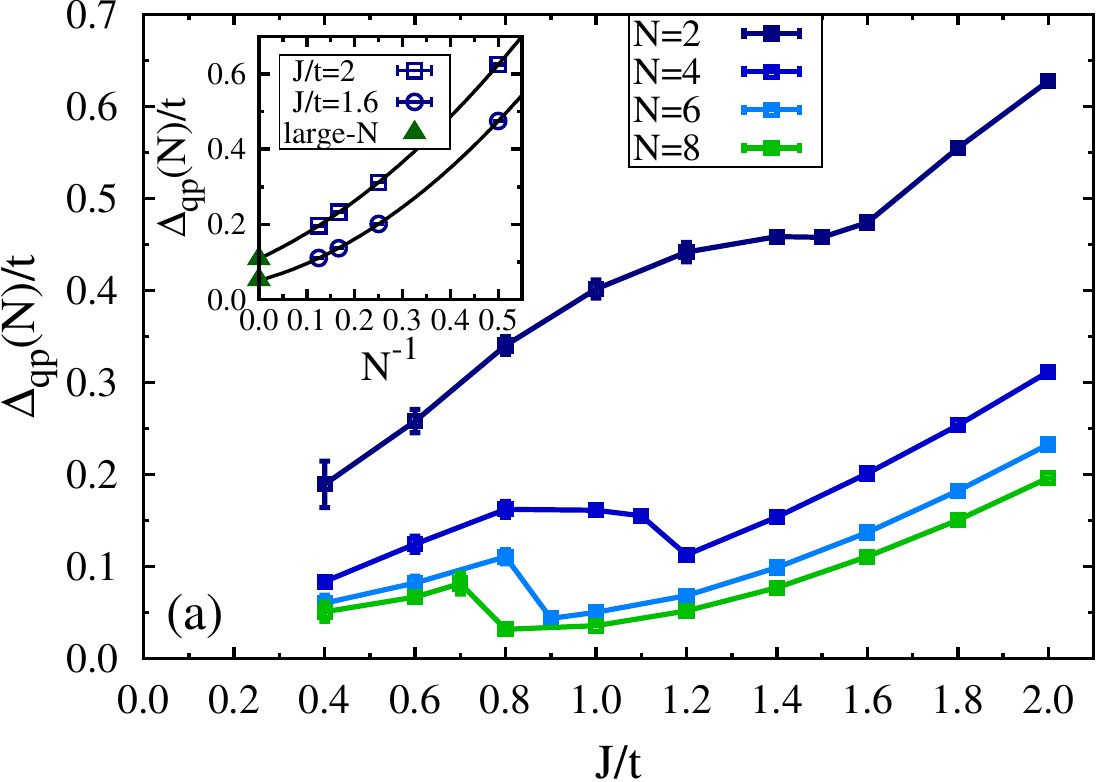}
\includegraphics[width=0.45\textwidth]{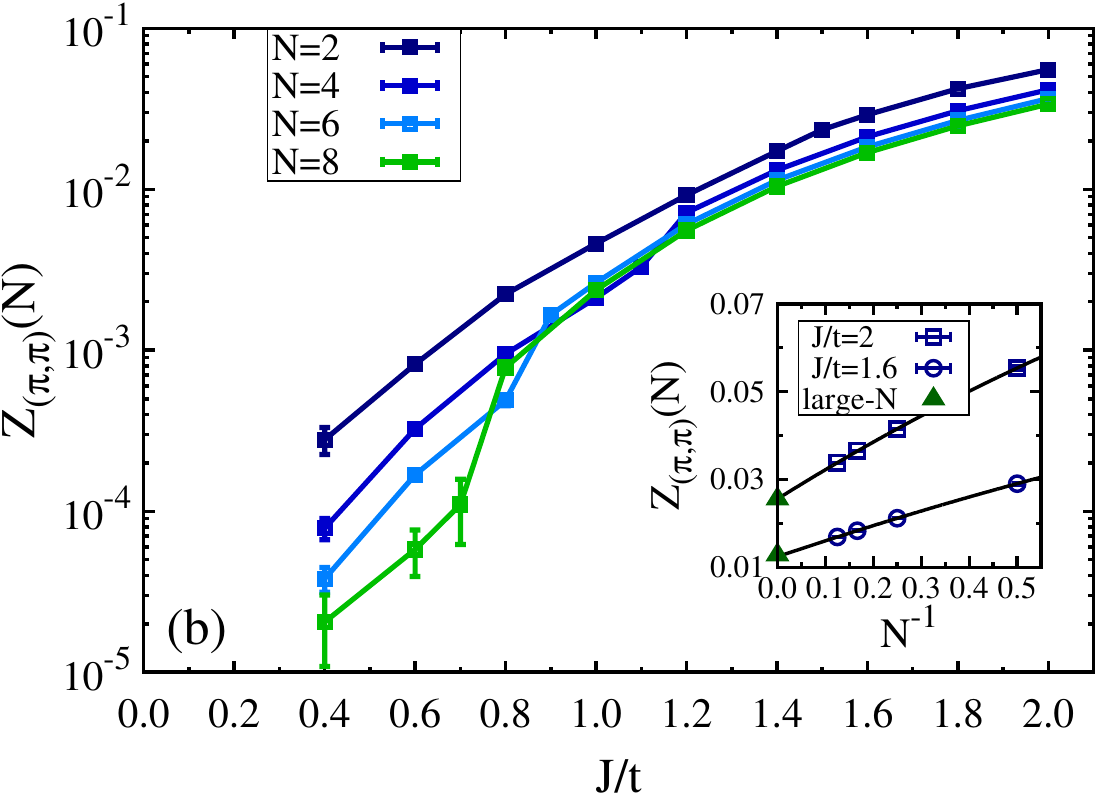}
\end{center}
\caption
{(a) Single-particle gap $\Delta_{qp}$ at momentum $\pmb k=(\pi,\pi)$  and (b)  the corresponding QP residue $Z_{(\pi,\pi)}$ 
after extrapolation the QMC data to the thermodynamic limit, see Appendix~\ref{app:delta}.
Insets show second-order polynomial fits to the QMC data in order to extract $\Delta_{qp}$ and  $Z_{(\pi,\pi)}$  in the $N\to\infty$ limit; 
the extrapolated values $\Delta_{qp}^{N\to\infty}=0.109(3)$ and  $Z_{(\pi,\pi)}^{N\to\infty}=0.0256(3)$  at $J/t=2$ 
and $\Delta_{qp}^{N\to\infty}=0.051(2)$  and $Z_{(\pi,\pi)}^{N\to\infty}=0.0125(2)$ at $J/t=1.6$ match well those obtained using the 
large-$N$ approximation (triangles).    
}
\label{Charge}
\end{figure}

Figure~\ref{Charge} plots $\Delta_{qp}$  and $Z_{(\pi,\pi)}$ as a function of $J/t$.   We first discuss the evolution of these quantities in the paramagnetic phase.   
As shown in the  insets of Fig.~\ref{Charge}    both quantities evolve smoothly from $N=2$ to $N=\infty$. 
The fact that we are able to recover the large-$N$ results  by extrapolating QMC data obtained by handling the constraint of no double 
occupancy on the localized  $f$-electron orbitals numerically exactly, validates the large-$N$ approximate treatments of the constraint and
confirms that the large-$N$ theory is the correct saddle point of the SU(2) KLM.
Furthermore, by comparing the $N$-dependence of  the single-particle gap $\Delta_{qp}$ in Fig.~\ref{Charge}(a) with that of the spin gap $\Delta_s$ in Fig.~\ref{Spin}(c), 
we conclude that upon increasing $N$ both quantities evolve in  the paramagnetic phase (i.e. $J/t = 2$)  to the asymptotic limit 
$\Delta_s=\Delta_c$ with $\Delta_c=2\Delta_{qp}$ being the charge gap, corresponding to the band insulator in the noninteracting 
periodic Anderson model.  Our results hence provide a text-book numerical demonstration that the $N=2$ Kondo lattice in the paramagnetic phase  is adiabatically connected 
to the large-$N$ saddle point.

Across the magnetic transition,   $\Delta_{qp}$  and $Z_{(\pi,\pi)}$    show a very strong  $N$ dependence. 
In contrast to the $N=2$ case with a smooth  evolution of both quantities  across the quantum critical point at $J_c/t \simeq 1.47$, a nonmonotonic behavior of the single-particle 
gap accompanied by an abrupt reduction of the QP weight on the magnetically ordered side is apparent. 
Although $Z_{(\pi,\pi)}$  shows an abrupt change, it remains finite in the magnetic phase. Hence we find the continued existence of  the heavy-fermion band for all the considered 
values of $N$ down to the smallest $J$.  Assuming a rigid-band scenario this implies that, in contrast to an Ansatz with frozen $f$-spins, 
the emergent heavy-fermion metal at small coupling $J$ is characterized by   a large Fermi surface containing both conduction and localized electrons (see Sec.~\ref{sec:discuss} ). 
As a consequence,  the coherence  temperature  is expected to drop abruptly  across the transition. 
On the magnetically ordered side, the QP gap tracks $J/N$ in the small $J$ limit. 
Finally, a notable feature in Fig.~\ref{Charge}(a) is a broad plateau in the $J$-dependence 
of $\Delta_{qp}$ at $N=4$. 
It is interesting to point out that a similar  plateau was found in quantum cluster theories allowing for SU(2) symmetry-breaking AF order~\cite{Bercx10,Lenz17}
as well as in the bond fermion theory~\cite{Eder18}.

To provide a  theoretical framework for the above,  we introduce in  Appendix \ref{app:SDW}  a mean-field theory.     Here we describe how this mean-field theory  and fluctuations around the 
corresponding saddle point  can account for the QMC results. Our numerics shows that  for any fixed value of $N$   the  paramagnetic state  is unstable to an RKKY driven magnetic instability and that  
deep in the magnetic phase the $f$-local moment is next to saturated. The N\'eel  state of  Eq.~(\ref{SUN_Neel.eq}) is hence a  good starting point to  formulate a mean-field theory.   
This wave function breaks the U($N$) symmetry  down to U($N$/2) $\times $ U($N$/2).    The mean-field Hamiltonian  derived in Appendix~\ref{app:SDW}   possesses  a U($N$/2) $\times $ U($N$/2)  
symmetry and is a generalization of  the mean-field theory of Ref.~\cite{MF00}  that captures both Kondo screening and magnetism  to the SU($N$)  group. 
In the mean-field Hamiltonian  the RKKY interaction scales as $1/N$. As a consequence, and owing to the nesting properties of  the conduction-electron band, the magnetically induced QP gap   
scales as $J/N$.  Our QMC results support this.

Following  Ref.~\cite{Assaad03}  one can define a model Hamiltonian that reduces to the KLM model  at $N=2$,  that has the U($N$/2) $\times $ U($N$/2) symmetry  beyond $N=2$, and that reproduces 
the saddle point of   Eq.~(\ref{SDW_SUN.eq}) in the large-$N$ limit.  It is very tempting to interpret our QMC results in terms of fluctuations  --   that are suppressed as a function of $N$ -- around 
this magnetic saddle point.  In the limit $N \rightarrow \infty$  \cite{Capponi01},  and deep in the magnetically ordered phase, the $f$-spins are frozen and  $Z_{(\pi,\pi)}$  vanishes.

\subsection{\label{spin_dynamic} Spin excitation spectrum}

\begin{figure*}[t]
\begin{center}
\includegraphics[width=0.32\textwidth]{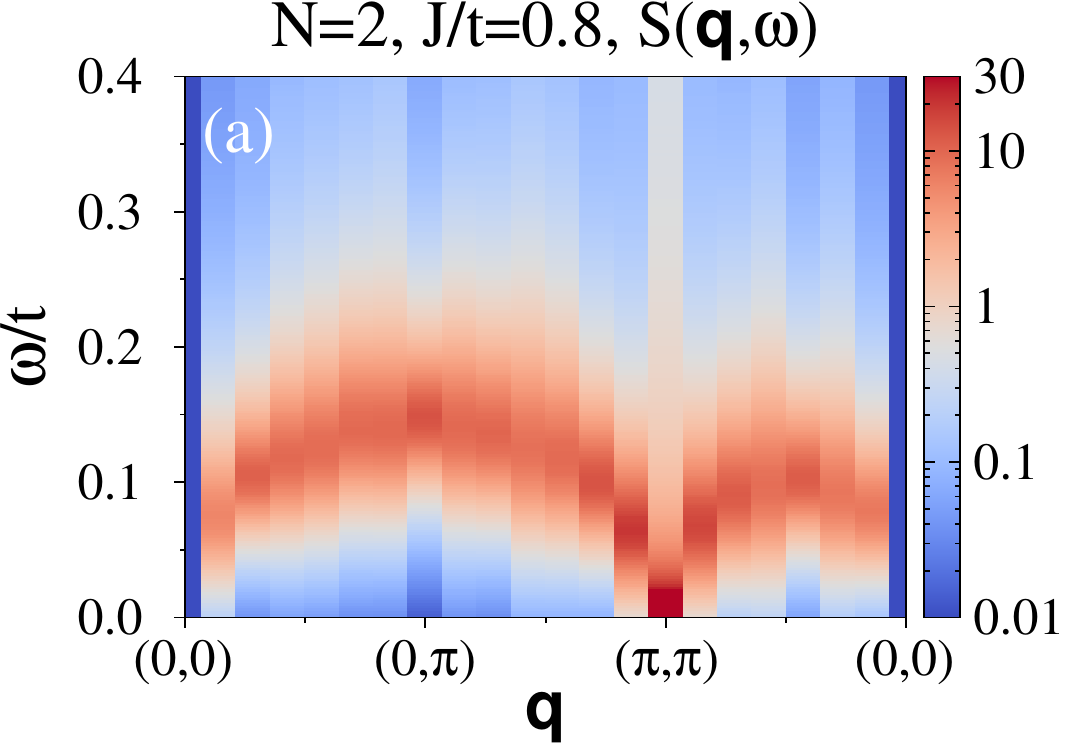}
\includegraphics[width=0.32\textwidth]{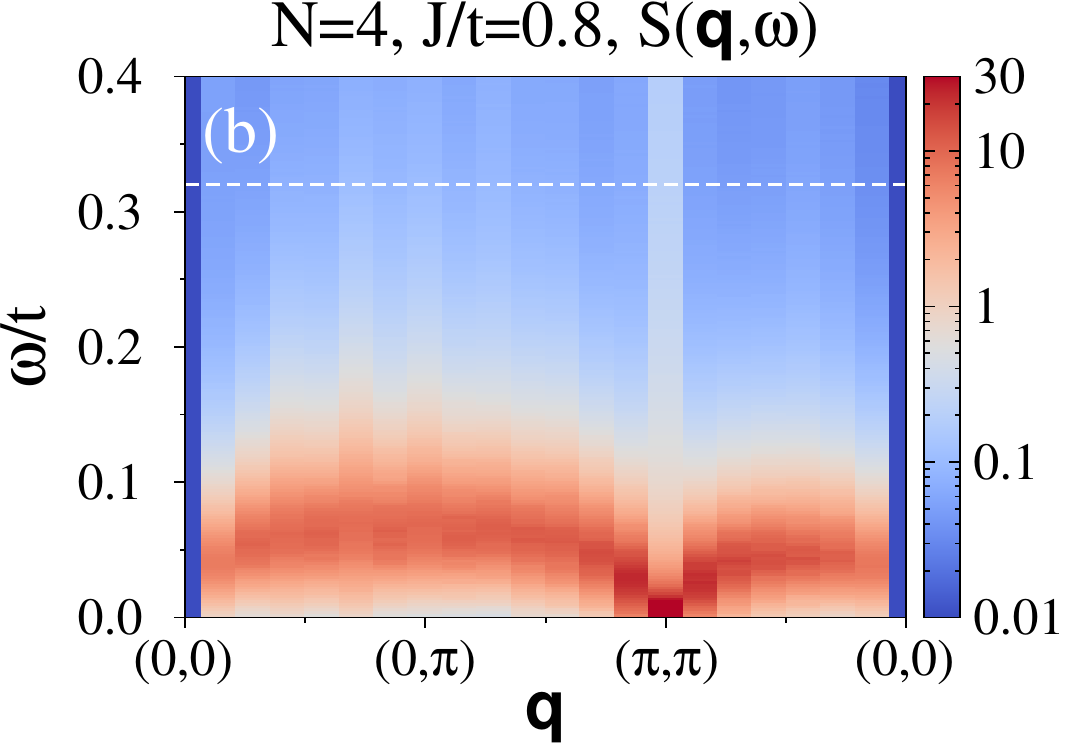}
\includegraphics[width=0.32\textwidth]{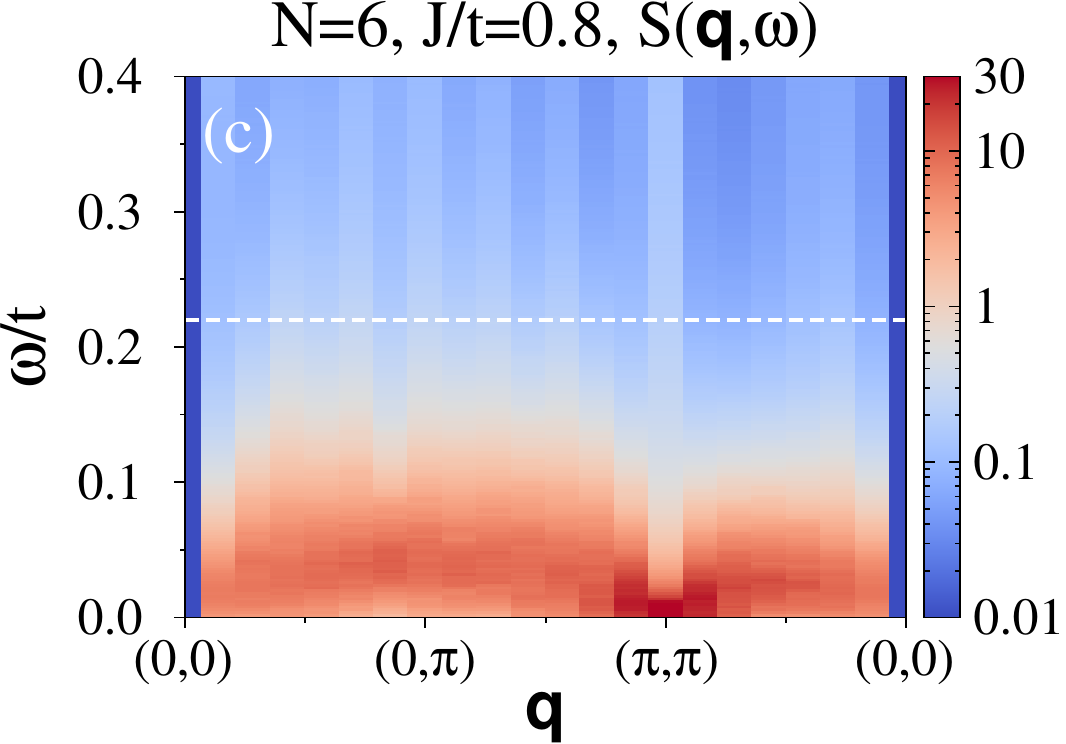}\\
\includegraphics[width=0.32\textwidth]{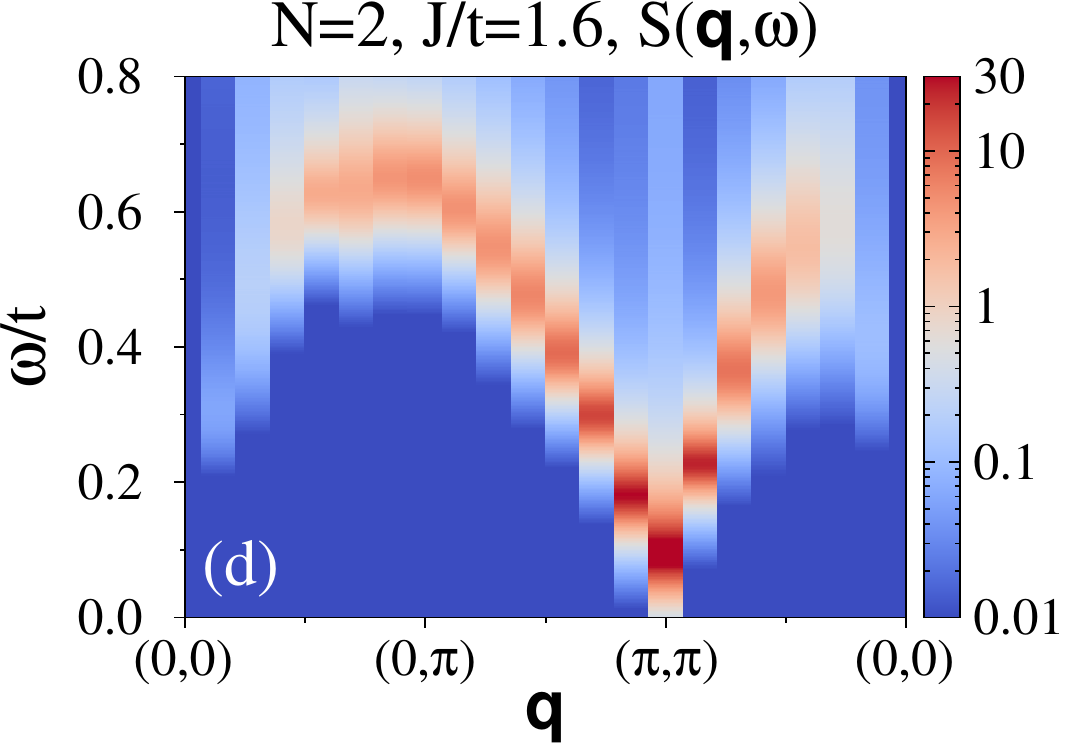}  
\includegraphics[width=0.32\textwidth]{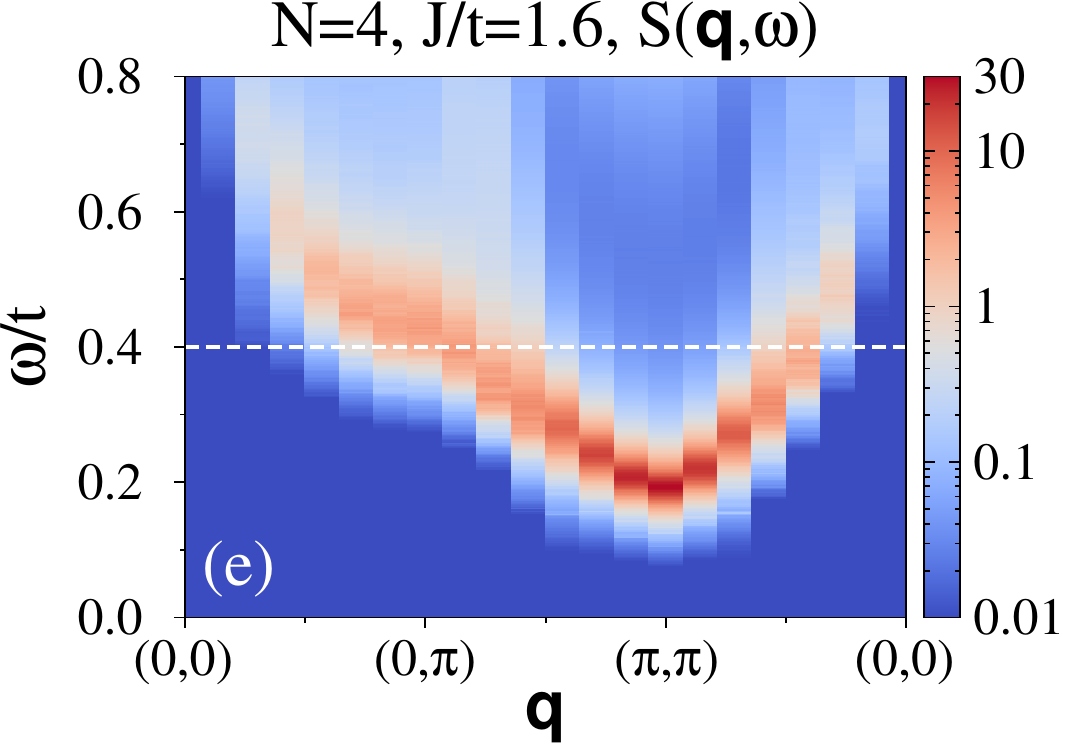}
\includegraphics[width=0.32\textwidth]{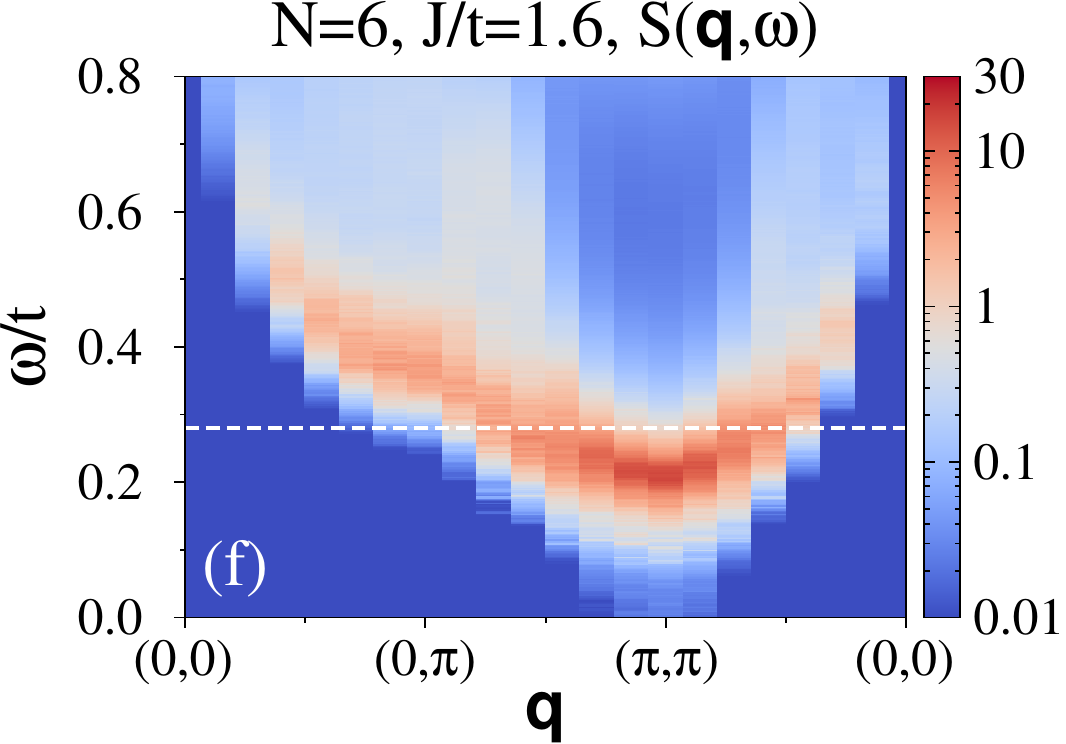}
\end{center}
\caption
{Dynamical spin structure factor $S(\pmb q,\omega)$ along a high symmetry path in the Brillouin zone in the: (a-c) AF phase at $J/t=0.8$  and (d-f)   KI phase at $J/t=1.6$
obtained on the $14\times 14$ lattice with increasing $N$ (from left to right).  
In panels (a-c), the particle-hole continuum threshold (dashed)  lies above the Goldstone modes $\omega_{\textrm{ph}}/t\simeq 0.68$, 0.32, and 0.22, respectively, 
while in panels (d-f) it drops from $\omega_{\textrm{ph}}/t\simeq 1$ through $0.4$ to $0.28$. } 
\label{Sq_fig}
\end{figure*}

To get further insight into the nature of AF and KI phases in the SU($N$) symmetric situation, we consider 
the dynamical spin structure factor $S(\pmb q,\omega)$. We have extracted this quantity from the QMC imaginary-time 
displaced spin correlation functions defined in Eq.~(\ref{S_tau}), 
\begin{equation}
S(\pmb{q},\tau) = \frac{1}{\pi} \int_0^{\infty} {\rm d} \omega\; e^{-\tau \omega} S(\pmb{q}, \omega),  
\label{Sq}
\end{equation}
by using the stochastic analytic continuation method~\cite{Beach04a}.     In the above, we consider the total spin.

The  spin-density-wave approximation presented in Appendix~\ref{app:SDW}  breaks explicitly the SU($N$) symmetry  and hence it fails to  capture Goldstone modes.  Since the N\'eel  state has 
the U($N/2$) $ \times$ U($N/2$)   symmetry  but the Hamiltonian has a  U($N$)   one, we expect a total of $ \text{dim }    \frac{ U(N) }  {   U(N/2) \times U(N/2) }    = N^2/2 $    
Goldstone modes~\cite{Goldstone62,Watanabe19} that should become apparent in the dynamical  spin structure  factor. 
One expects that  this large number of Goldstone modes  will destabilize the ordering and in this respect, it is interesting  to see that in the KLM an AF state can be stabilized for each $N$. 
Concerning the energetics,  and as argued in Appendix~\ref{app:scales}, the effective RKKY coupling $J_{\textrm{RKKY}}$ scales as $\tfrac{J^2 N(\epsilon_f) }{N}$  and the single-particle gap as 
$\tfrac{J}{N} $.   Hence in the \emph{small} $J$ limit, the Goldstone modes are expected  to be  located  well \emph{below} the particle-hole continuum.  

We demonstrate the above in  Figs.~\ref{Sq_fig}(a-c) where substantial slowing down of the spin-wave velocity $v_s\propto J_{\textrm{RKKY}}\sim \tfrac{J^{2} N(\epsilon_f) }{N}$ in the AF phase 
with $J/t=0.8$ is clearly seen for $N>2$.  For each $N$ considered in   Figs.~\ref{Sq_fig}(a-c),   the particle-hole continuum lies above the Goldstone modes  such 
that we should  interpret the data solely in terms of an  SU($N$) quantum spin model.   
Adopting this point of view,  the relevant energy scale is the spin-wave velocity that at fixed $J$ scales as $1/N$.  In terms of this energy scale, it becomes apparent that  
the  width of the dynamical spin structure factor at say wavevector  $\ve{q} = \left( 0, \pi \right) $  grows as a function of $N$.  We interpret  this as a consequence of  scattering 
between a growing  number  of Goldstone modes.  One should also mention that as a function of growing values of $N$, the distance to the magnetic order-disorder transition point is 
suppressed.    Although for all considered values of  parameters the magnetic moment is well developed, this could certainly play a role in the interpretation of the $N$-dependence of the spectrum.

Figures~\ref{Sq_fig}(d-f)   plot the dynamical spin structure factor at $J/t=1.6$  as a function of $N$.   At $N=2$, we are close to the quantum critical point, and  the triplon mode shows a 
minimal gap at  the AF wavevector, ${\pmb Q}=(\pi,\pi)$.  Triplons will condense at the transition  to generate the magnetic order. 
 In this regime triplons are bound  electron-hole pairs and the binding originates from vertex corrections.   Enhancing $N$ from this point,  damps vertex corrections such that the bound  triplon 
mode will progressively  merge  in the particle-hole  continuum.   Precisely  this effect is seen in Figs.~\ref{Sq_fig}(d-f).

\subsection{\label{charge_dynamic} Single-particle spectral function}

\begin{figure*}[t!]
\begin{center}
\includegraphics[width=0.32\textwidth]{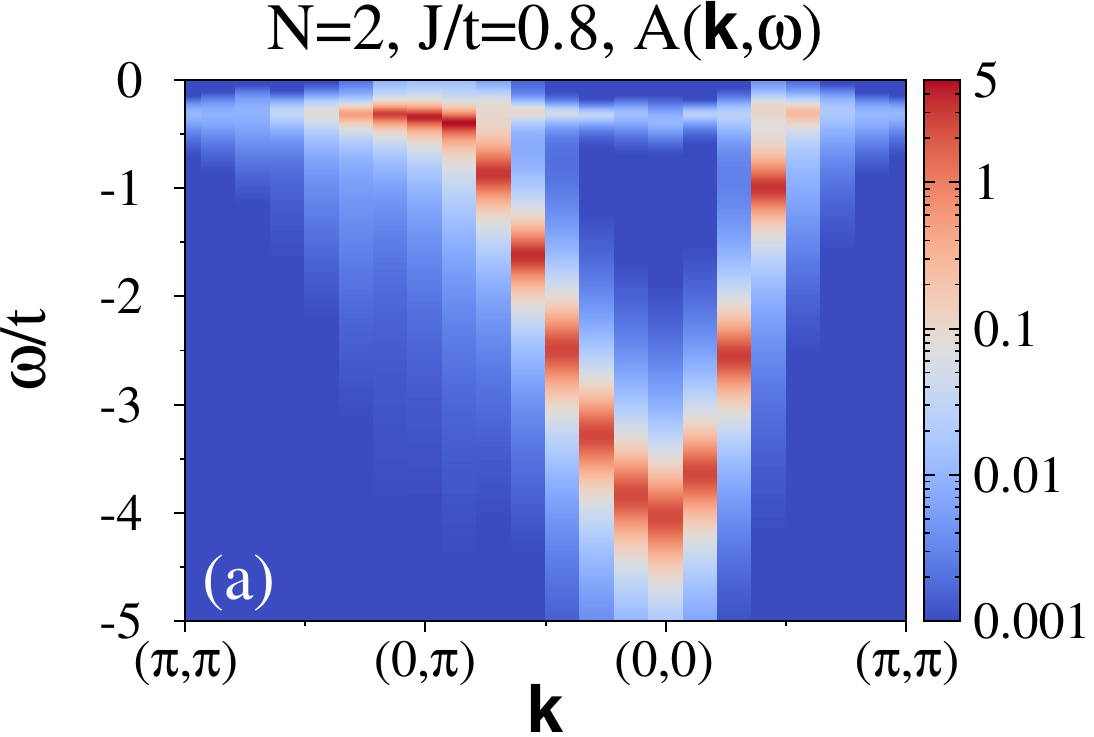}
\includegraphics[width=0.32\textwidth]{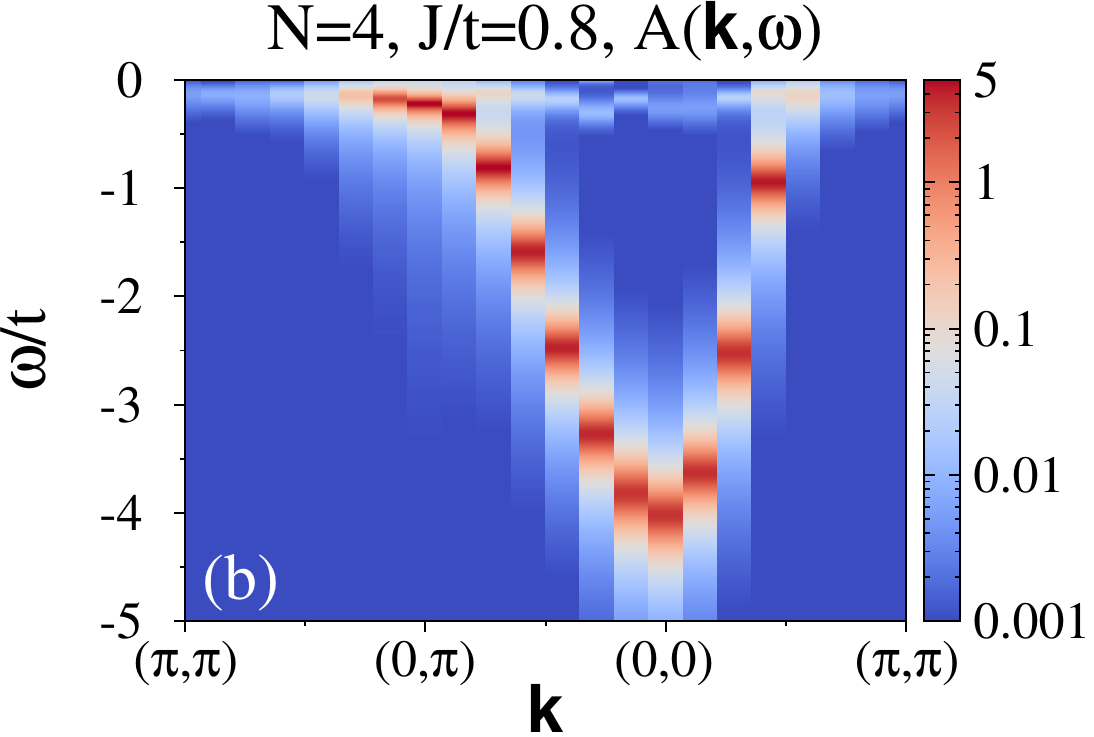}
\includegraphics[width=0.32\textwidth]{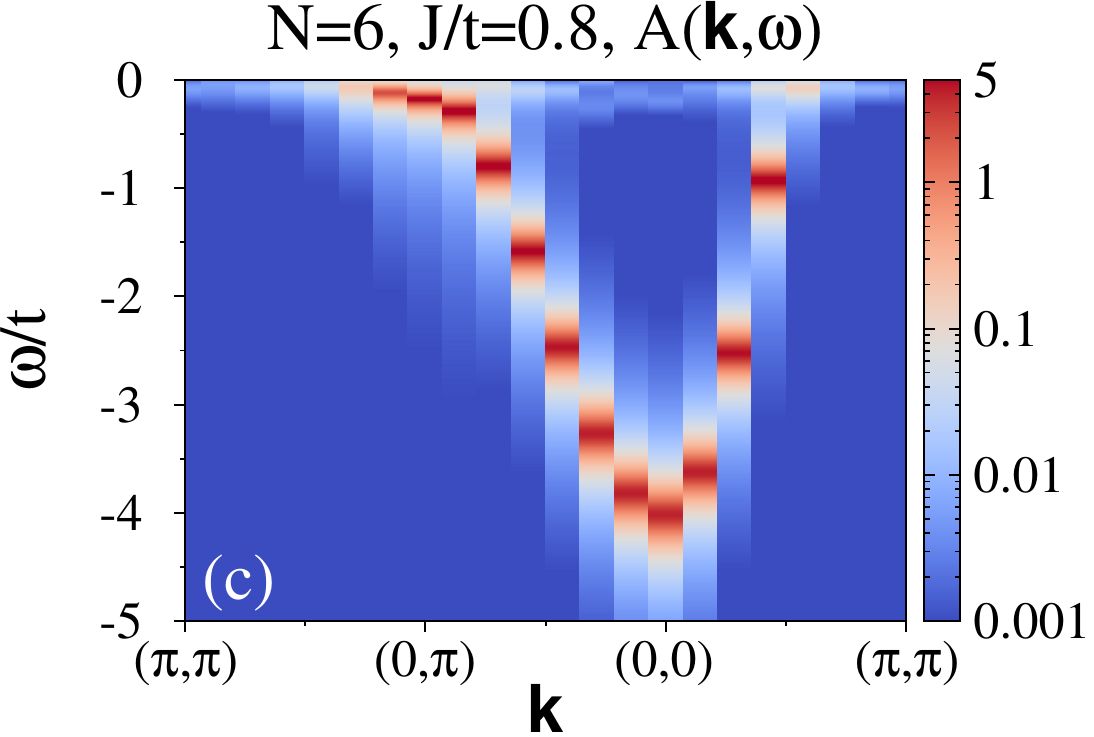}\\
\includegraphics[width=0.32\textwidth]{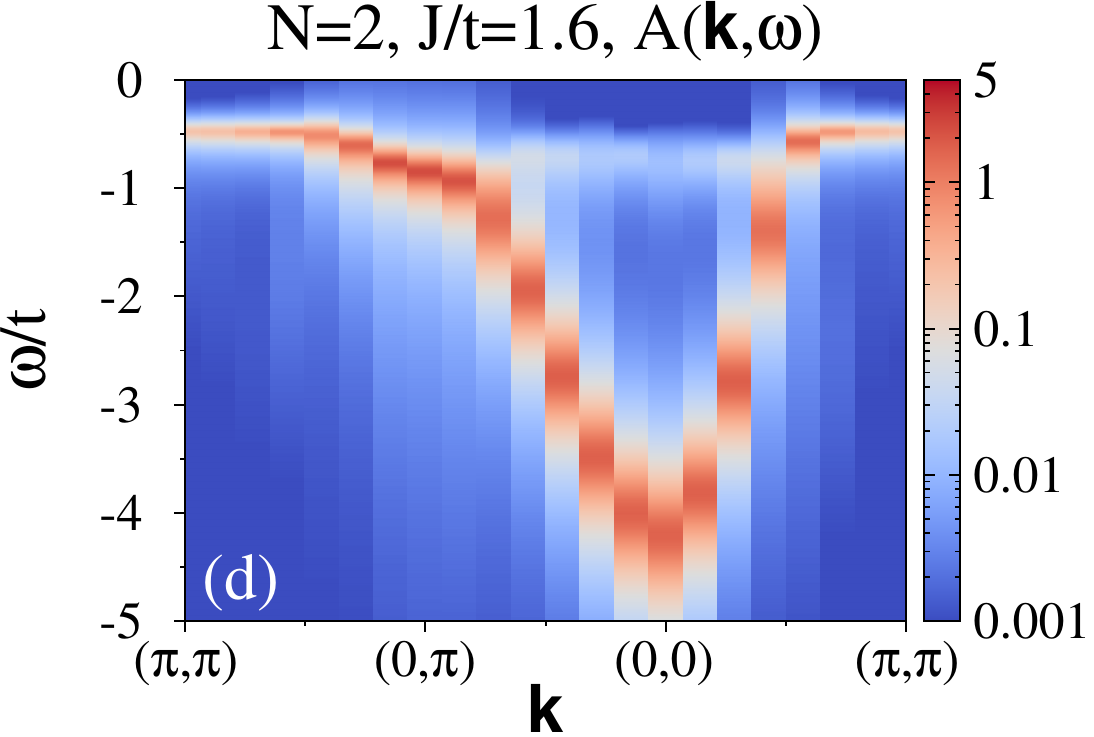}
\includegraphics[width=0.32\textwidth]{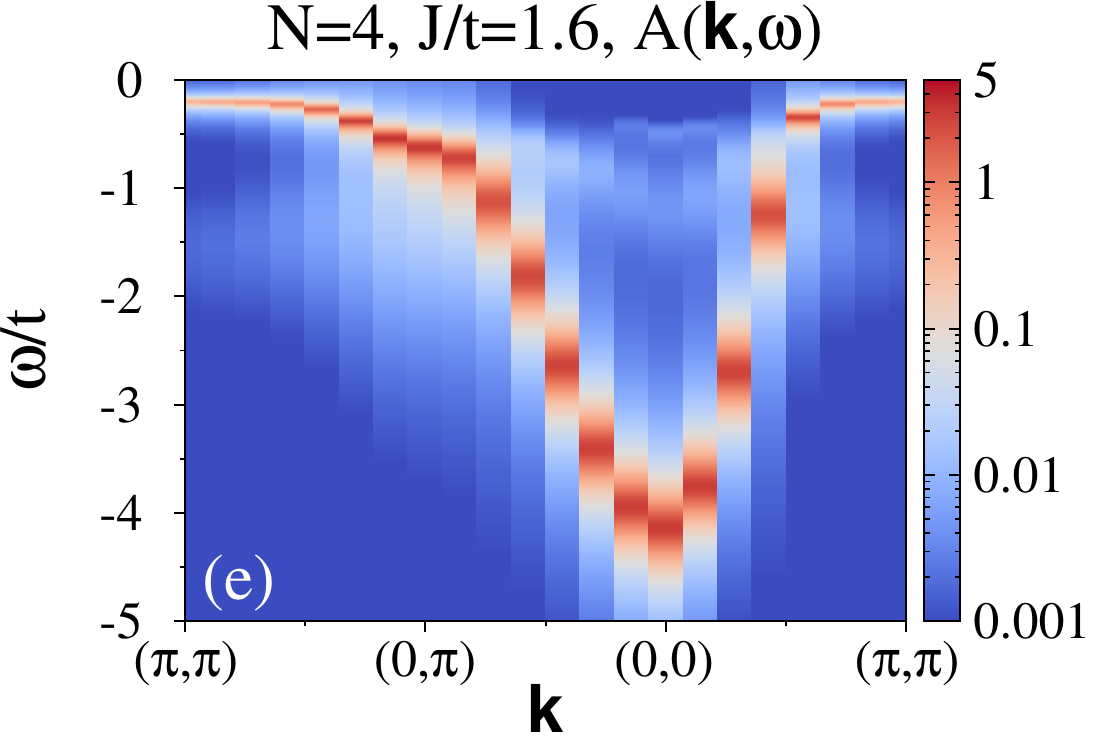}
\includegraphics[width=0.32\textwidth]{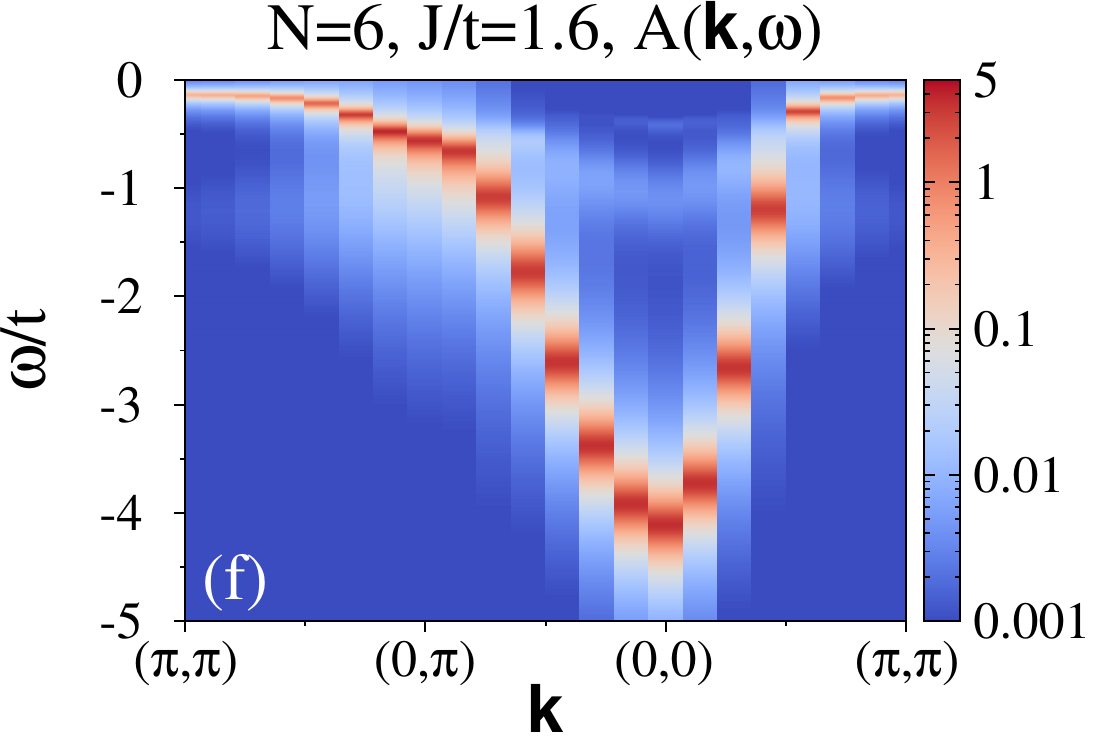}
\end{center}
\caption
{Single-particle spectral function $A(\pmb k,\omega)$  of the conduction electrons in the: (a-c) AF phase at $J/t=0.8$ 
and  (d-f)  KI phase at $J/t=1.6$  obtained on the $14\times 14$ lattice with increasing $N$ (from left to right).   
}
\label{Ak_fig}
\end{figure*}

We move on to discuss in more detail  the single-particle dynamics. To this end,  we have computed the single-particle spectral 
function $A(\pmb{k},\omega)$  of the conduction electrons. It  is related to the imaginary-time Green's function  defined in 
Eq.~(\ref{Green})  via:
\begin{equation}
       G(\pmb{k},\tau) = \frac{1}{\pi} \int_{0}^{\infty} {\rm d} \omega \; e^{-\tau \omega} A(\pmb{k}, -\omega).
\label{Ak}
\end{equation}
Again, we use the stochastic analytic continuation method to extract $A(\pmb{k},\omega)$.

The evolution of $A(\pmb{k},\omega)$ upon increasing $N$ in the AF phase at $J/t=0.8$ is shown in Figs.~\ref{Ak_fig}(a-c).
As expected for the half-filled case, all the spectra display a clear hybridization gap which, in agreement with findings in Sec.~\ref{charge},  
becomes gradually smaller at larger $N$. Furthermore, the spectral function features a flat heavy-fermion band extending to the 
$\pmb{k}=(\pi,\pi)$ point with relatively low spectral weight. The continued presence of this band around $\pmb{k}=(\pi,\pi)$  even  in 
the AF phase shows that the heavy fermions undergo a magnetic instability such that Kondo screening is still present in the ordered phase. 

A direct consequence of the magnetic ordering is a  back-folding of the Brillouin zone and the emergence of additional low-energy 
spectral feature around the $\pmb{k}=(0,0)$ momentum. It arises due to the scattering of the heavy QP off spin fluctuations with 
the AF wavevector $\pmb{Q}=(\pi,\pi)$  and thus it corresponds to the shadow of the band in the vicinity of the 
$\pmb{k}=(\pi,\pi)$ point.   As apparent, the shadow band becomes less pronounced  at larger $N$.   This can be traced back  to the combined effects  that the 
$Z$-factor at  $\pmb{k}=(\pi,\pi)$   drops as a function of $N$ and that the  magnetic moment is reduced as $N$ grows from $2$ to $6$ at $J/t=0.8$.

\begin{figure}[b!]
\begin{center}
\includegraphics[width=0.36\textwidth]{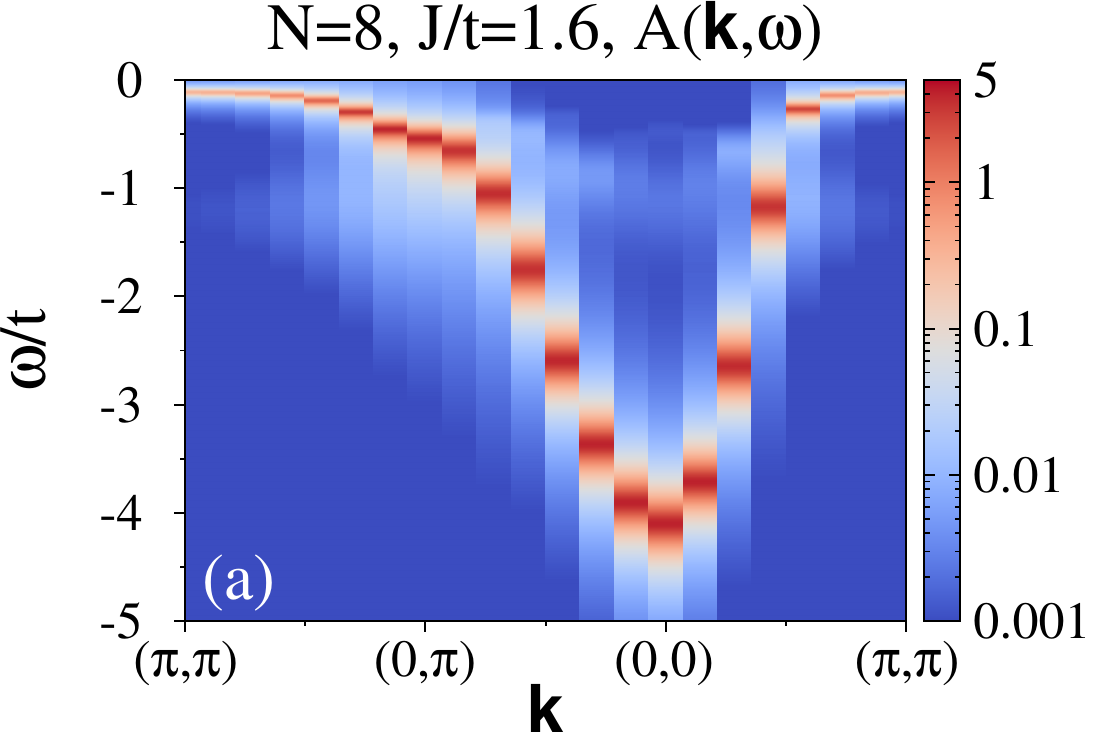}
\includegraphics[width=0.36\textwidth]{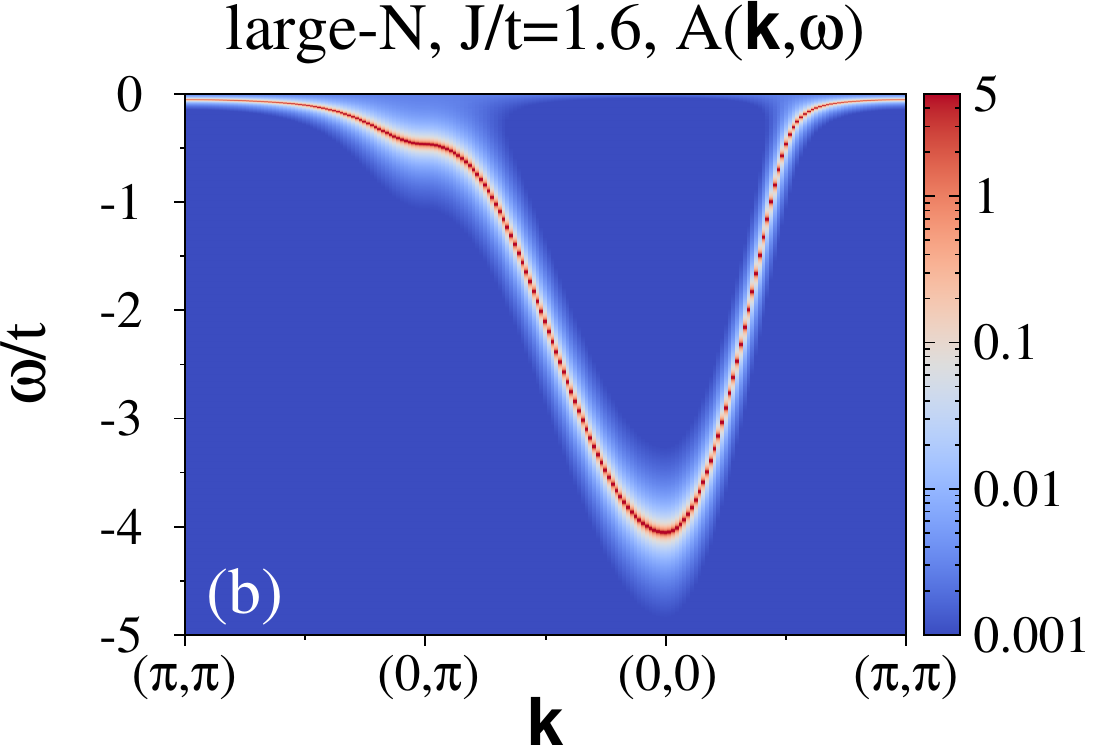}
\end{center}
\caption
{Single-particle spectral function $A(\pmb k,\omega)$ in the KI phase at $J/t=1.6$ from: (a) QMC simulations with $N=8$  and (b) the large-$N$ approach.
}
\label{Ak_cmp}
\end{figure}

In Figs.~\ref{Ak_fig}(d-f) we show the evolution of $A(\pmb{k},\omega)$ upon increasing $N$ in the KI phase at $J/t=1.6$. 
In the disordered phase with $N=2$, only a precursive feature of the shadow band is visible, see Fig.~\ref{Ak_fig}(d): 
despite a much larger spectral weight of the heavy QP band at $\pmb{k}=(\pi,\pi)$ point with respect to $J/t=0.8$, 
the precursive feature has relatively low intensity and it is shifted by the energy corresponding roughly to the spin gap $\Delta_s$.  
As shown in Figs.~\ref{Ak_fig}(e,f), this feature becomes broad and  consequently more difficult to resolve at larger $N$.  This can be traced back to the fact that as a function of 
$N$, the triplon mode approaches the particle-hole continuum broadens, and ultimately  disappears.

Finally, in Fig.~\ref{Ak_cmp} we plot $A(\pmb k,\omega)$ in the KI phase at $J/t=1.6$ for our largest $N=8$ together with that 
obtained in the large-$N$ approach. As apparent, the large-$N$ approximation produces a single-particle spectrum  which compares favorably 
with the  QMC spectral function.    One of the key properties of the large-$N$ self-energy, is its locality: 
\begin{equation}
	\Sigma(\ve{k}, i \omega_m) = \frac{ (JV)^2 }{ 4  i \omega_m}.
\end{equation} 
Thereby,   despite all the caveats of the   large-$N$ approximation -- finite hybridization order parameter which breaks the local gauge symmetry -- it can be considered 
to be well suited to account for the essence of Kondo screening deep in the KI phase.

\begin{figure*}[t!]
\begin{center}
\includegraphics[width=0.45\textwidth]{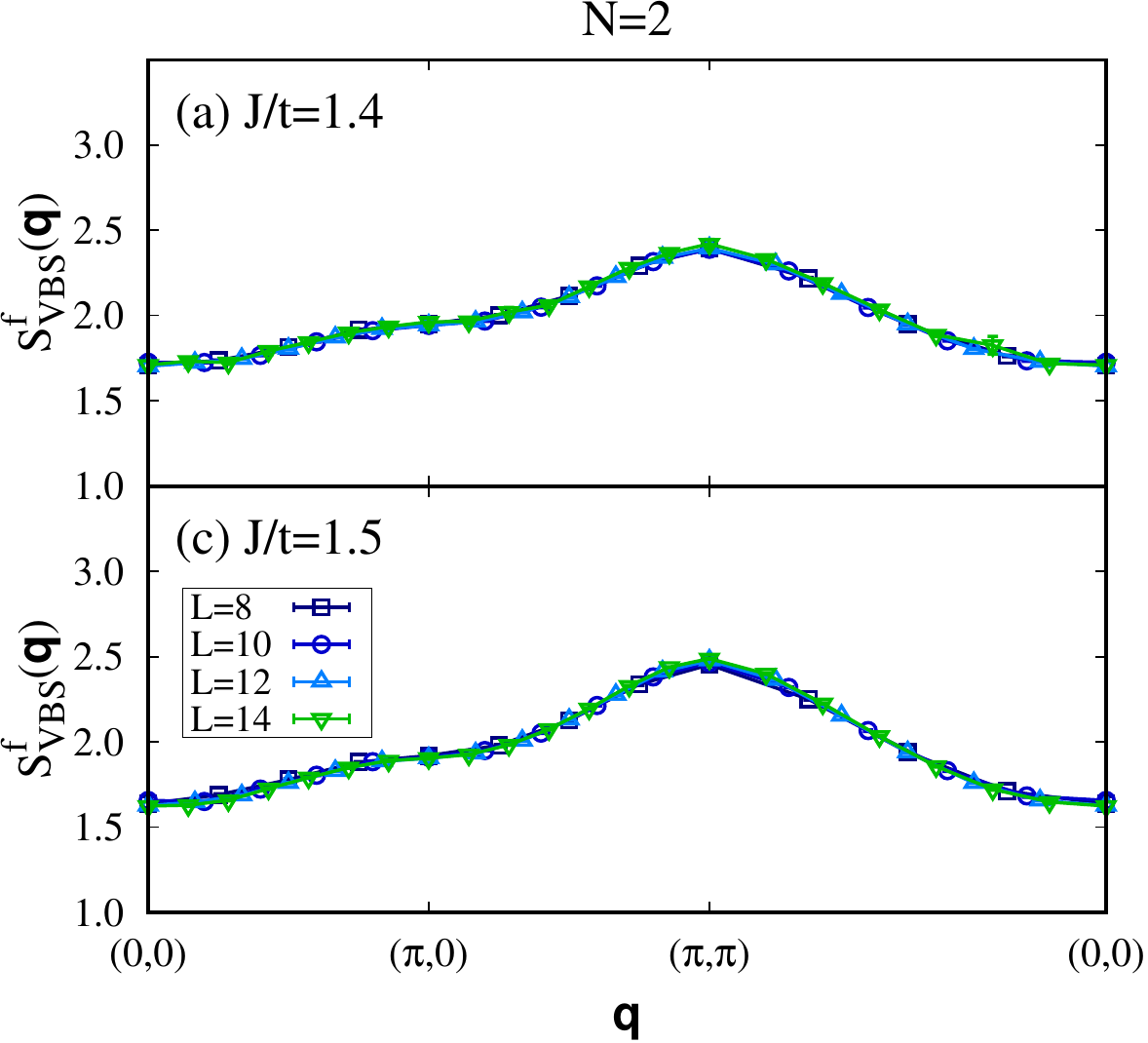}
\includegraphics[width=0.45\textwidth]{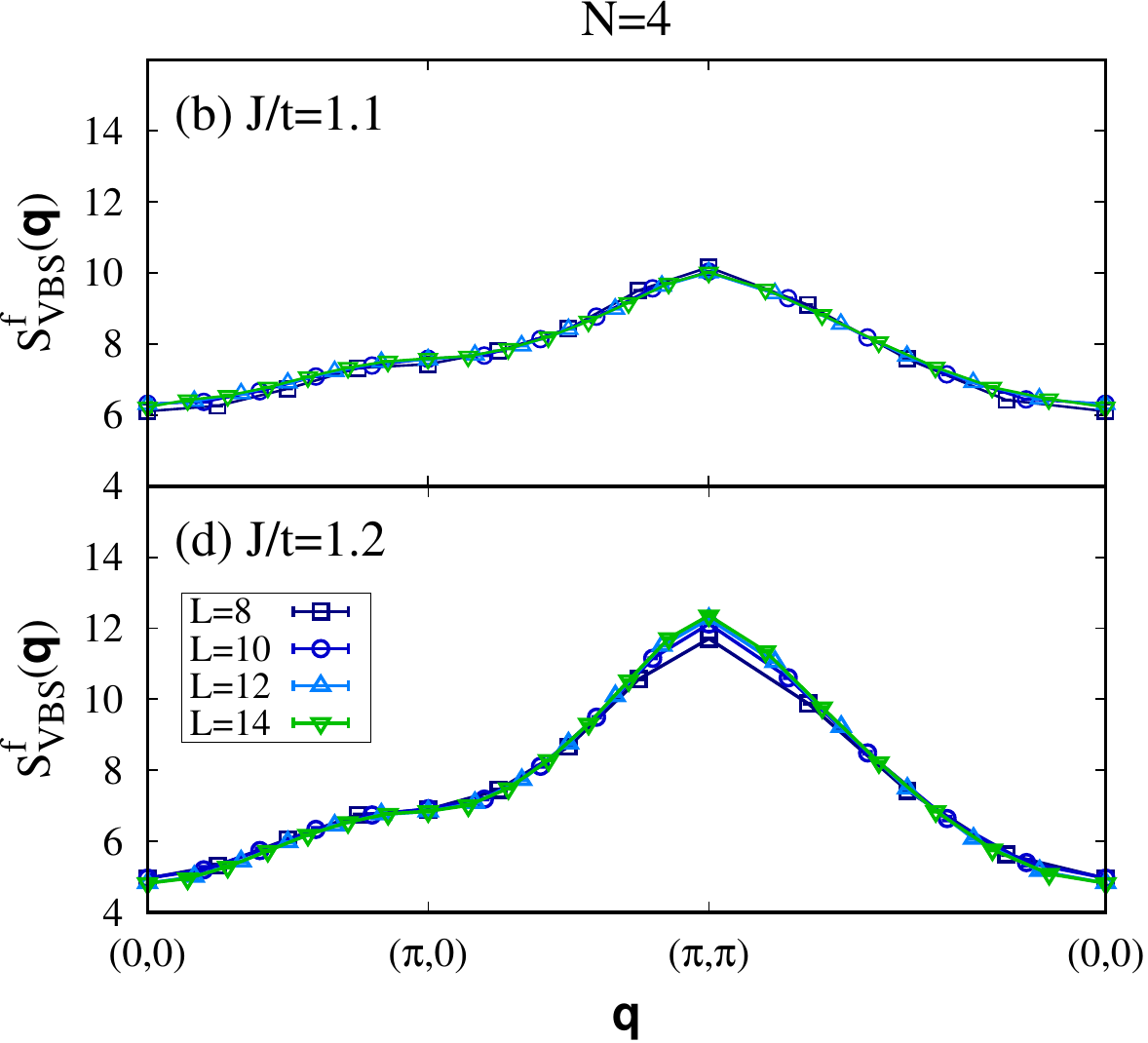}
\end{center}
\caption
{VBS correlation function $S^f_{\rm VBS}(\pmb{q})=\sum_{\pmb{\delta}}[S^f_{\rm VBS}(\pmb{q})]_{\pmb{\delta},\pmb{\delta}}$
for the $f$-electrons for various lattice sizes measured across the magnetic order-disorder transition point in the:  (a,b) AF and (c,d)  KI phases for $N=2$ (left) and $N=4$ (right). 
}
\label{VBS}
\end{figure*}

\begin{figure*}[t!]
\begin{center}
\includegraphics[width=0.45\textwidth]{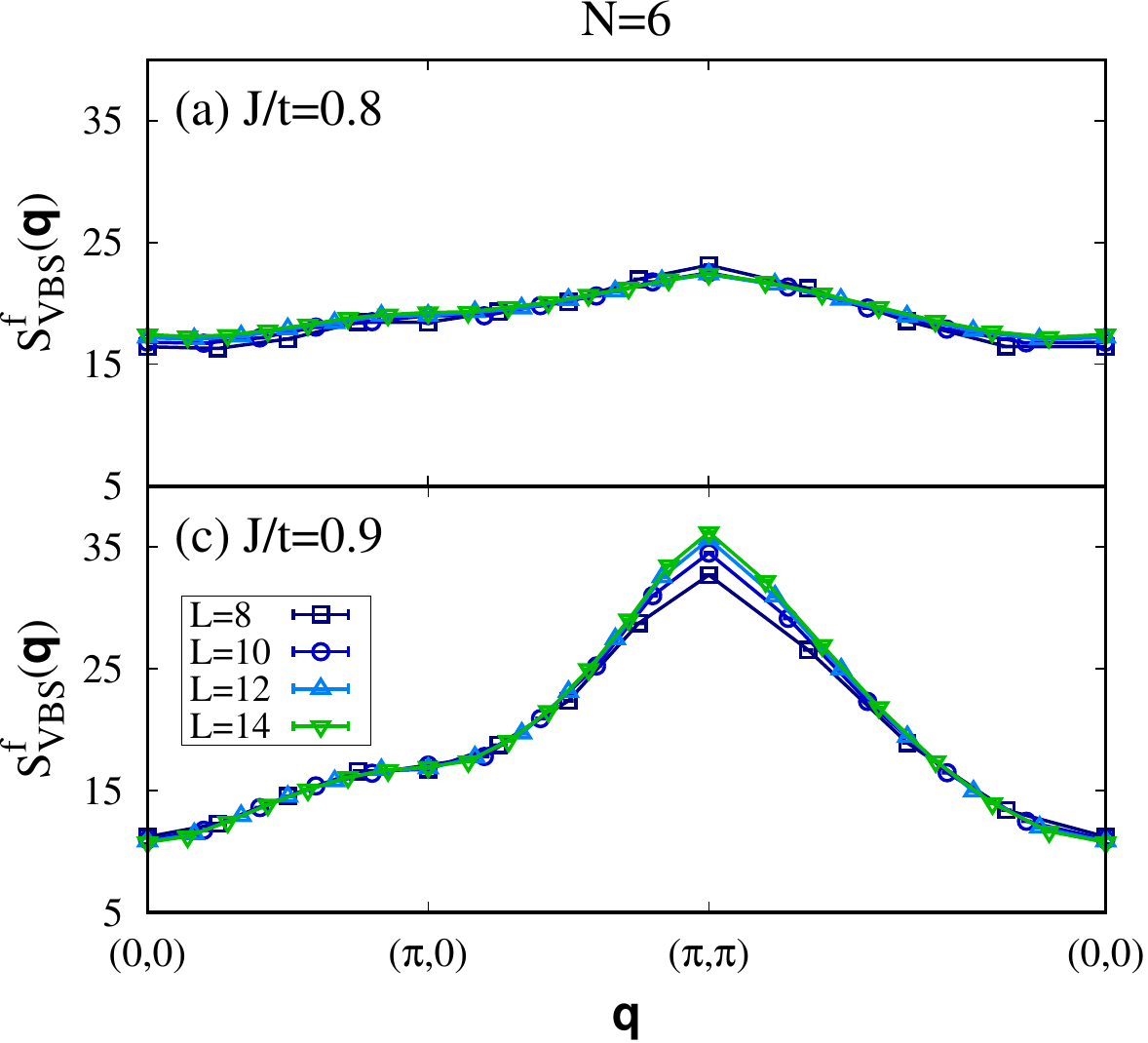}
\includegraphics[width=0.45\textwidth]{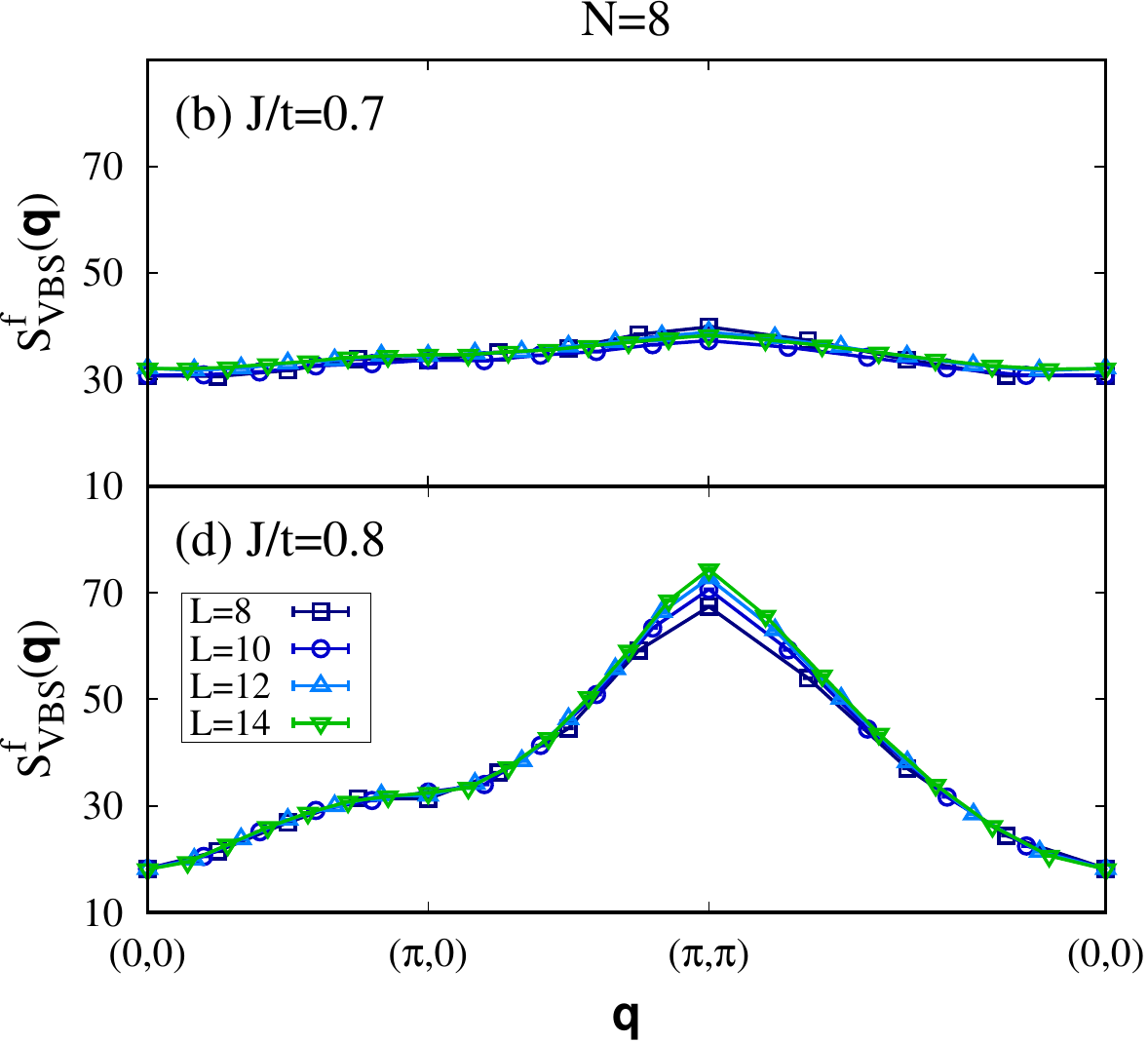}
\end{center}
\caption
{Same as in Fig.~\ref{VBS} but for $N=6$ (left) and $N=8$ (right).
}
\label{VBS2}
\end{figure*}

\subsection{\label{vbs} VBS correlation function}

Generically, enhancing  the symmetry group from SU(2) to SU($N$) leads to VBS orders.  To confirm the absence of this instability in  the SU($N$) KLM, we have computed the VBS correlation function 
for the $f$-spins: 
\begin{align}
    \left[ S^{f}_{\text{VBS}}  (\pmb{q})\right]_{\pmb{\delta},\pmb{\delta'} }  &  = 
    \frac{1}{L^2} \sum_{\pmb{i},\pmb{j}}  e^{i \pmb{q} \cdot \left( \pmb{i} - \pmb{j} \right) }  \times  \nonumber \\   
  &   \left( \langle \hat{\Delta}_{\pmb{i},\pmb{i}+\pmb{\delta}} \hat{\Delta}_{\pmb{j},\pmb{j}+\pmb{\delta'}} \rangle  
     - \langle \hat{\Delta}_{\pmb{i},\pmb{i}+\pmb{\delta}} \rangle \langle \hat{\Delta}_{\pmb{j},\pmb{j}+\pmb{\delta'} } \rangle \right),
\end{align}
 with $\hat{\Delta}_{\pmb{i},\pmb{i}+\pmb{\delta}} = 
     \sum_{\mu,\nu}  \hat{S}^{f,\mu}_{\pmb{i},\nu} \hat{S}^{f,\nu}_{\pmb{i}+\pmb{\delta},\mu}$.

In Fig.~\ref{VBS} we plot this quantity for various lattice sizes along a high symmetry path in the Brillouin zone
across the magnetic order-disorder transition for $N=2$ [Figs.~\ref{VBS}(a,c)]  and $N=4$ [Figs.~\ref{VBS}(b,d)]. 
As expected,  the VBS correlation function $S^f_{\rm VBS}(\pmb{q})$ for $N=2$ is featureless and lattice-size independent  
throughout the transition confirming that the SU(2) KLM is dominated by magnetic fluctuations.  
Given numerical evidence for enhanced columnar ${\pmb q}=(\pi,0)$ dimer correlations in SU($N$) Hubbard and Heisenberg 
models~\cite{Paramekanti07,Kawashima03,Assaad05,Kawashima07,Beach09,Congjun14,Kawashima15,Lang13,Congjun16,Congjun17}, one could 
expect that  the same physics shows up in the SU($N$) KLM. In contrast, even though the lineshape of $S^f_{\rm VBS}(\pmb{q})$
gets sharper at $N=4$, a dominant cusp feature is found at the AF wavevector ${\pmb Q} =(\pi,\pi)$, see Figs.~\ref{VBS}(b,d).
The same behavior is observed across the magnetic order-disorder transition for larger $N=6$ and $N=8$ shown in Figs.~\ref{VBS2}(a,c) 
and \ref{VBS2}(b,d), respectively.
Thus, we conclude that there are no significant columnar dimer fluctuations in the phase diagram and we interpret
the cusp feature at ${\pmb Q}=(\pi,\pi)$ as a fingerprint of the perfectly nested conduction-electron Fermi surface in the noninteracting limit.


\section{\label{sec:discuss}Discussion}

\subsection{Experimental relevance}

Heavy-fermion systems are prototype materials to study quantum criticality of the magnetic order-disorder transition.
Given the complexity of the problem, theoretical and experimental studies on the quantum criticality in 
heavy-fermion systems explore various routes to approach the quantum critical point (QCP)~\cite{RMP07,Gegenwart08,Si10}. 
One possibility is to modify the strength of the Kondo coupling, e.g., by varying chemical or external pressure.  
Another route is to tune intersite interactions between the $f$-moments by considering  systems with different 
dimensionality or with geometrical frustration. 

In some materials, e.g., Ce$_{1-x}$La$_x$Ru$_2$Si$_2$~\cite{Knafo09}, the data are consistent with predictions of the 
conventional spin-density-wave theory~\cite{Hertz76,Millis93}, which considers the $f$-electrons as itinerant on both sides 
of the QCP. In this case, the dominant critical AF fluctuations modify neither the shape nor size of the large Fermi surface 
which incorporates both conduction electrons and the $f$-electron states. 
In other compounds such as CeCu$_{6-x}$Au$_x$~\cite{Schroder00} and  YbRh$_2$Si$_2$~\cite{Custers03,Paschen04,Pfau12}, 
there are indications for the breakup of composite heavy-fermion QPs and the concomitant collapse of the large 
Fermi surface driven by local critical magnetic  fluctuations~\cite{Coleman01,Si01}. 
Moreover, the reconstruction of the Fermi surface may also occur away from the QCP --- within the 
magnetically ordered phase~\cite{Custers12}  or even more exotically --- outside~\cite{Friedemann09}, 
paving the way for an intervening phase where the local $f$-moments are neither Kondo screened nor antiferromagnetically 
ordered.

Here, in order to gain novel insight into the quantum criticality in heavy-fermion systems, we have considered the SU($N$) 
generalization of the KLM. Given that increasing $N$ changes the degree of quantum fluctuations of the local $f$-moments, 
it allows one to investigate the impact of magnetic fluctuations on the coherent Kondo-lattice formation in a single setup. 
Importantly, we do not observe a breakdown of Kondo screening which continues to exist on the magnetically ordered side of the phase diagram. 
However, our findings show that increasing $N$ strongly modifies the behavior of the QP residue $Z_{(\pi,\pi)}$  across the magnetic phase transition.  
As such they have important implications for the interpretation of experimental data. 
Considering that in reality  experiments are performed at small but finite temperatures, a rapid decrease of the QP residue  resolved 
for $N=8$, see Fig.~\ref{Charge}(b), could be easily mistaken in the isothermal measurement of Hall coefficient as that 
in Ref.~\cite{Paschen04}, as evidence for a collapse of the large Fermi surface at the QCP.

\input{discuss}


\section{\label{sec:conclude}Conclusions}

Our major findings can be summarized by the following points. 

(i) A serious caveat of the large-$N$ approximation is that it introduces a finite hybridization order parameter. 
It breaks the local gauge symmetry and implies that the constraint of single occupancy on the $f$-sites is fulfilled only on average~\cite{Burdin00}. 
Here, we have handled the constraint of no double occupancy  numerically exactly with QMC simulations. 
By extrapolating finite-$N$ QMC data to the $N\to\infty$ limit, we were able to recover the large-$N$ results in the KI phase.  
This  validates large-$N$ approximate treatments of the constraint and confirms that the large-$N$ theory is the correct saddle point 
of the SU(2) KLM.

(ii)  Up to $N=8$ we observe  a  magnetically ordered phase.    The RKKY interaction scales as  $1/N$ and   the Kondo   energy is $N$-independent such that matching the two energy scales gives  
$J_c(N) \propto \frac{1}{\ln(N) N (\epsilon_F) } $.  This  form is consistent with our data.  Since the  charge degrees of freedom are gapped throughout the phase diagram,  the magnetically 
ordered state should be understood in terms of an SU($N$) quantum antiferromagnet on a bilayer square lattice.   Let us  consider the representations  discussed in Ref.~\cite{Read90}, consisting of a 
Young tableau of $m$ ($N-m$)  rows and one column on sublattice A (B).  The N\'eel broken symmetry phase of  the model has Lorentz symmetry and accordingly  $ 2 Nm -2m^2 $ Goldstone 
modes~\cite{Goldstone62,Watanabe19}.  This count matches the dimension of the manifold on which the NL$\sigma$ model of  Sec.~\ref{QFT.Sec} is defined.   The number of Goldstone modes is a measure  
of the fluctuations around  the N\'eel  state and is maximal  for the representation   $m=N/2$     considered here.   It is hence   interesting to compare  our result to that of Ref.~\cite{Kaul12b}   
for the SU($N$) bilayer Heisenberg model  with nearest-neighbor couplings at $m=1$: at $N=8$ no magnetic ordering is present.  
To reconcile this apparent contradiction we have to take into account the range of the RKKY interaction.   To a first approximation,  it is given by the inverse single-particle gap in the KI phase 
at a value of $J$  just above $J_c(N)$.   With the above form for $J_c(N)$ and  large-$N$ form for the   single-particle gap, $\Delta_{qp}  \propto e^{-\frac{1}{J  N(\epsilon_F) }}  $,   
we find that the range of the  RKKY interaction grows as a power of $N$.     We believe that this enhanced range of the interaction  -- very specific to the KLM -- is the key to stabilize  
antiferromagnetism at large $N$.    

(iii) We have argued in Sec.~\ref{QFT.Sec}   that the Berry phase  could be omitted, such that the KLM provides a unique possibility 
to study  the critical phenomena associated with the  NL$\sigma$ model  of Eq.~(\ref{NLs.eq})  at $m=N/2$.  To the best of our knowledge,   and as discussed in Sec.~\ref{QFT.Sec}, 
this  universality class has never been studied. Our results suggest  however that as a function of $N$, the transition  does not sustain a quantum  critical point~\cite{Assaad99,Held19} and  becomes  first order.     
Remarkably for bilayer geometries,  choosing the fundamental representation on one sublattice and  the adjoint on the other,  as in Ref.~\cite{Kaul12b}, one also  observes  that  for $N=4$  and beyond   
the order-disorder transition  is a first order one.

(iv)   Since the range of the  RKKY interaction grows as a function of $N$ we expect that the interplay between charge and spin degrees of freedom will become more mean-field-like.   In fact, 
at large $N$ we observe an abrupt reduction of the QP residue $Z_{(\pi,\pi)}$ upon entering the AF phase.  This behavior is very reminiscent of that observed in  mean-field calculations of 
the SU(2) KLM that take into account both antiferromagnetism and Kondo screening~\cite{Capponi01}.  Within a rigid band shift
assumption  to describe the  heavy-fermion metallic state at small doping,   this means that the coherence temperature drops by up to an order of magnitude across the  magnetic transition.   
Isothermal measurement of the Hall coefficient  made below the coherence  temperature in the paramagnetic heavy-fermion phase and above it in the magnetically ordered state, would   be  interpreted 
as a breakdown of Kondo screening~\cite{Paschen04}.  
Nevertheless, we find the signature of the heavy-fermion band for all the considered values of $N$ down to the smallest $J$. With the aforementioned rigid  band shift,  the emergent heavy-fermion metal 
at small coupling $J$ is characterized by a large Fermi surface containing both conduction and localized electrons.  In the magnetically ordered phase, back-folding  of the Fermi surface accounts for 
the reduced translation symmetry.   
This abrupt  reduction of $Z_{(\pi,\pi)}$  upon entering the AF phase accompanied by a jump in the free-energy derivative $\frac{\partial F}{\partial J }$ could be interpreted as a sign of 
a first order transition in the high SU($N$) symmetric case.

\begin{acknowledgments}
We would like to acknowledge  discussions with M. Vojta. 
This work was supported by the German Research Foundation (DFG) through Grant No. RA 2990/1-1.
FFA acknowledges financial support from the DFG through the W\"urzburg-Dresden Cluster of Excellence on Complexity and Topology in Quantum Matter - ct.qmat (EXC 2147, project-id 39085490) 
as well as  through the SFB 1170 ToCoTronics.
The authors gratefully acknowledge the Gauss Centre for Supercomputing e.V. (www.gauss-centre.eu) 
for funding this project by providing computing time through the John von Neumann Institute for Computing (NIC) 
on the GCS Supercomputer JUWELS~\cite{juwels}   at J\"ulich Supercomputing Centre (JSC).
\end{acknowledgments}

\appendix 

\input{append.tex}

\section{\label{app:data} Supplemental data}

\begin{figure*}[t!]
\begin{center}
\includegraphics[width=0.32\textwidth]{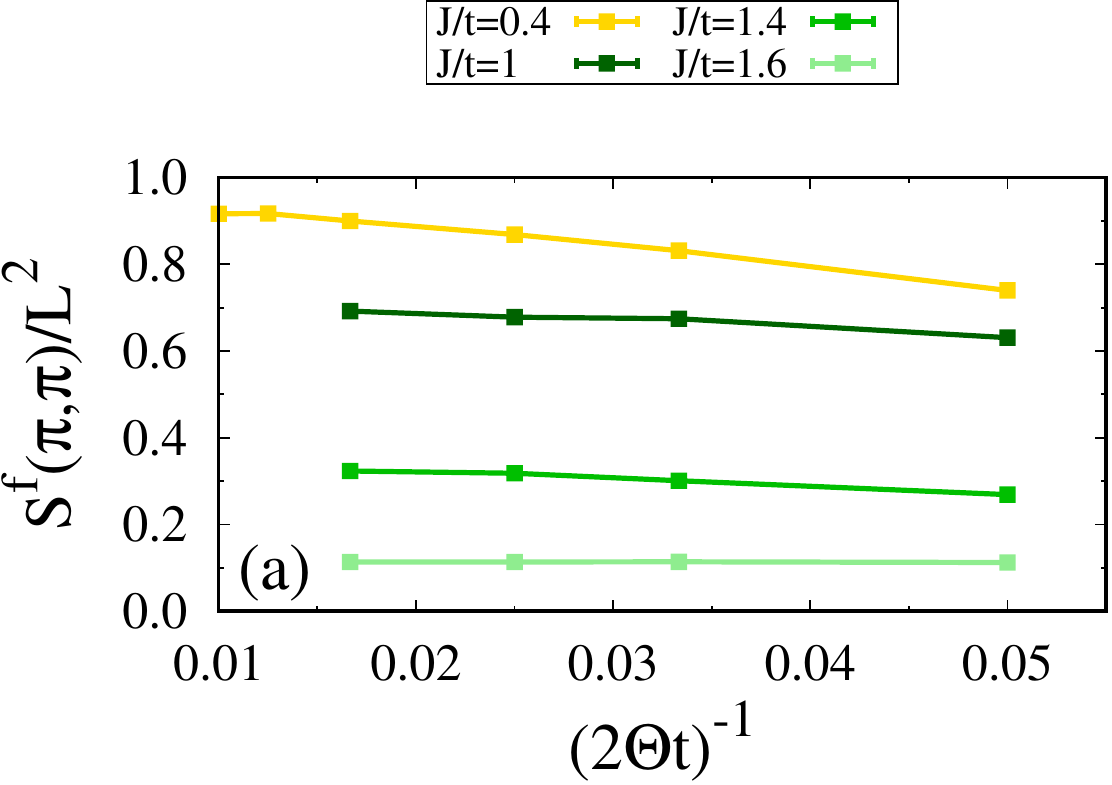}
\includegraphics[width=0.32\textwidth]{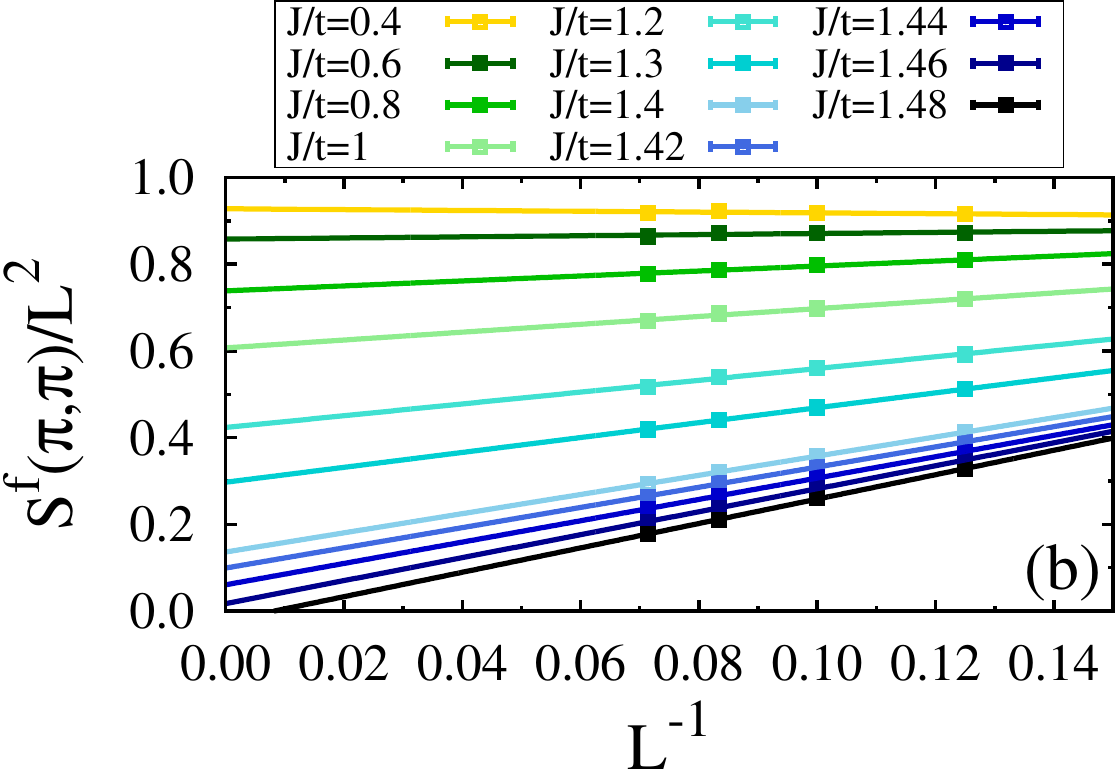}
\includegraphics[width=0.32\textwidth]{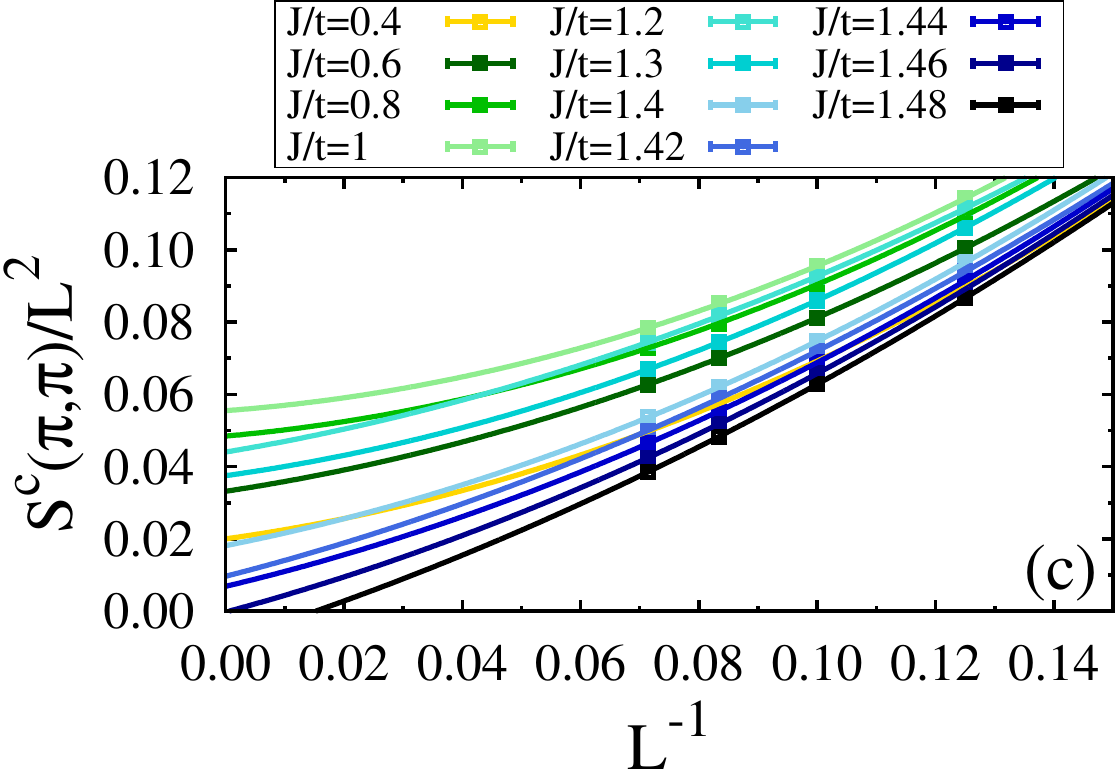}\\
\includegraphics[width=0.32\textwidth]{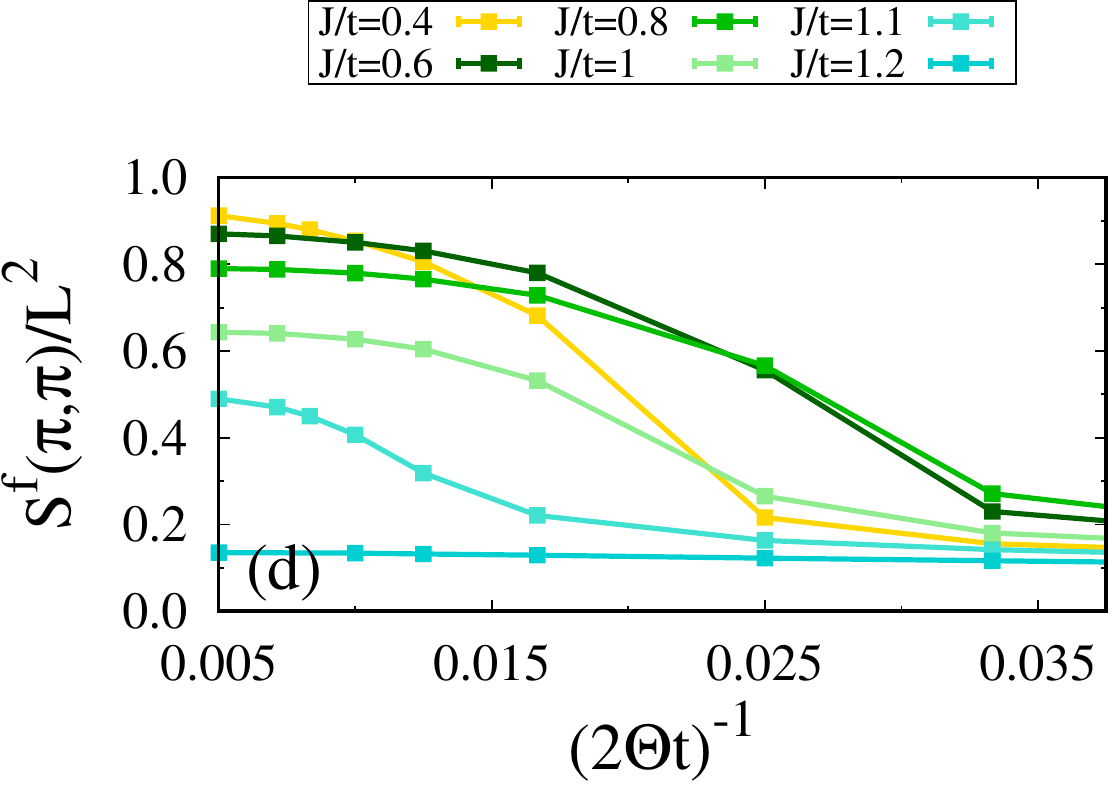}
\includegraphics[width=0.32\textwidth]{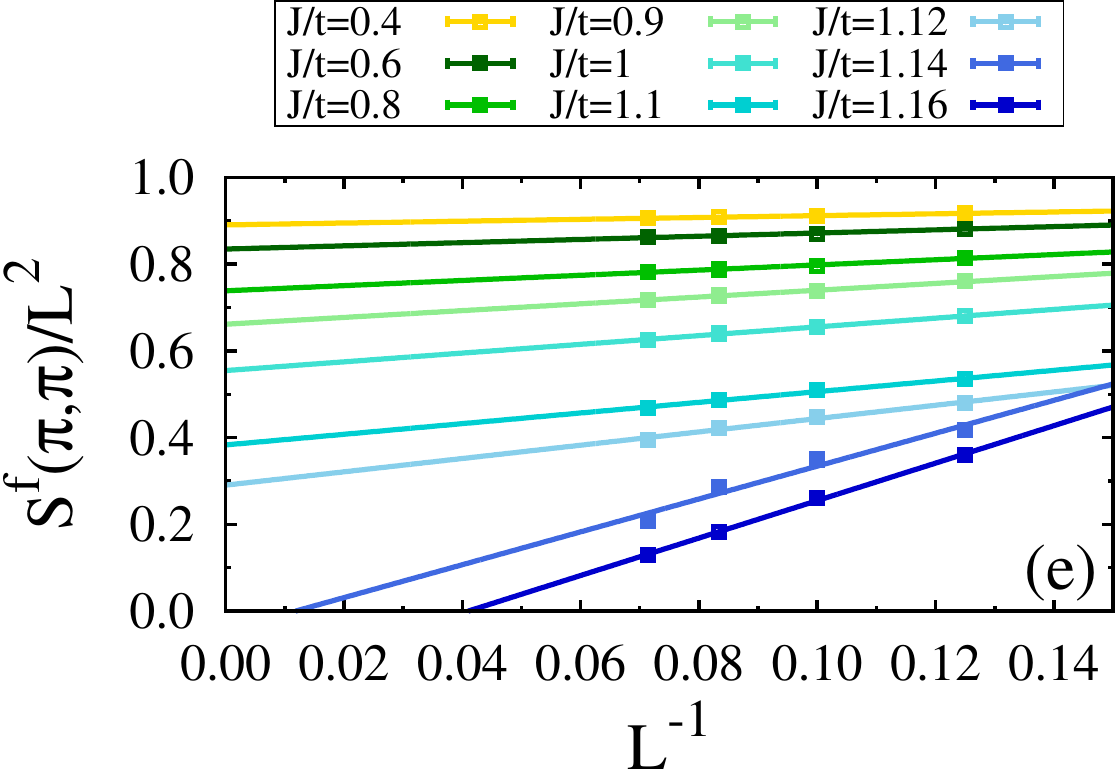}
\includegraphics[width=0.32\textwidth]{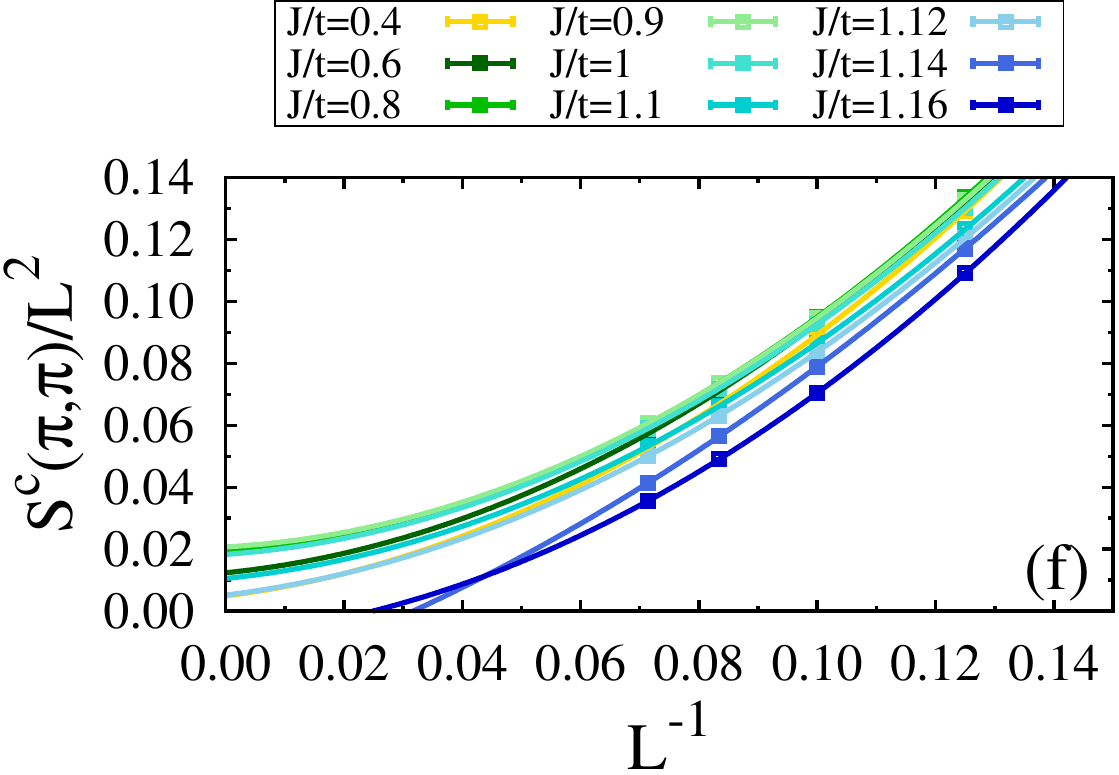}\\
\includegraphics[width=0.32\textwidth]{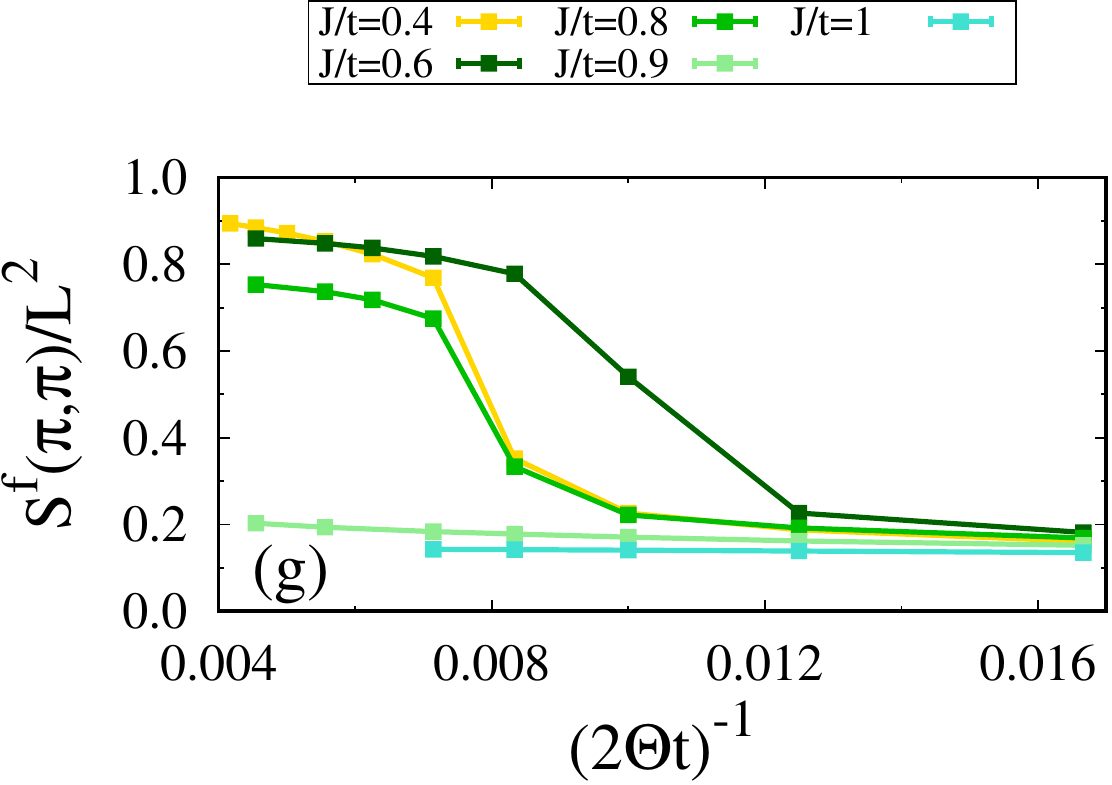}
\includegraphics[width=0.32\textwidth]{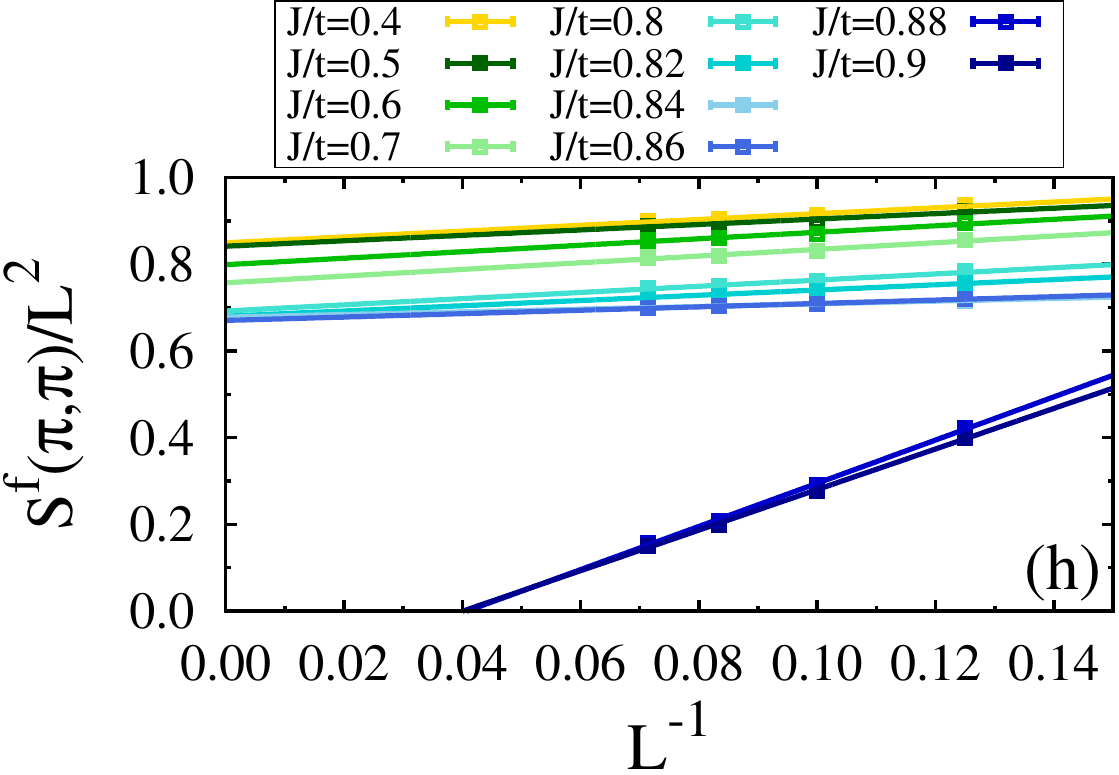}
\includegraphics[width=0.32\textwidth]{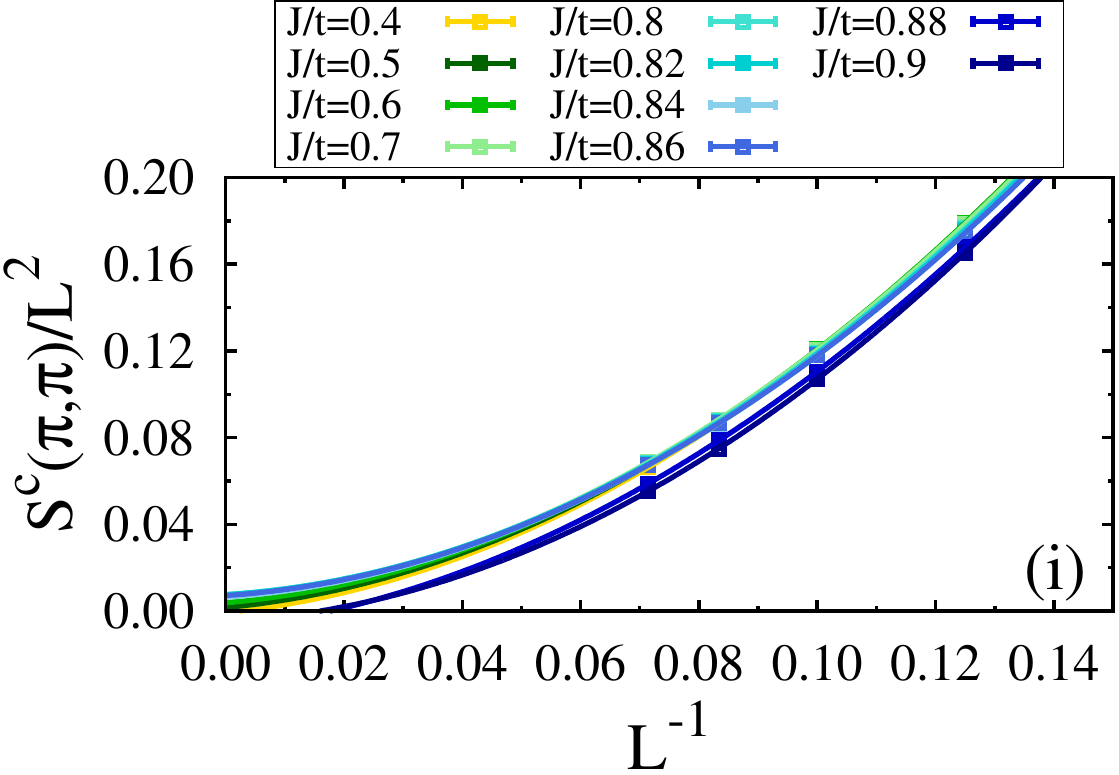}\\
\includegraphics[width=0.32\textwidth]{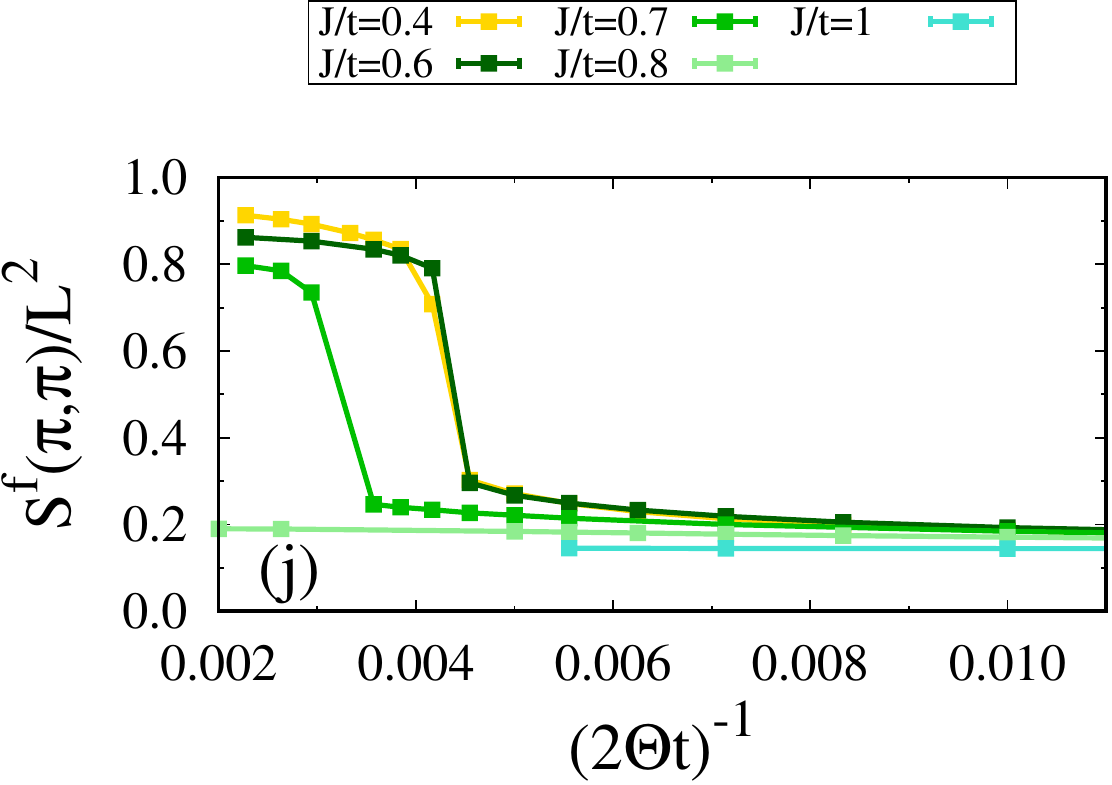}
\includegraphics[width=0.32\textwidth]{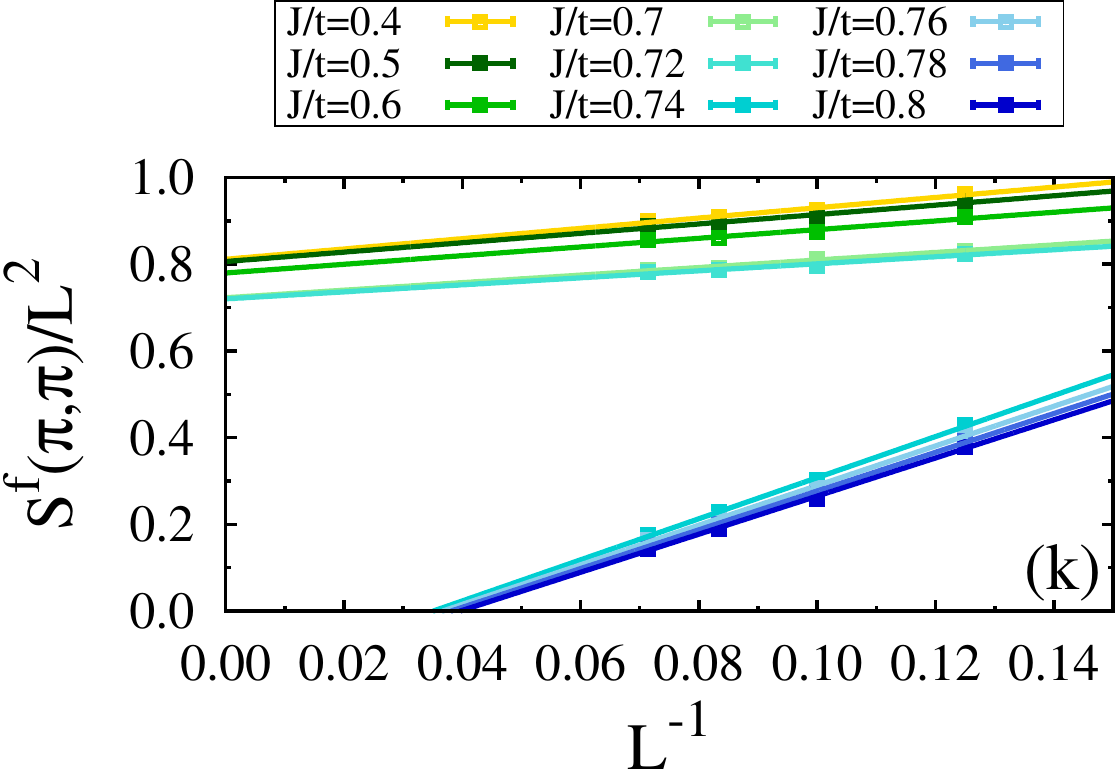}
\includegraphics[width=0.32\textwidth]{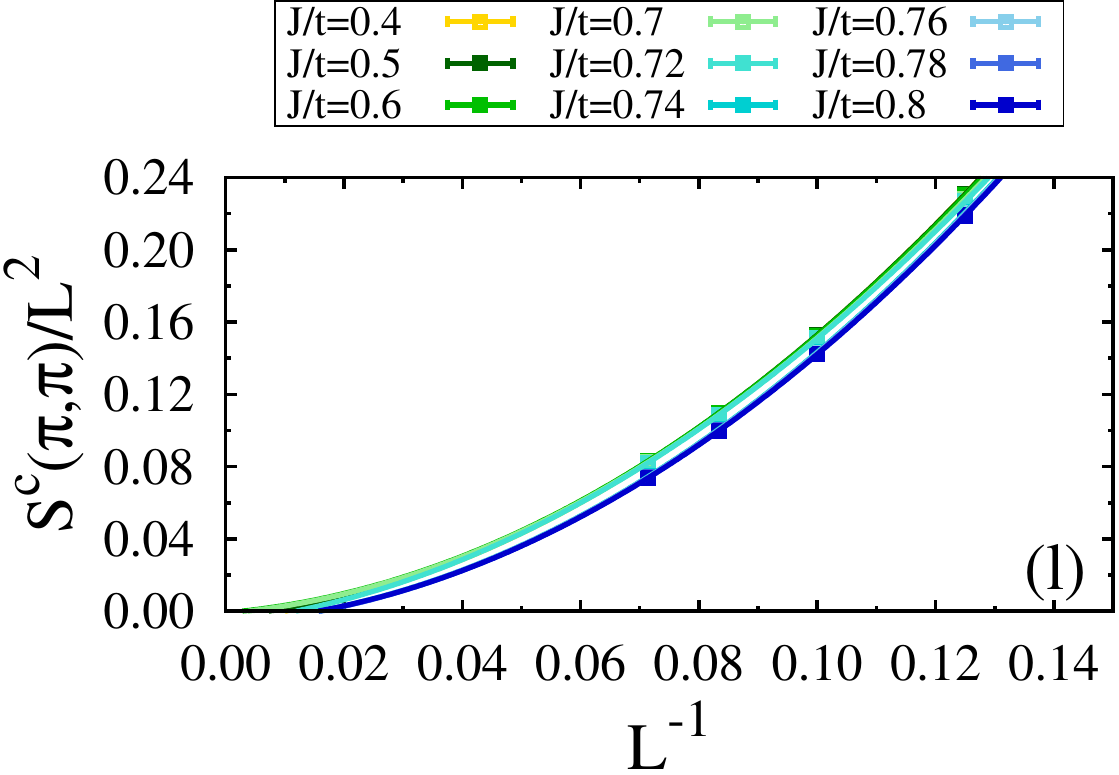}
\end{center}
\caption
{(a,d,g,j) Convergence of the spin structure factor $S^f(\pmb{Q}=(\pi,\pi))$ for the $f$-electrons at representative values of $J/t$
as a function of the projection parameter $\Theta t$ for the $L=12$ system. 
Panels (b,e,h,k) and (c,f,i,l) show finite-size extrapolation of the spin structure factor  for the $f$- and $c$-electrons, respectively, 
on approaching the magnetic order-disorder transition point; solid lines are linear and second-order polynomial fits to the 
QMC data. From top to bottom: $N=2,4,6,\text{and } 8$.}
\label{App_one}
\end{figure*}

Here we provide further details about the QMC simulation results  discussed  in the main text. 

\subsection{\label{app:conv}  Convergence to the ground state}

In this appendix we check the dependence of the QMC results on the projection parameter $\Theta$. In order to ensure that a given result 
corresponds to the ground state we have performed test simulations on the $12\times 12$ system at a variety of projection parameters $\Theta$. 
The energy scales of the KLM, the single-ion Kondo temperature, coherence temperature,  and the RKKY scale, they all become smaller on decreasing $J/t$. 
The calculations become more expensive in the SU($N$) case  since as shown in Appendix~\ref{app:scales},  the RKKY scale $\propto \tfrac{1}{N}$. 
Consequently, increasingly large projection parameters are required to reach the AF ground state and the issue becomes particularly severe for 
small values of $J$,  see Figs.~\ref{App_one}(a,d,g,j).

\begin{figure*}[t!]
\begin{center}
\includegraphics[width=0.32\textwidth]{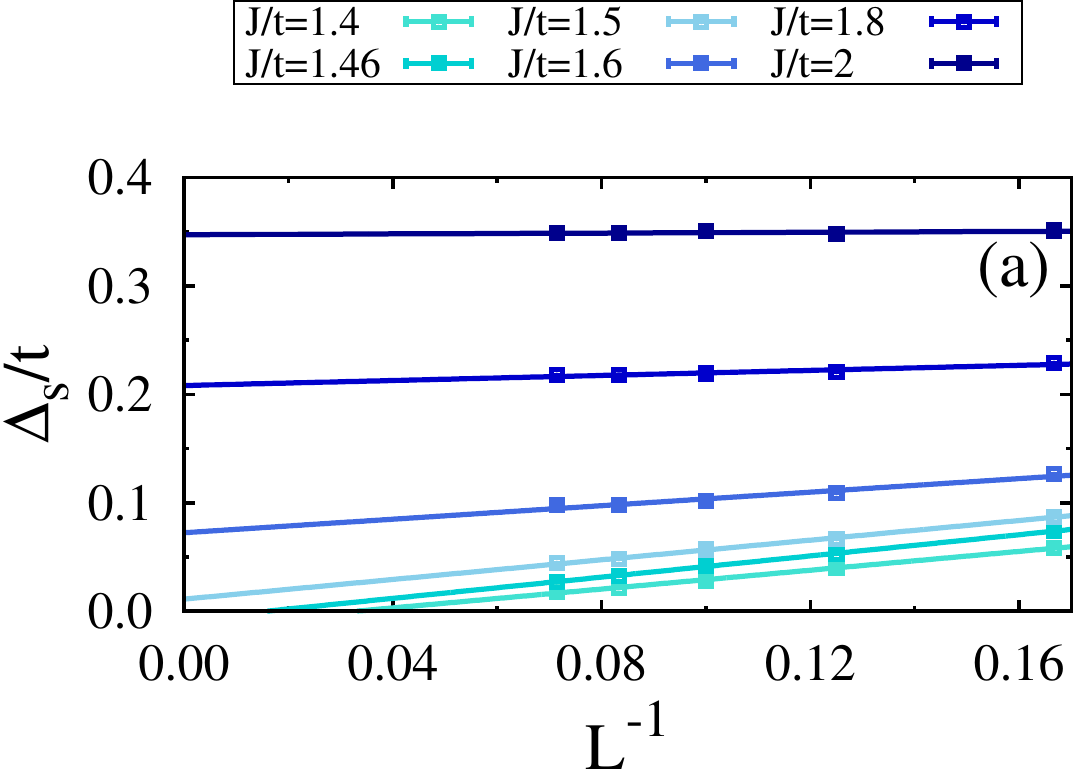}
\includegraphics[width=0.32\textwidth]{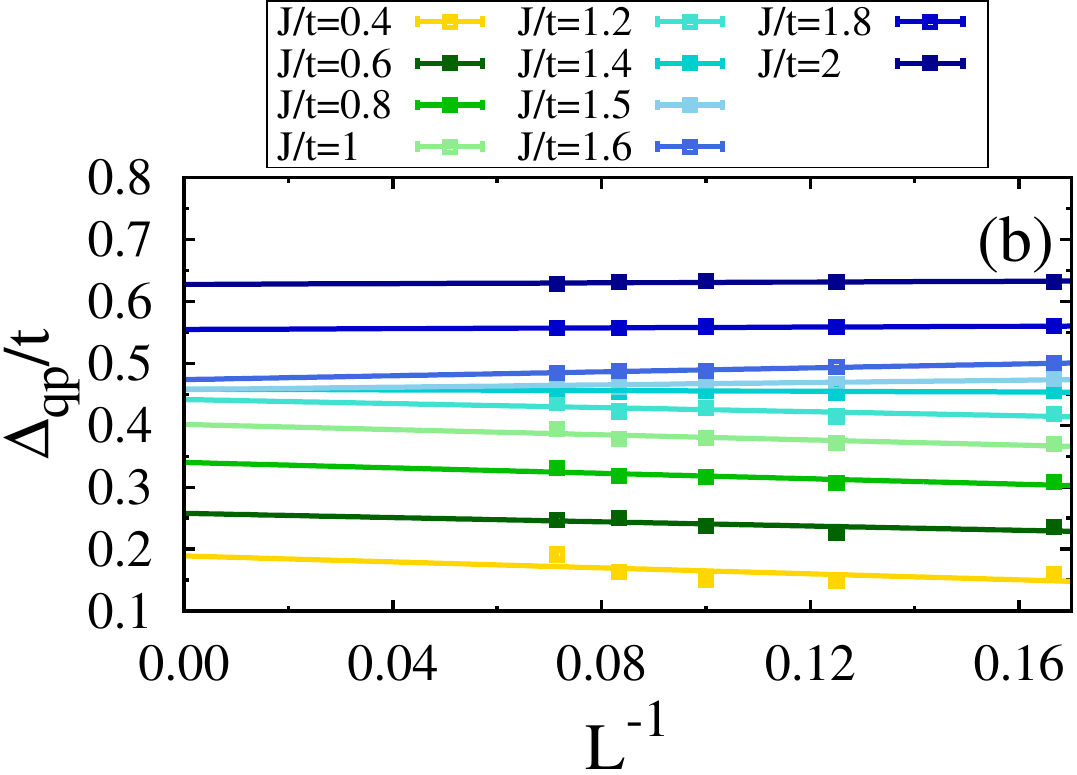}
\includegraphics[width=0.32\textwidth]{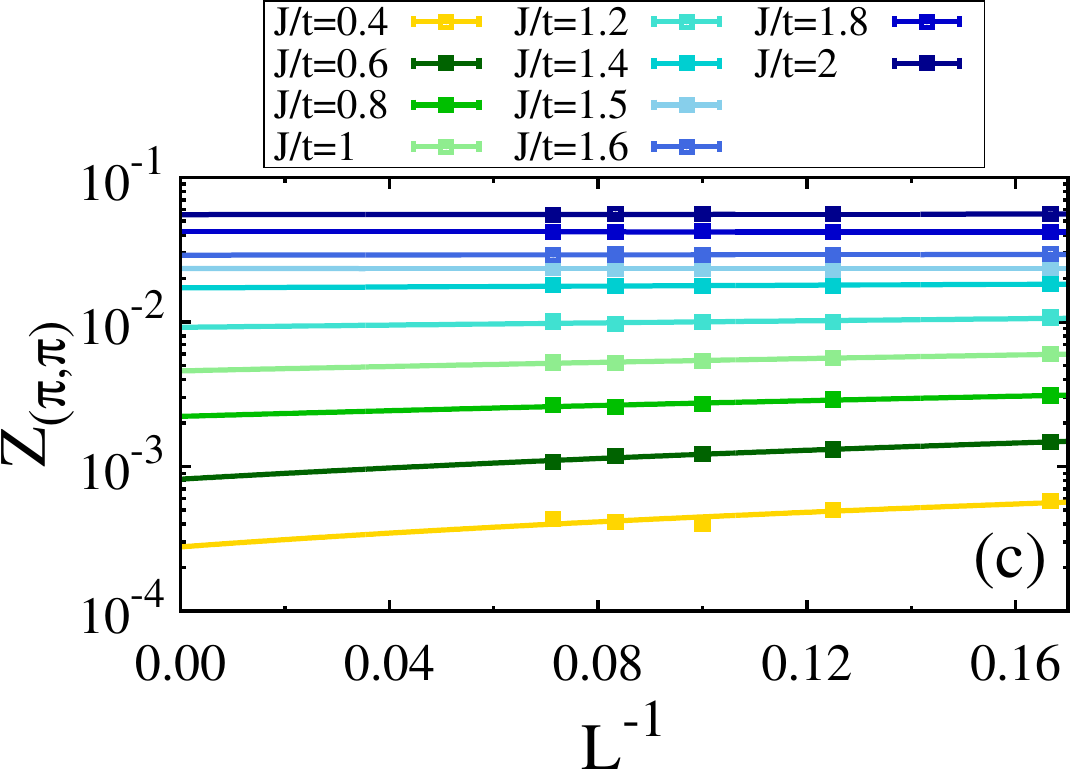}\\
\includegraphics[width=0.32\textwidth]{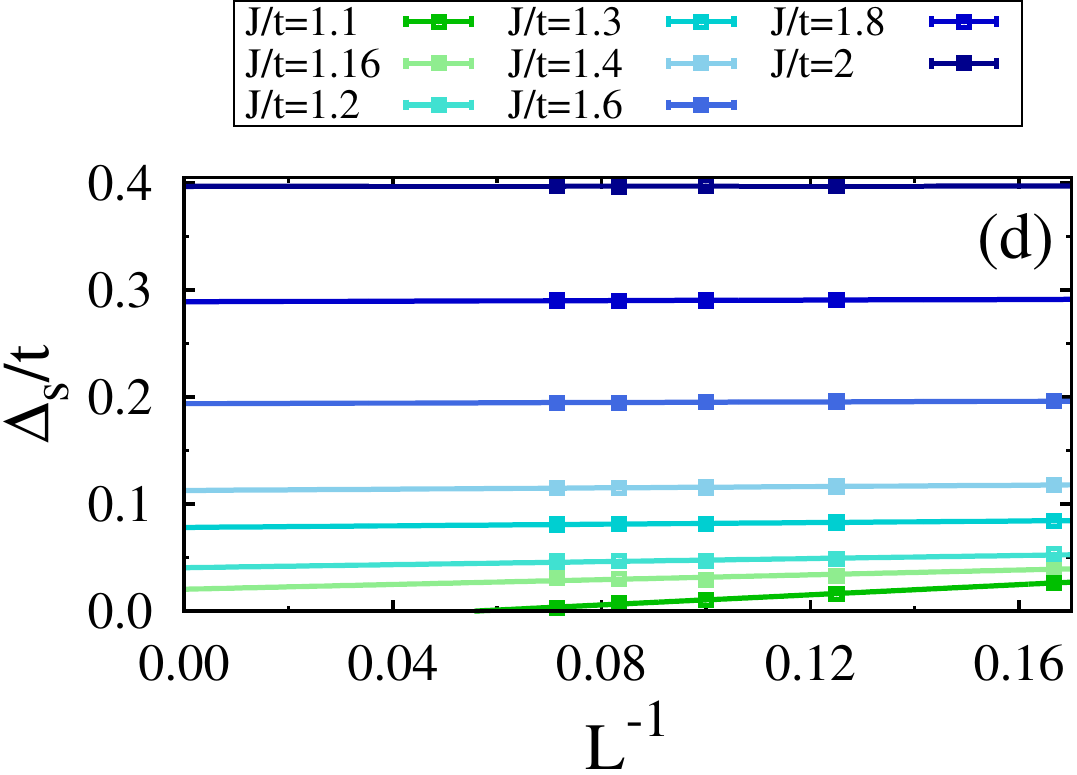}
\includegraphics[width=0.32\textwidth]{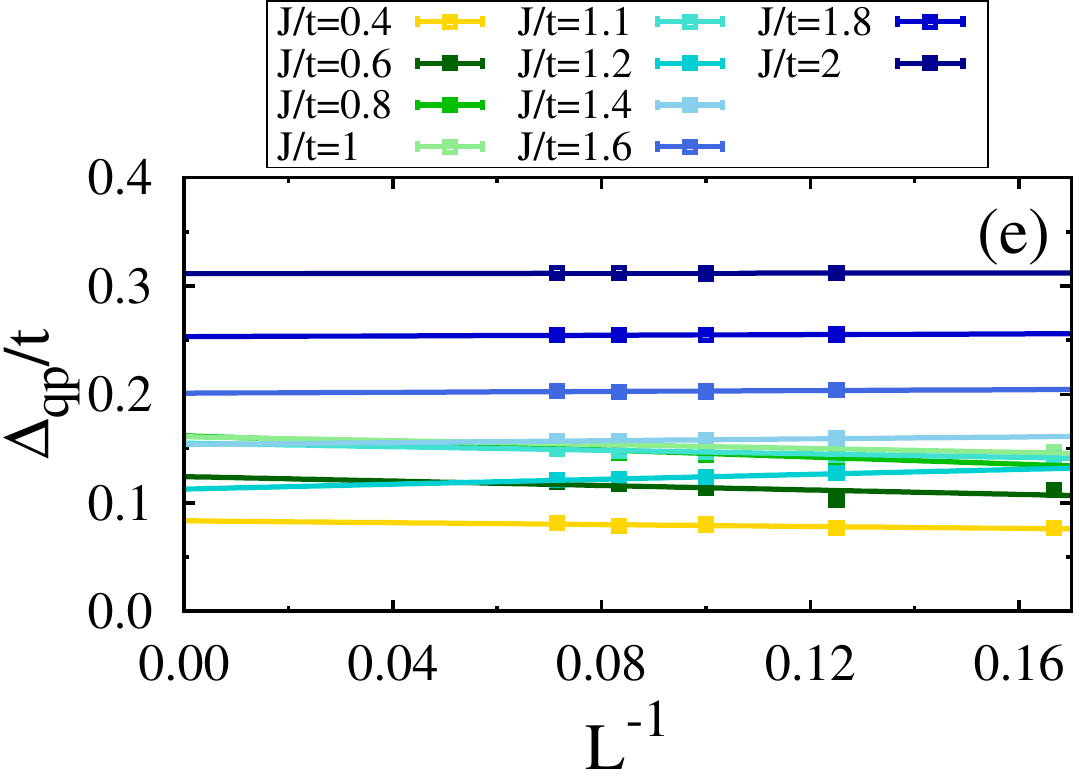}
\includegraphics[width=0.32\textwidth]{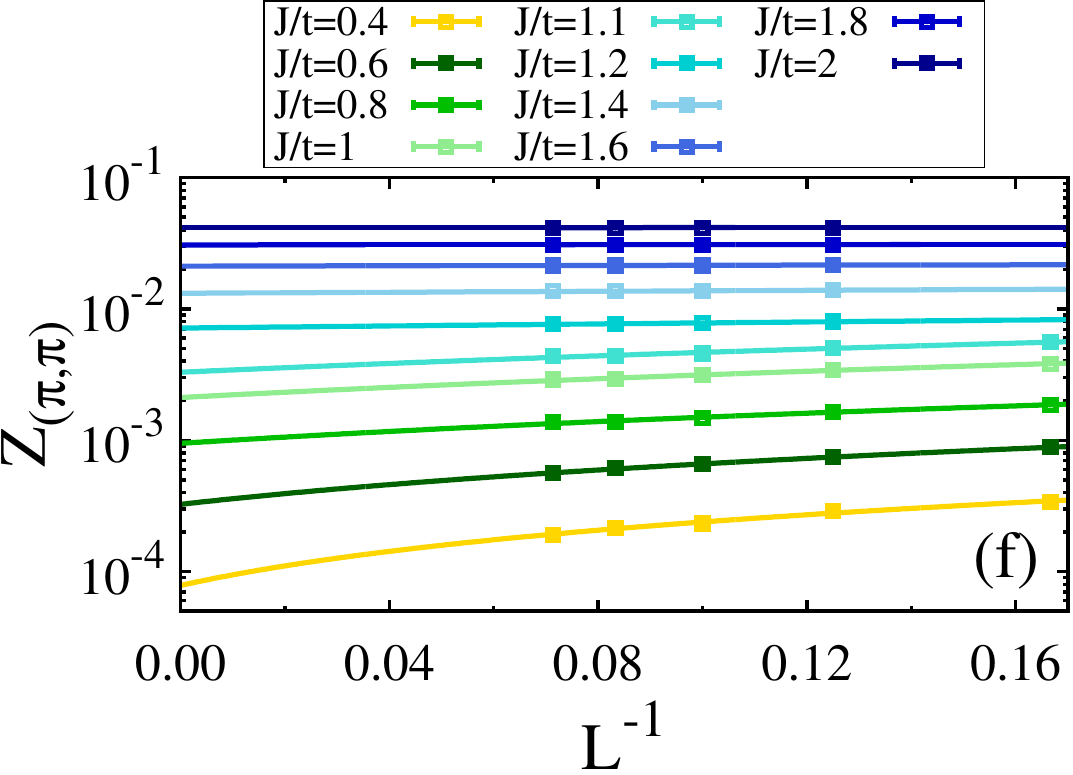}\\
\includegraphics[width=0.32\textwidth]{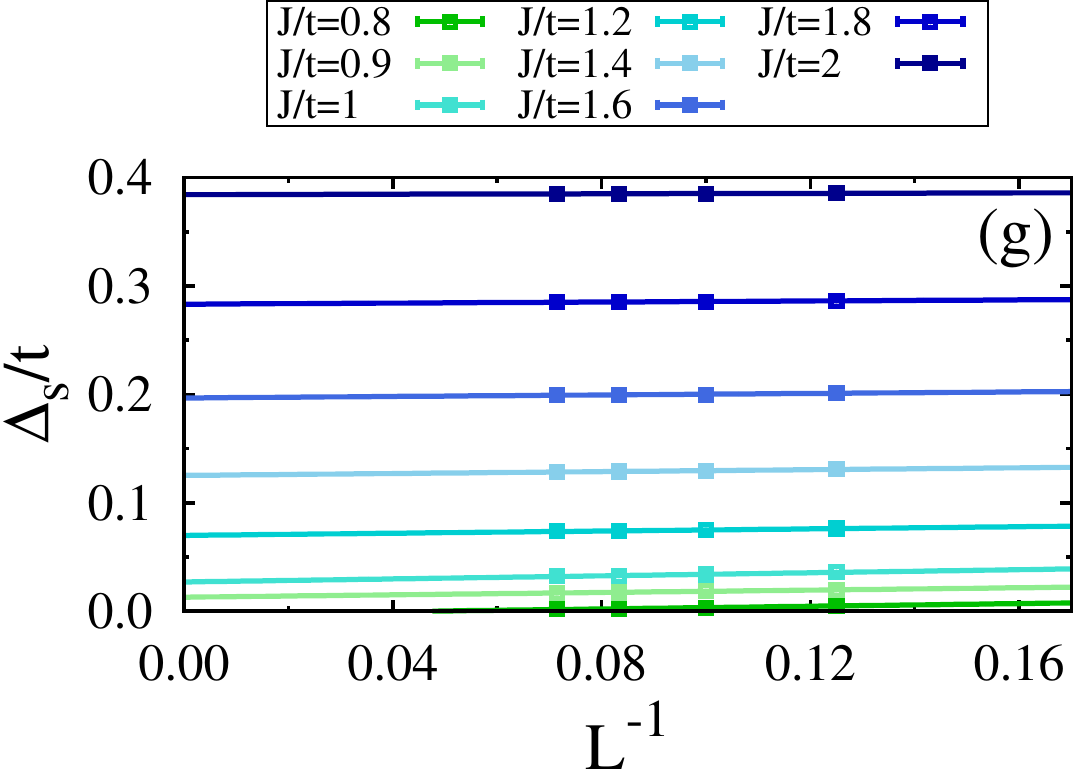}
\includegraphics[width=0.32\textwidth]{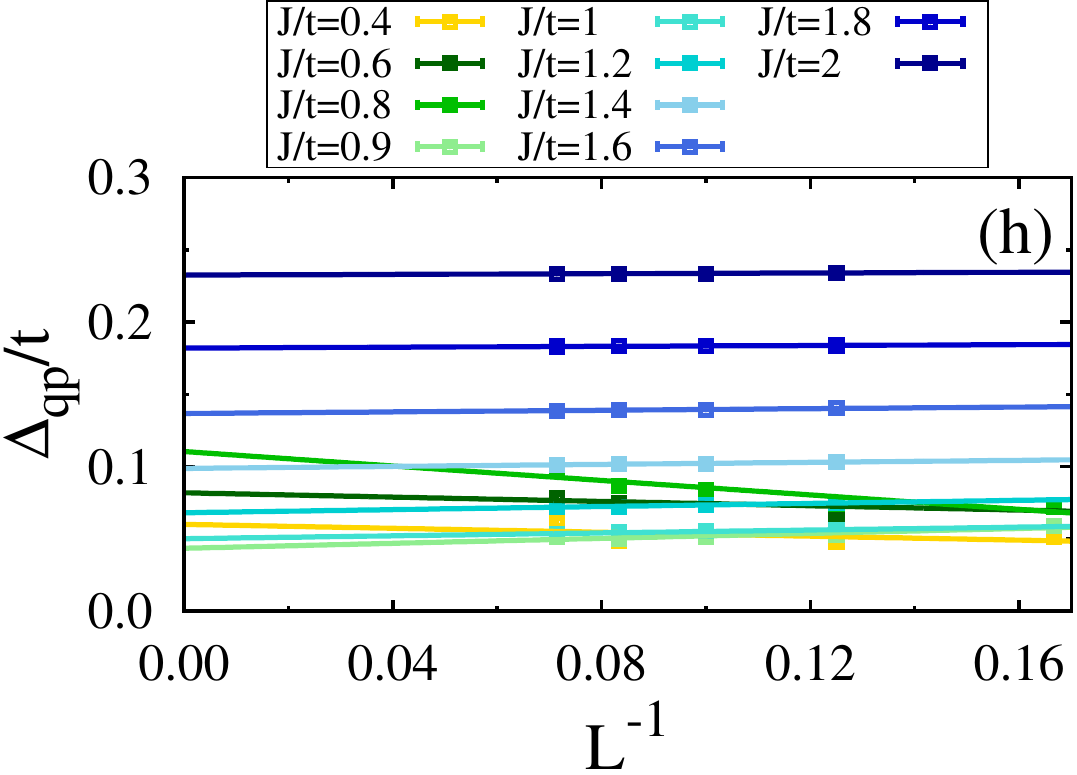}
\includegraphics[width=0.32\textwidth]{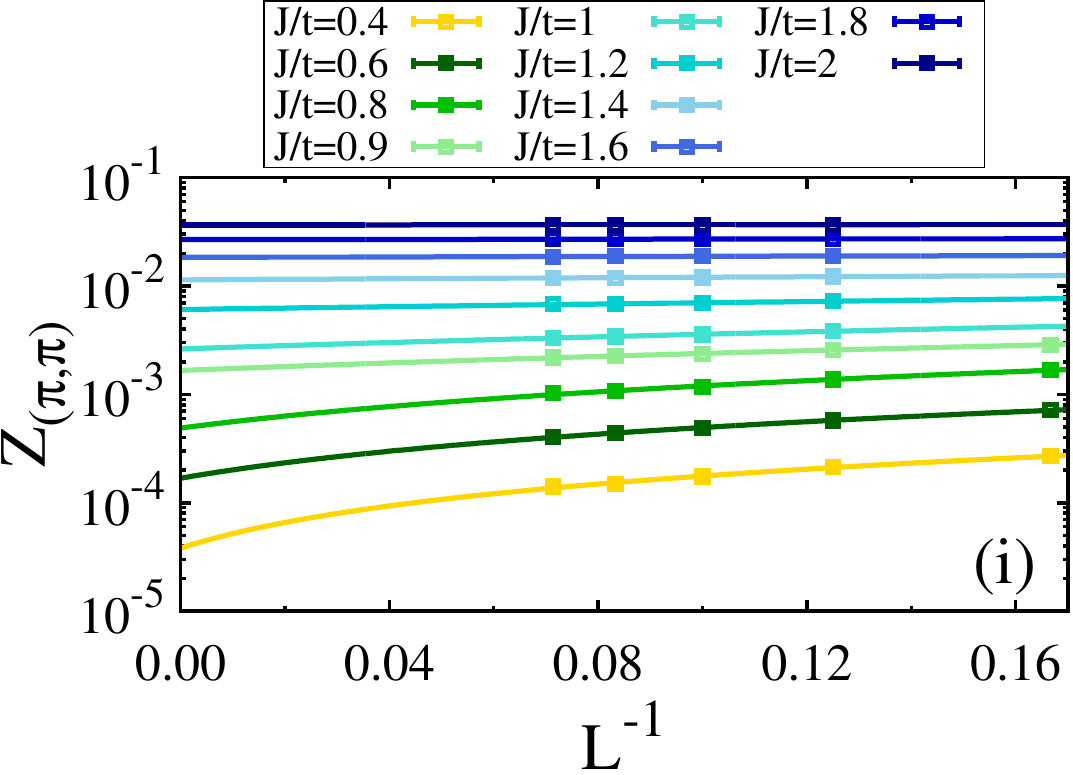}\\
\includegraphics[width=0.32\textwidth]{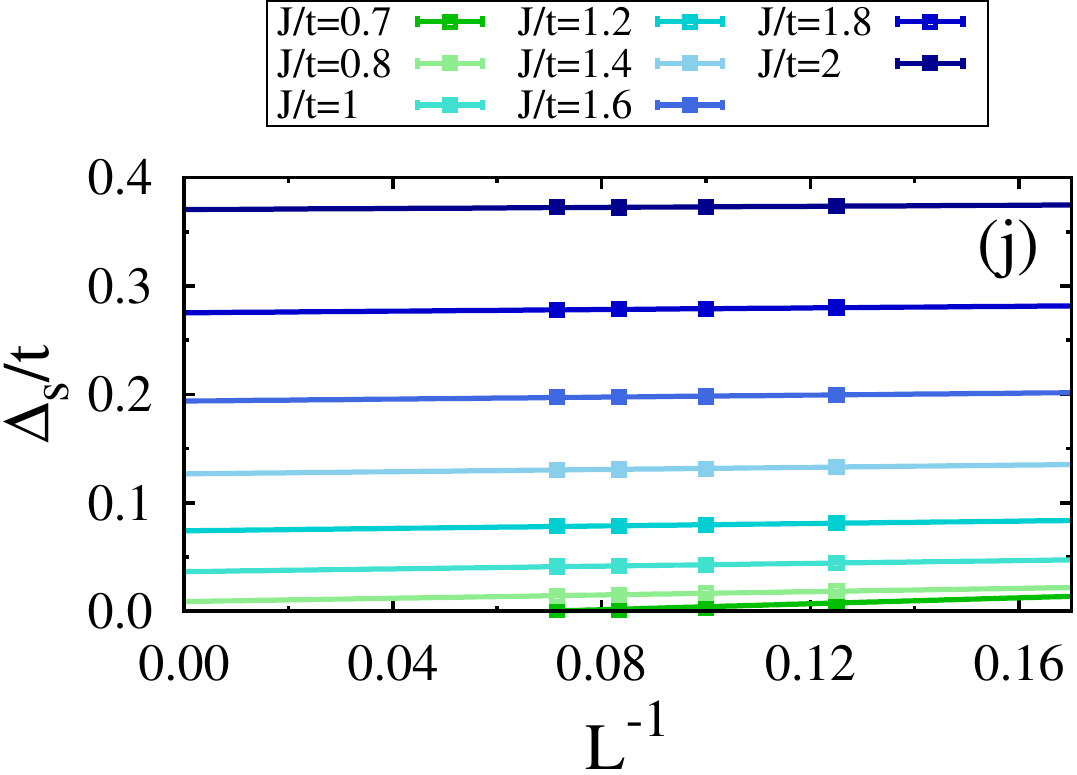}
\includegraphics[width=0.32\textwidth]{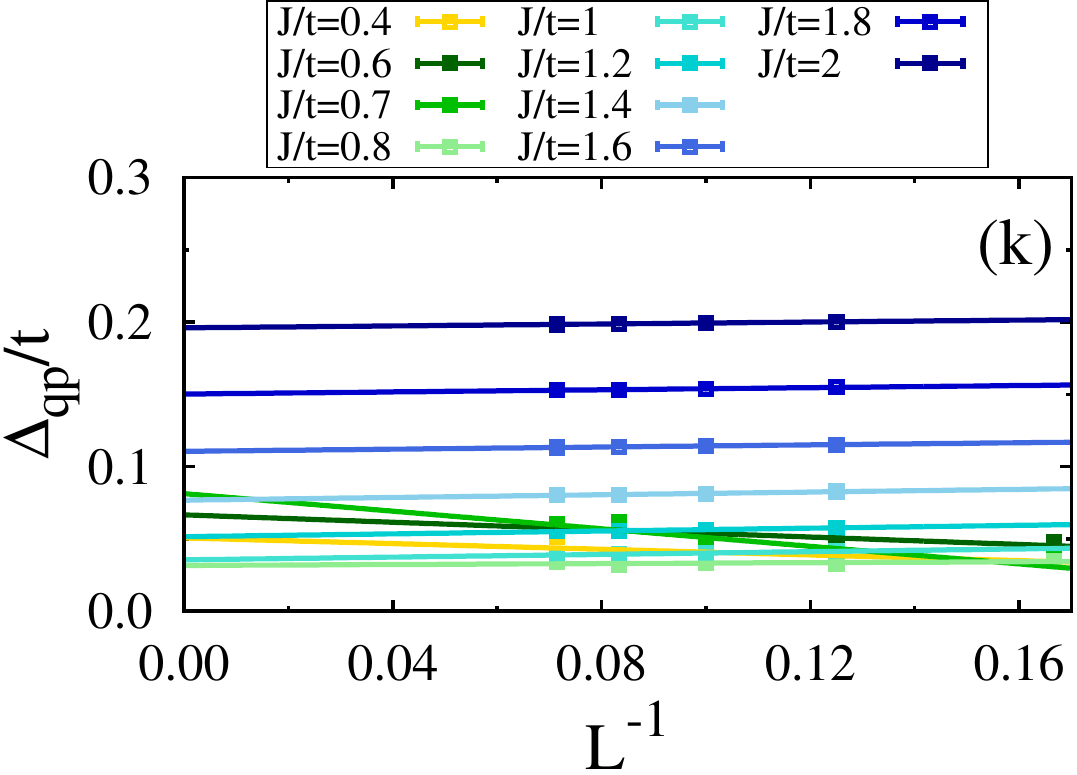}
\includegraphics[width=0.32\textwidth]{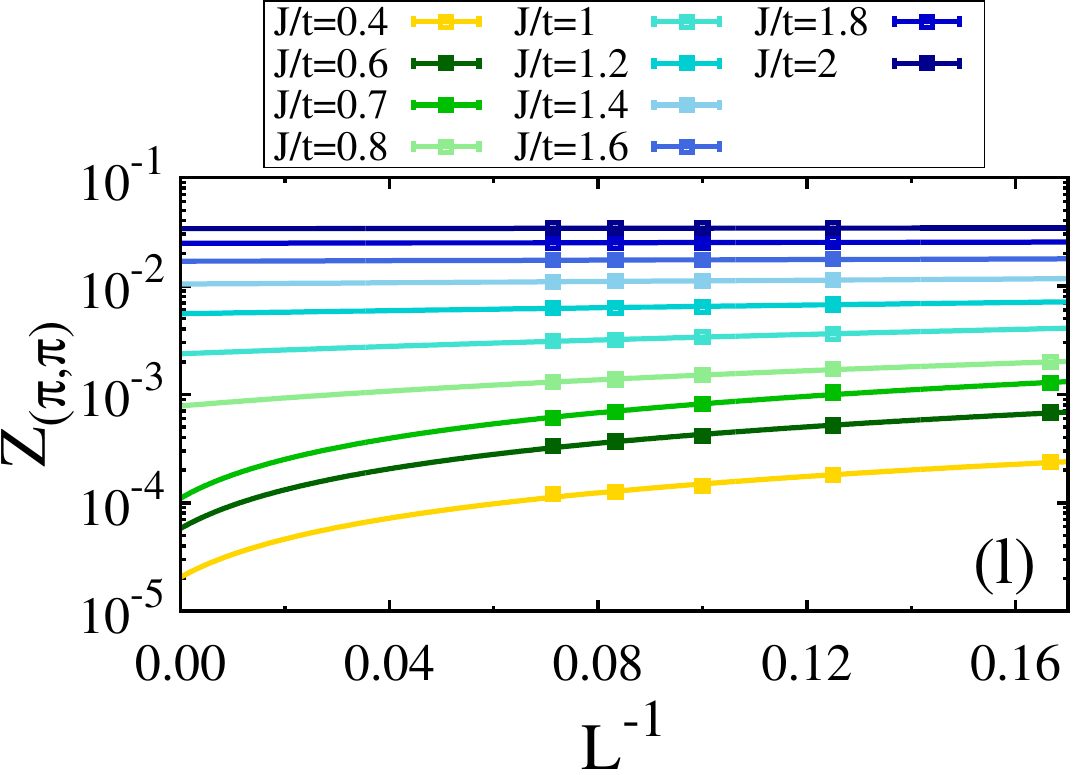}
\end{center}
\caption
{Finite-size extrapolation of the: (a,d,g,j) spin gap $\Delta_s$, (b,e,h,k)  single-particle gap $\Delta_{qp}$ at momentum $\pmb k=(\pi,\pi)$,
and (c,f,i,l) QP residue $Z_{(\pi,\pi)}$  at representative values of $J/t$. Solid lines are linear fits to the QMC data. From top to bottom: 
	$N=2,4,6,\text{and } 8$.
}
\label{App_two}
\end{figure*}

\begin{figure*}[t!]
\begin{center}
\includegraphics[width=0.32\textwidth]{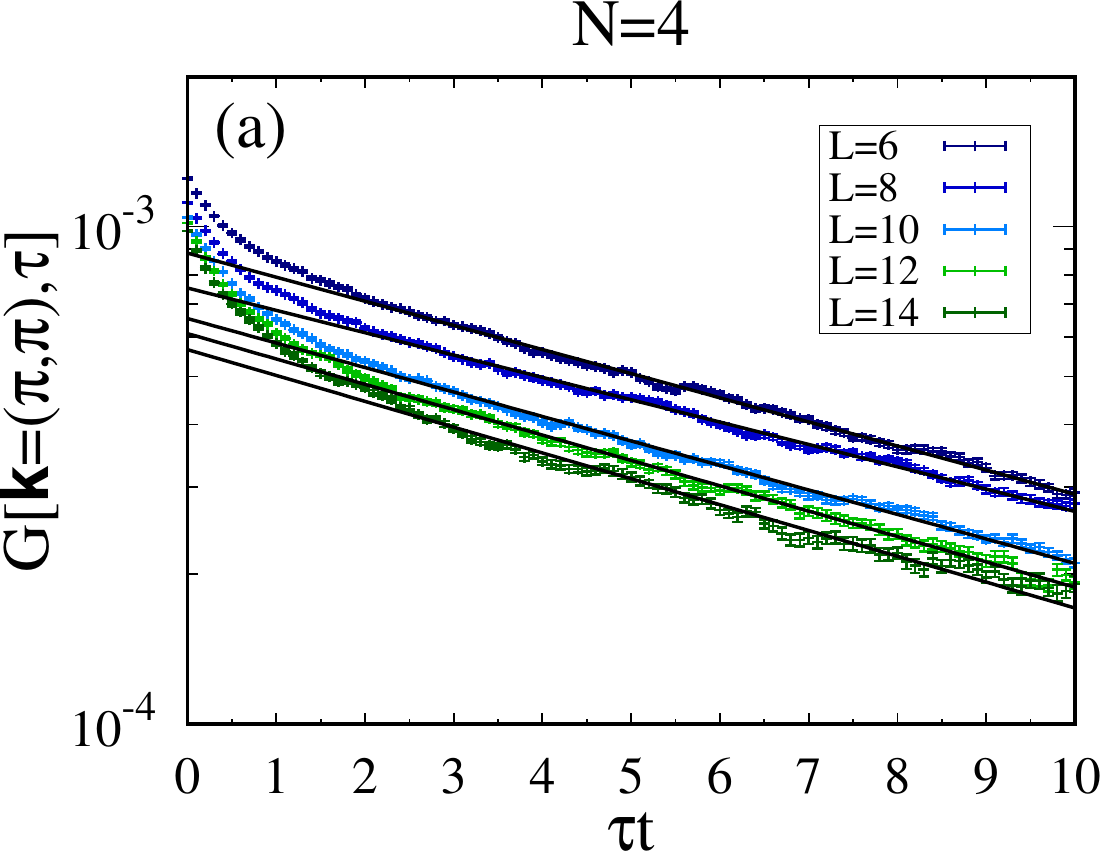}
\includegraphics[width=0.32\textwidth]{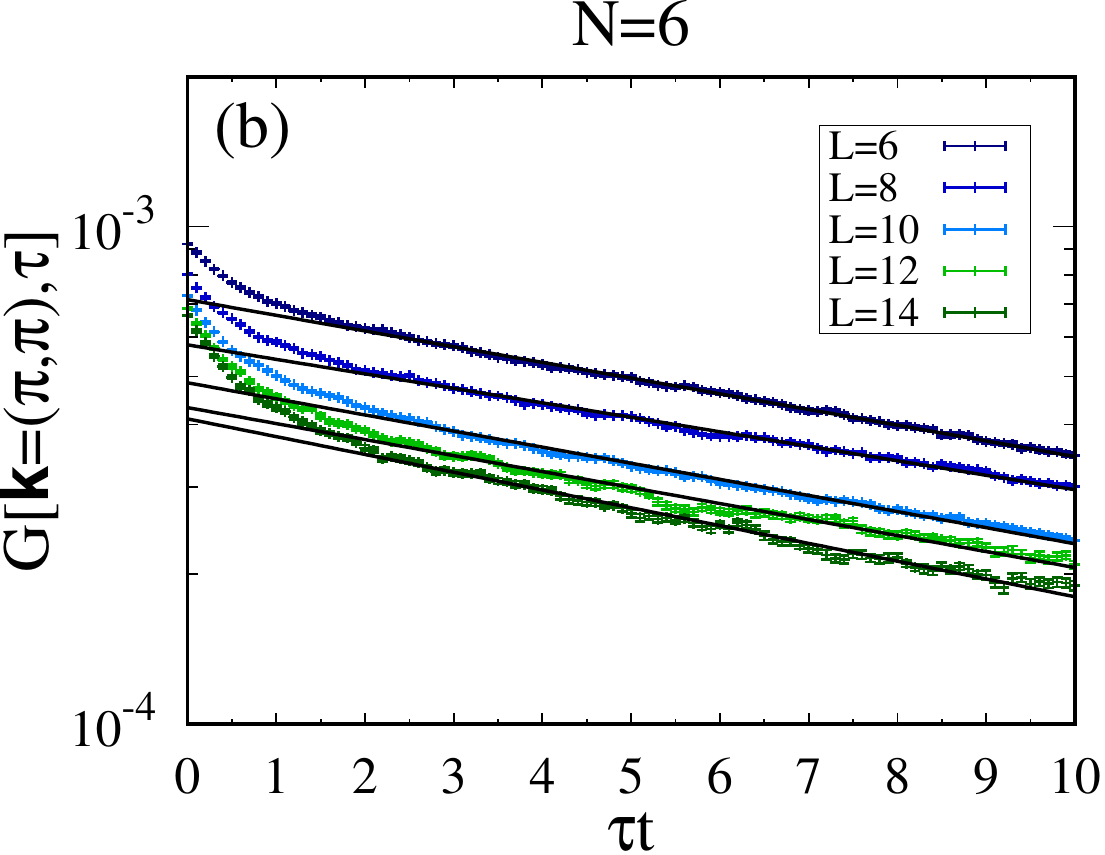}
\includegraphics[width=0.32\textwidth]{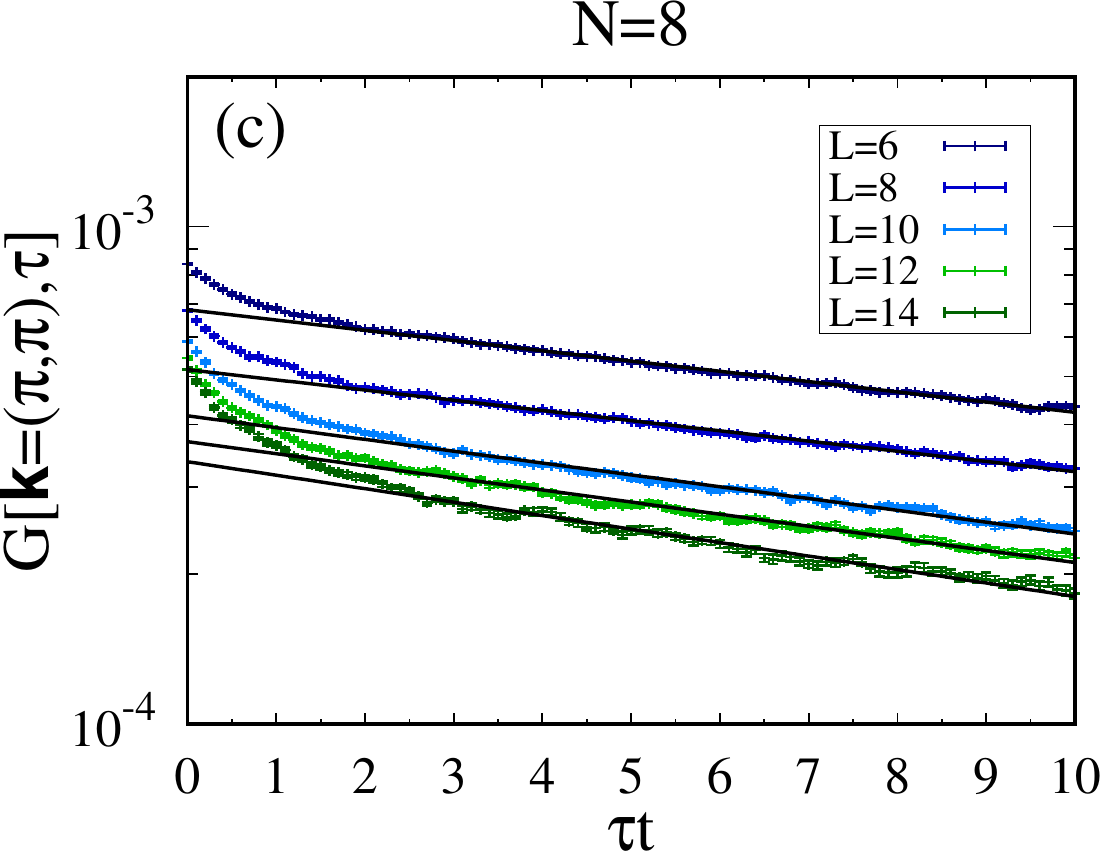}
\end{center}
\caption
{Imaginary-time Green's function for conduction electrons at momentum $\pmb{k}=(\pi,\pi)$ 
obtained from QMC simulations for a given linear system size $L$ at $J/t=0.6$ for: 
(a) $N=4$, (b) $N=6$, and (c) $N=8$.  Finite-size estimates of the single-particle gap $\Delta_{qp}(\pmb{k})$ 
and QP residue $Z_{\pmb{k}}$ are extracted by fitting the tail of the Green's function to 
the form $Z_{\pmb{k}}e^{-\Delta_{qp}(\pmb{k}) \tau}$  
(solid lines). 
}
\label{App_raw}
\end{figure*}

\begin{figure*}[t!]
\begin{center}
\includegraphics[width=0.32\textwidth]{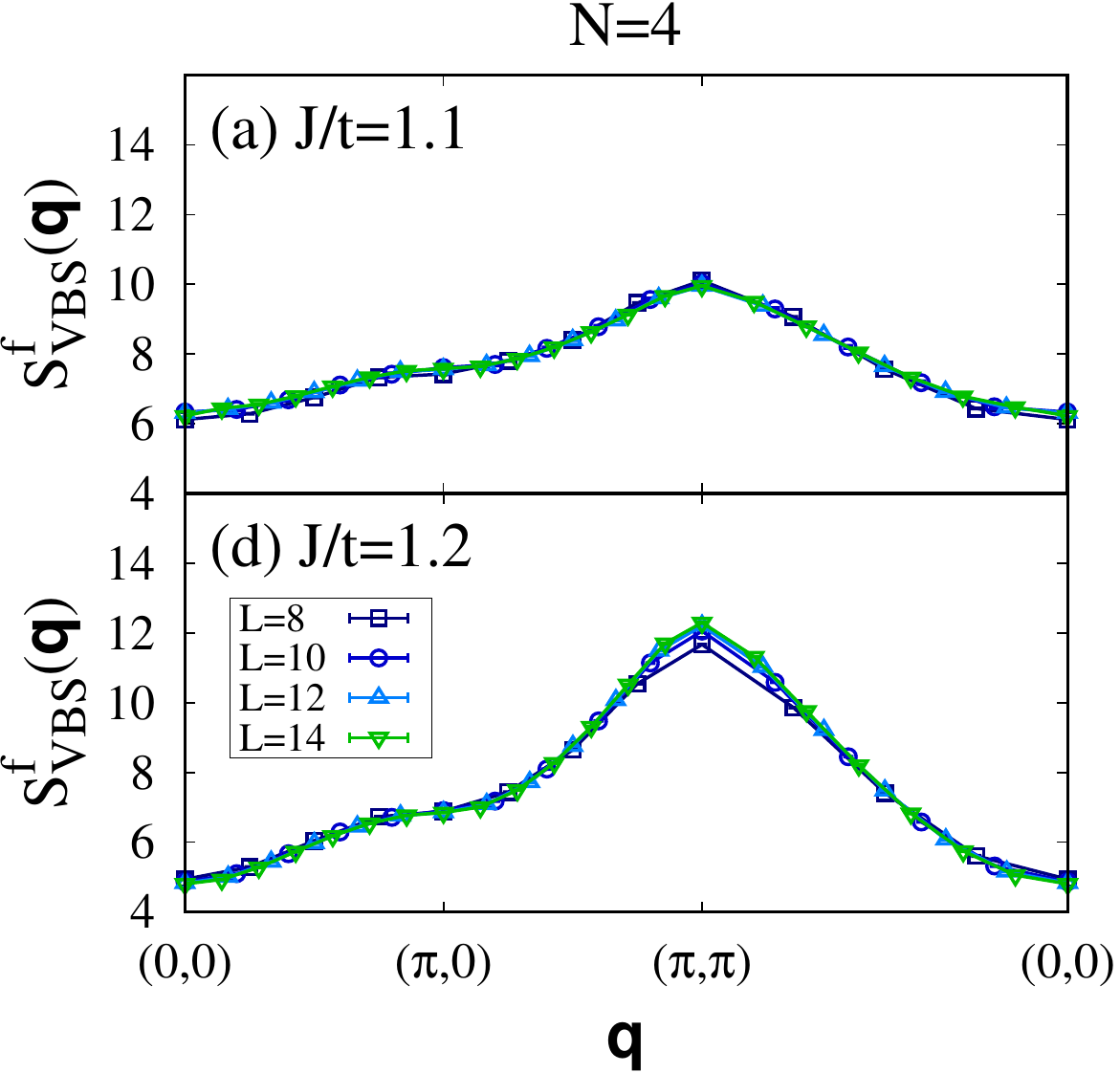}
\includegraphics[width=0.32\textwidth]{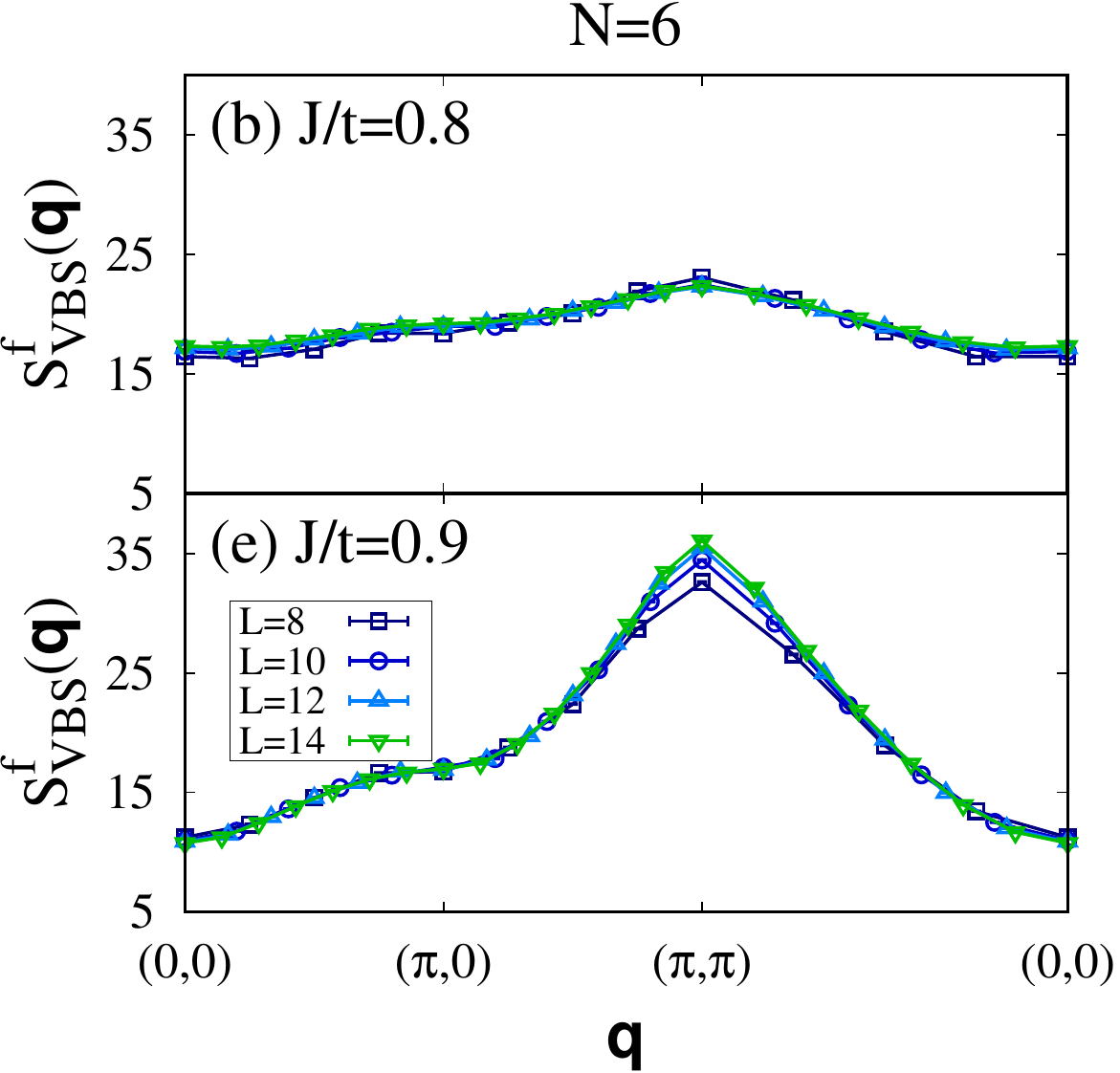}
\includegraphics[width=0.32\textwidth]{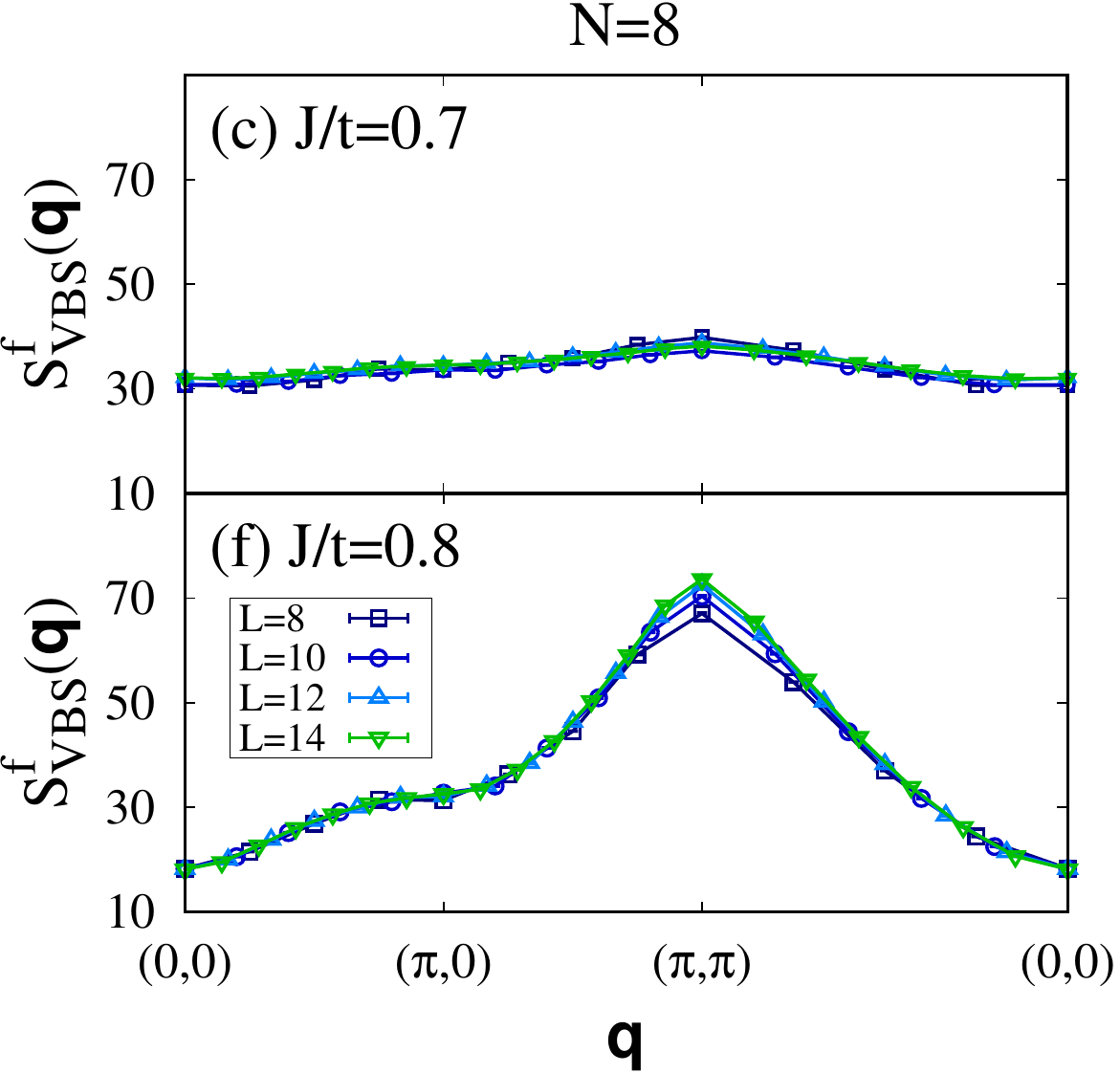}\\
\end{center}
\caption
{VBS correlation function $S^f_{\rm VBS}(\pmb{q})=\sum_{\pmb{\delta}}[S^f_{\rm VBS}(\pmb{q})]_{\pmb{\delta},\pmb{\delta}}$
for the $f$-electrons for various lattice sizes along a high symmetry path in the Brillouin zone
across the magnetic order-disorder transition point in the: (a-c) AF and (d-f)  KI phases for increasing $N$ (from left to right).
Here, a twice smaller imaginary-time step $\Delta \tau t=0.05$ in the Trotter-Suzuki decomposition in Eq.~(\ref{TS}) 
is used as compared to Figs.~\ref{VBS} and \ref{VBS2}.
}
\label{VBStau}
\end{figure*}

\subsection{\label{app:moment}  Spin structure factor}

As discussed in Sec.~\ref{spin}, we have estimated the onset of long-range magnetic order from the 
behavior of the staggered magnetic moment: 
\begin{equation}
	m_{}^{\alpha} =  \sqrt{  \lim_{L\to \infty} \frac{S^{\alpha}(\pmb{Q}=(\pi,\pi))}{L^2}},  
\end{equation}
extracted separately for the $f$- and $c$-electrons. The corresponding finite-size scaling analysis of the spin structure factor  $S^{\alpha}(\pmb{Q})$ 
is shown in Figs.~\ref{App_one}(b,e,h,k)  and \ref{App_one}(c,f,i,l), respectively. 
Long-range AF order is present when $\lim_{L\to \infty} S^{\alpha}(\pmb{Q})/{L^2}$ extrapolates to a finite value.

\subsection{\label{app:spin}  Spin gap}

In Sec.~\ref{spin}, the gap for spin excitations  $\Delta_s(\pmb{q})$ was  obtained by considering  the imaginary-time displaced spin correlation functions,
\begin{equation}
S(\pmb{q},\tau) = \sum_{\mu,\nu}\langle \hat{S}^{\mu}_{\nu}(\pmb{q},\tau) \hat{S}^{\nu}_{\mu} (-{\pmb{q}})\rangle,
\end{equation}
where $\hat{S}^{\mu}_{\nu}(\pmb{q},\tau)= \hat{S}^{f,\mu}_{\nu}(\pmb{q},\tau) + \hat{S}^{c,\mu}_{\nu}(\pmb{q},\tau)$ is the total spin.
The spin gap $\Delta_s$ for a given linear system size $L$ has been extracted  from the asymptotic behavior of 
$S(\pmb Q=(\pi,\pi),\tau) \propto \exp \left( -\tau  \Delta_s \right)$ at large imaginary time $\tau$. Extrapolating to the thermodynamic limit, one finds 
for each $N$ that the spin gap scales to a finite value in the KI phase and vanishes inside the AF state due to the emergence of Goldstone modes of the broken 
continuous SU($N$) symmetry group, see Figs.~\ref{App_two}(a,d,g,j).

\subsection{\label{app:delta}  Single-particle dynamics}

As described in Sec.~\ref{charge}, to probe the single-particle dynamics we have  measured the imaginary-time displaced Green's function for the 
conduction electrons, 
\begin{equation} 
\label{app:Green}
        G(\pmb{k},\tau) = \frac{1}{N}\sum_{\sigma}
\langle \Psi_0^n | c^{\dagger}_{\pmb{k},\sigma}(\tau) c^{}_{\pmb{k}, \sigma} | \Psi_0^n \rangle.  
\end{equation}
The single-particle gap $\Delta_{qp}$ at momentum $\pmb{k}=(\pi,\pi)$ and the corresponding QP weight $Z_{(\pi,\pi)}$ were extracted by fitting the tail of 
the Green's function to the form  $Z_{\pmb{k}}e^{-\Delta_{qp}(\pmb{k}) \tau }$. 
As an example, Fig.~\ref{App_raw} shows raw data of $G[\pmb{k}=(\pi,\pi),\tau]$  for $N=4$, 6, and 8 obtained from QMC simulations with different system sizes 
$L$ at $J/t=0.6$. The good quality of the data allowed us to determine finite-size estimates of  $\Delta_{qp}$ and $Z_{(\pi,\pi)}$  directly on the imaginary-time 
axis. The corresponding extrapolation of both quantities to the thermodynamic limit is performed in Figs.~\ref{App_two}(b,e,h,k) and \ref{App_two}(c,f,i,l), respectively.
Note the enhanced finite-size effects in vicinity of the magnetic transition point.

\subsection{\label{app:vbs} Imaginary-time discretization ${\boldsymbol\Delta\tau}$ }

In Sec.~\ref{vbs}, we have calculated the VBS correlation function $S^f_{\rm VBS}(\pmb{q})$. 
We used the imaginary-time step $\Delta \tau t=0.1$ in the discrete Trotter-Suzuki decomposition in Eq.~(\ref{TS}) which  yields an error  $\mathcal{O}(\Delta\tau^2)$. 
In order to exclude that the cusp feature at the AF wavevector ${\pmb Q}=(\pi,\pi)$  is an artifact related to the Trotter-Suzuki decomposition, 
we have repeated QMC simulations for $N=4$, 6, and 8  with a twice smaller imaginary-time step $\Delta \tau t=0.05$.  
The corresponding dimer structure factors $S^f_{\rm VBS}(\pmb{q})$ shown in Fig.~\ref{VBStau} 
look qualitatively very similar to those in Figs.~\ref{VBS}(b,d) and \ref{VBS2} in the main text.

\bibliographystyle{bibstyle}
\bibliography{marcin_refs}

\end{document}

%% file: discuss.tex
\subsection{Quantum-field-theoretic perspective}
\label{QFT.Sec}
Throughout  the $J$-$N$    plane, the charge degrees of freedom are gapped. Hence charge fluctuations around  half-filling --  that mix SU($N$) spin representations -- will not contribute in the low-energy effective field theory and can be safely  omitted.   The remaining degree of freedom is an  SU($N$) spin  in the totally antisymmetric representation corresponding  to a Young tableau consisting of a single column and $N$/2 rows.    Since we observe AF phases,  the low-energy effective model is that of an SU($N$)  quantum antiferromagnet: 
\begin{equation}
	\hat{H}_{eff}   = \frac{J}{N} \sum_{\langle  \ve{i}, \ve{j} \rangle , \mu, \nu }    \hat{S}^{\mu}_{\ve{i},\nu}   \hat{S}^{\nu}_{\ve{j},\mu},
\end{equation}
in the aforementioned representation.   The generalization of Haldane's   SU(2) spin coherent state path  integral formulation~\cite{Haldane88}  to the  SU($N$) 
group has been carried by Read and Sachdev~\cite{RS89}.   It is beyond the   scope of this paper to review  the derivation, and we will only cite  the final result.    We consider a  square lattice that can be decomposed into two subattices, A and B, such that the nearest neighbors of one sublattice belong to the other.   
 As for the SU(2) case   the SU($N$) spin coherent state  $| q \rangle$   is obtained by an SU($N$) rotation of the N\'eel state~\cite{Perelomov72,RS89}.  
It satisfies the relation: 
\begin{eqnarray}
	& & \langle q |  \hat{S}^{\mu}_{\ve{i},\nu} | q \rangle   =   \pm Q^{\mu}_{\nu} (\ve{i})  \equiv \pm  \left( \hat{U}^{\dagger}(\ve{i})  \Lambda  \hat{U}(\ve{i}) \right)^{\mu}_{\nu}  \text{  with }  \nonumber \\
	& & \Lambda   =  
 	\begin{pmatrix} 
	1_{N/2\times N/2} & 0 \\
	0  & -1_{N/2\times N/2}
	\end{pmatrix}, 
\end{eqnarray}
and the $\pm$ sign refers to the A and B sublattices.   $ Q^{\mu}_{\nu} (\ve{i}) $ hence corresponds to the  N\'eel order parameter, that owing to the sign convention is uniform in space, and whose low-energy fluctuations are governed by the action: 
\begin{equation}
	  S =  S_B+  S_{NL\sigma},
\end{equation}
  with Berry phase  $S_B$ \cite{RS89} and non-linear $\sigma$ (NL$\sigma$) model,  
\begin{equation}
\label{NLs.eq}
S_{NL\sigma} = \int d \tau d^2{\ve{x}}  \frac{\rho_s}{2}  \text{Tr} \left(  \left[ \partial_{\ve{x}} Q(\ve{x},\tau) \right]^2  + \frac{1}{c^2} \left[ \partial _\tau Q(\ve{x},\tau) \right]^2 \right). 
\end{equation}
In the above $\rho_s$ corresponds to the spin stiffness and $c$ to the velocity. For the SU(2) case, we can write $ Q = \ve{n}  \cdot \ve{\sigma} $ with $\ve{n}$ a unit vector and  $ \ve{\sigma} $ the vector of Pauli spin matrices.  With this parametrization, the above reduces to   the well known O(3) NL$\sigma $ model with Berry phase.     In contrast to the one-dimensions, smooth  space-time variations of the N\'eel  order parameter  have a vanishing Berry phase~\cite{Haldane88}.     For the above U($N$) model,   the order parameter manifold corresponds to  $\frac{U(N)}{U(N/2) \times U(N/2) } $. Since the second homotopy group of this space is given by  $\mathbb{Z}$, skyrmions are well defined, and  one can carry over  the  arguments put forward by Haldane for the SU(2) case. In particular, skyrmion  number changing  field configurations  (hedgehogs) carry a non-trivial Berry phase  and quadruple hedgehog insertions carry no  Berry phase and hence do  not interfere destructively.  This suggests that the Hilbert space splits  into four  distinct classes   corresponding to the skyrmion number modulo 4.    Proliferation of quadruple hedgehog configurations has been argued to correspond to the VBS state~\cite{Senthil04b}   
and is the essence of the notion of deconfined quantum criticality. 

With this background that links the Berry phase to a four-fold degenerate  VBS state  and  the lack of any $(0, \pm \pi) $  and $(\pm \pi, 0 )$   singularities in the  VBS order parameter  across the magnetic order-disorder transition point  in the QMC data, see Sec.~\ref{vbs}, we conclude that the Berry phase can be  omitted in the effective field theory.  A similar result holds for the  SU(2) bilayer Heisenberg model~\cite{Sandvik94,Sandvik95,Zaanen97,Wang06}. The SU($N$) KLM hence provides a lattice realization of NL$\sigma$ model  of Eq.~(\ref{NLs.eq}).   To the best of our knowledge, the critical exponents as well as the very nature of the transition as a function of $N$ are unknown.  A  $1/N$ expansion study of  the critical  exponents has been carried out in Ref.~\cite{Wang3}   for the general  representation corresponding to Young tableau of $m$ ($N-m$) rows and one column on sublattice A (B).  As pointed out in the paper~\cite{Wang3},  the results require $m/N$ to be a small number and cannot be carried over to the self-conjugate representation where $m=N/2$.

%% file: append.tex
\section{\label{app:scales} Energy scales of the SU($\boldsymbol N$) KLM}

\subsection{The RKKY scale}
The SU($N$)  generalization of the KLM of Eq.~(\ref{su2})  reads:
\begin{equation}
\hat{\mathcal{H}} = -t\sum_{\langle\pmb{i},\pmb{j}\rangle, \sigma =1}^{N} 
                  \hat{c}^{\dagger}_{\pmb{i},\sigma} \hat{c}^{}_{\pmb{j},\sigma} +
         \frac{2 J}{N} \sum_{\pmb{i}, a=1  }^{N^2 -1}  \hat{T}^{a,c}_{\pmb{i}}  \hat{T}^{a,f}_{\pmb{i}}. 
\label{suN}
\end{equation}
Here,  
\begin{equation}
	 \hat{T}^{a,c}   = \hat{\ve{c}}^{\dagger} T^{a}  \hat{\ve{c}}^{\phantom\dagger}, \; \; 
	  \hat{T}^{a,f}   = \hat{\ve{f}}^{\dagger} T^{a}  \hat{\ve{f}}^{\phantom\dagger},
\end{equation}
and the  $N^2 -1  $ generators of SU($N$)   satisfy the normalization condition: 
\begin{equation}
	\text{Tr}  \left[ T^{a} T^{b} \right]   = \frac{1}{2}\delta_{a,b}.
\label{Normalization_condition.eq}
\end{equation}
To estimate the energy scale of the RKKY   interaction,  we will first consider  a single  impurity at the origin   with a frozen  $f$-spin:
\begin{equation}
\hat{\mathcal{H}} = -t\sum_{\langle\pmb{i},\pmb{j}\rangle, \sigma =1}^{N} 
                  \hat{c}^{\dagger}_{\pmb{i},\sigma} \hat{c}^{}_{\pmb{j},\sigma} +
         \frac{2 J}{N} \sum_{ a=1  }^{N^2 -1}  \hat{T}^{a,c}_{\pmb{0}} \langle \hat{T}^{a,f}_{\pmb{0}} \rangle. 
\end{equation}
Within first order perturbation theory in $J$  the frozen $f$-spin at the origin produces ripples in the spin texture that follow the spin susceptibility of the conduction electrons, $\chi^{c}$, 
\begin{equation}
 \langle \hat{T}^{a,c}_{\pmb{r}} \rangle = - \frac{2J}{N} \langle \hat{T}^{a,f}_{\pmb{0}} \rangle \chi^{c}(\ve{r}, i\Omega_m = 0). 
\end{equation}
Here, 
\begin{equation}
  \chi^{c}(\ve{q}, i\Omega_m = 0)  = \frac{1}{2L^2} \sum_{\ve{k}}   \frac{ f(\ve{k} -  \ve{q} )  -  f(\ve{k}) } { \epsilon(\ve{k}) - \epsilon(\ve{k} - \ve{q})},
\end{equation} 
with $ \epsilon(\ve{k} )  = -2 t  \left( \cos(k_x) +  \cos(k_y)  \right) $,  $f(\ve{k}) = \frac{1}  { 1 + e^{\beta \epsilon(\ve{k}) }  } $ and 
$\chi^{c}(\ve{r}, i\Omega_m = 0)  = \frac{1}{L^2} \sum_{\ve{q}} e^{- i\ve{q} \cdot \ve{r}}  \chi^{c}(\ve{q}, i\Omega_m = 0) $. 
We  now consider a second  impurity at position $\ve{r}$ that Kondo couples to the conduction electrons  according to  Eq.~(\ref{suN}).  At the mean-field level, the interaction energy  between the two spins, reads: 
\begin{equation}
	\frac{J}{2N}  \langle \hat{T}^{a,c}_{\pmb{r}} \rangle \hat{T}^{a,f}_{\pmb{r}} \equiv  - \left(\frac{2J}{N} \right)^2 \chi^{c}(\ve{r}, i\Omega_m = 0)    \langle \hat{T}^{a,f}_{\pmb{0}}  \rangle  \hat{T}^{a,f}_{\pmb{r}}.
\end{equation}
Comparing the above expression to the RKKY  Hamiltonian: 
\begin{equation}
	  \hat{H}_{RKKY}  = \frac{1}{2N}  \sum_{a, \ve{r} \neq \ve{r}'}  J_{RKKY}(\ve{r} - \ve{r}')    \hat{T}^{a,f}_{\pmb{r}}  \hat{T}^{a,f}_{\pmb{r}'},
\end{equation}
that  describes the effective  SU($N$)   Heisenberg interaction between the  impurity spins  gives:
\begin{equation}
	J_{RKKY}(\ve{r})   =  - \frac{8J^2}{N}  \chi^{c}(\ve{r}, i\Omega_m = 0). 
\end{equation}
Hence,   the RKKY  interaction measured relative to the kinetic energy scales as $\frac{1}{N}$. 

\subsection{The Kondo scale }
In contrast, we now argue that the Kondo scale is $N$-independent in the large-$N$ limit.   To formulate the large-$N$ mean-field saddle-point, we use the completeness relation, 
\begin{equation}
	\sum_{a} T^{a}_{\alpha,\beta} T^{a}_{\alpha',\beta'} = \frac{1}{2} \left(  \delta_{\alpha,\beta'}  \delta_{\alpha',\beta} - \frac{1}{N} \delta_{\alpha,\alpha'} \delta_{\beta, \beta'} \right), 
\end{equation}
to  show that up to a constant: 
\begin{equation}
	\hat{\mathcal{H}} = -t\sum_{\langle\pmb{i},\pmb{j}\rangle, \sigma =1}^{N} 
                  \hat{c}^{\dagger}_{\pmb{i},\sigma} \hat{c}^{}_{\pmb{j},\sigma}  - \frac{J}{2N} \sum_{\pmb{i}}  \left( 
                \hat{D}^{\dagger}_{\ve{i}} \hat{D}^{\phantom\dagger}_{\ve{i}}   +    \hat{D}^{\phantom\dagger}_{\ve{i}} \hat{D}^{\dagger}_{\ve{i}}    \right),
\end{equation}
with, 
\begin{equation}
	   \hat{D}^{\dagger}_{\ve{i}}   =  \sum_{\sigma=1}^{N} \hat{c}^{\dagger}_{\ve{i},\sigma}  \hat{f}^{\phantom\dagger}_{\ve{i},\sigma}.
\end{equation}
Using the mean-field Ansatz  $V=  2 \langle \hat{D}^{\dagger}_{\ve{i}} \rangle  /N $  and imposing the constraint  $ \sum_{\sigma=1}^{N} \hat{f}^{\dagger}_{\ve{i},\sigma}  \hat{f}^{\phantom\dagger}_{\ve{i},\sigma}   = \frac{N}{2} $ on average yields the gap equation: 
\begin{equation}
     \frac{2}{J} \Delta   = \Delta \int d \epsilon N(\epsilon)  \frac{  f(\frac{\epsilon}{2} - E )  -  f(\frac{\epsilon}{2} + E) }{E},
\end{equation}
where  $\Delta = JV/2$, $E = \sqrt{ \left(\frac{\epsilon}{2} \right)^2  + \Delta^2 }$,  $N(\epsilon)  = \frac{1}{L^2} \sum_{\ve{k}} \delta( \epsilon - \epsilon(\ve{k})) $, and   $f$ the Fermi function.      For the particle-hole symmetric  case considered here,   the $f$-electron half-filling constraint is satisfied by  symmetry such that no  Lagrange multiplier has to be introduced.  At  the Kondo temperature, $T_K$,  $\Delta$ vanishes, such that $T_K $   is given by: 
\begin{equation}
	\frac{1}{J}  = \int_{0}^{\infty} d \epsilon N(\epsilon)  \frac{ \tanh\left( \frac{ \epsilon}{2 k_B T_K } \right) }{\epsilon}.
\end{equation}
In the above, we have used the particle-hole symmetric condition $N(\epsilon) =  N(-\epsilon) $. 
As  apparent,  the above equation is independent on $N$ such that at the mean-field level, the Kondo temperature  does not scale with $N$.
Finally we note that for a  density of states of width $W$   and in the small $J$ limit,  
\begin{equation}
   k_B T_K   \propto \frac{e W}{2} e^{-W/J}.
\end{equation}

\subsection{Functional form of the critical coupling $\boldsymbol{J_c(N)}$  } 

We can now compare scales to estimate the  the critical value of $J_c(N)$  where the Kondo effect gives way to magnetic ordering: 
\begin{equation}
	\frac{e}{2N(\epsilon_f)} e^{-1/N(\epsilon_f) J_c} =    \frac{8J_c^2}{N}  \chi^{c}. 
\end{equation}
In the above, we have used $ N(\epsilon_f) = \frac{1}{W} $ for the aforementioned flat density of states,  and for instance, considered the  value of the spin-susceptibility at a distance given by the lattice spacing.    In the large-$N$ limit where we expect $J_c$ to be small,    we obtain: 
\begin{equation}
	J_c(N)  \propto \frac{1}{\ln(N) N(\epsilon_f)}. 
\end{equation}

\vspace{0.5cm}
\section{\label{app:SDW}  Spin-density-wave approach for the SU($\boldsymbol N$) KLM}

The data presented in the main text suggests that for each $N$  magnetism and Kondo screening coexist,  and that in  the magnetically ordered phase, the $f$-local moment is  large since up to $N=8$ it  exceeds 80\% of  the N\'eel value.    In this appendix, we generalized the mean-field theory of Ref.~\cite{MF00}  that captures both Kondo screening and magnetism to the SU($N$)  group. To do so, we consider  the following explicit form of the SU($N$) generators.  For $\alpha > \beta  $  included in the set of $ \left[  1,N \right] $  we consider the   $N^2 - N $ off-diagonal generators: 
\begin{eqnarray}
  \hat{T}^{x, c, \alpha, \beta}_{\ve{i}}  & =  & \frac{1}{2} \left( \hat{c}^{\dagger}_{\ve{i},\beta}  \hat{c}^{\phantom\dagger}_{\ve{i},\alpha} +  \hat{c}^{\dagger}_{\ve{i},\alpha}  \hat{c}^{\phantom\dagger}_{\ve{i},\beta}   \right),  \nonumber \\
   \hat{T}^{y, c, \alpha, \beta}_{\ve{i}}  & =  & \frac{1}{2} \left( i \hat{c}^{\dagger}_{\ve{i},\beta}  \hat{c}^{\phantom\dagger}_{\ve{i},\alpha} - i \hat{c}^{\dagger}_{\ve{i},\alpha}  \hat{c}^{\phantom\dagger}_{\ve{i},\beta}   \right),
\end{eqnarray}
alongside the $N-1$ diagonal operators: 
\begin{equation}
	\hat{T}^{z,c,n}_{\ve{i}} = \frac{1}{\sqrt{2(n + n^2)}}   \left( \sum_{\alpha = 1}^{n} \hat{c}^{\dagger}_{\ve{i},\alpha} \hat{c}^{\phantom\dagger}_{\ve{i},\alpha}    -   n  \hat{c}^{\dagger}_{\ve{i},n+1} \hat{c}^{\phantom\dagger}_{\ve{i},n+1}  \right). 
\end{equation}
In the above, $n$ runs from  $1,\ldots, N-1 $. This definition of  the SU($N$) generators  satisfies the normalization condition   of Eq.~(\ref{Normalization_condition.eq}) and similar forms hold for the $f$-electrons.  As in  Ref.~\cite{MF00},  the off-diagonal operators will account for Kondo screening whereas the diagonal ones for   magnetism.  
With the above, the SU($N$) Kondo interaction reads: 
\begin{widetext}
\begin{eqnarray}
	\hat{H}_{K}   = \frac{2 J}{N} & &   \sum_{\pmb{i}}   \left(  \sum_{n=1}^{N-1}  \hat{T}^{z,n,c}_{\pmb{i}}  \hat{T}^{z,n,f}_{\pmb{i}}   + \sum_{\alpha > \beta = 1}^{N}   \hat{T}^{x,\alpha,\beta,c}_{\pmb{i}}  \hat{T}^{x,\alpha,\beta,f}_{\pmb{i}}  +   \hat{T}^{y,\alpha,\beta,c}_{\pmb{i}}  \hat{T}^{y,\alpha,\beta,f}_{\pmb{i}}    \right)  \nonumber   \\
	    = \frac{2 J}{N} & &   \sum_{\pmb{i}}   \left(  \sum_{n=1}^{N-1}  \hat{T}^{z,n,c}_{\pmb{i}}  \hat{T}^{z,n,f}_{\pmb{i}}    -\frac{1}{4} \sum_{\alpha > \beta = 1}^{N}   \left( \hat{c}^{\dagger}_{\ve{i},\alpha} \hat{f}^{\phantom\dagger}_{\ve{i},\alpha} +  
	                                                 \hat{f}^{\dagger}_{\ve{i},\beta} \hat{c}^{\phantom\dagger}_{\ve{i},\beta}  \right)^2    +  
	   	\left( \hat{c}^{\dagger}_{\ve{i},\beta} \hat{f}^{\phantom\dagger}_{\ve{i},\beta} + 
	       \hat{f}^{\dagger}_{\ve{i},\alpha} \hat{c}^{\phantom\dagger}_{\ve{i},\alpha}  \right)^2  \right).  
\end{eqnarray}
\end{widetext}
To account for the Kondo effect,  we adopt the mean-field Ansatz: 
\begin{equation}
	     \langle \hat{c}^{\dagger}_{\ve{i},\alpha} \hat{f}^{\phantom\dagger}_{\ve{i},\alpha} +  
	                                                 \hat{f}^{\dagger}_{\ve{i},\beta} \hat{c}^{\phantom\dagger}_{\ve{i},\beta}  \rangle =  \langle  \hat{c}^{\dagger}_{\ve{i},\beta} \hat{f}^{\phantom\dagger}_{\ve{i},\beta} + 
	       \hat{f}^{\dagger}_{\ve{i},\alpha} \hat{c}^{\phantom\dagger}_{\ve{i},\alpha}  \rangle = - V.                                           
\end{equation}
For the magnetism, it  is convenient to carry out an orthogonal transformation of the diagonal operators: 
\begin{equation}
	 \tilde{\hat{T}}^{z,c,m}_{\ve{i}}  = \sum_{n=1}^{N-1}  O_{m,n}\hat{T}^{z,c,n}_{\ve{i}}, 
\end{equation}
such that: 
\begin{equation}
	 \tilde{\hat{T}}^{z,c,1}_{\ve{i}}  =  \frac{1}{\sqrt{2N}}  \left( \sum_{n=1}^{N/2}  \hat{c}^{\dagger}_{\ve{i},n} \hat{c}^{\phantom\dagger}_{\ve{i},n}
	    - \sum_{n=N/2+1}^{N}  \hat{c}^{\dagger}_{\ve{i},n} \hat{c}^{\phantom\dagger}_{\ve{i},n} \right). 
\end{equation}
Identical forms hold for the $f$-electrons.   In the N\'eel state, 
\begin{equation}
	| \Psi_{\text{N\'eel}} \rangle  = \prod_{\ve{i}\in A}  \hat{f}^{\dagger}_{\ve{i},1} \cdots   \hat{f}^{\dagger}_{\ve{i},N/2}  
	\prod_{\ve{i} \in B} \hat{f}^{\dagger}_{\ve{i},N/2+1} \cdots   \hat{f}^{\dagger}_{\ve{i},N}  | 0\rangle,
\end{equation} 
we have, 
\begin{equation}
\langle  \Psi_{\text{N\'eel}} | \tilde{\hat{T}}^{z,f,1}_{\ve{i}} |  \Psi_{\text{N\'eel}} \rangle  = \frac{\sqrt{N}}{2\sqrt{2}} e^{i \ve{Q}\cdot \ve{i}}, 
\end{equation}
with the AF wave vector $\ve{Q} = \left( \pi,\pi \right) $. 
This motivates the Ansatz: 
\begin{eqnarray}
	\langle  \tilde{\hat{T}}^{z,f,n}_{\ve{i}}    \rangle  =     \delta_{n,1} \frac{\sqrt{N}}{2\sqrt{2}} e^{i \ve{Q}\cdot \ve{i}}   m_f,  \nonumber \\
	\langle  \tilde{\hat{T}}^{z,c,n}_{\ve{i}}    \rangle  = -   \delta_{n,1} \frac{\sqrt{N}}{2\sqrt{2}} e^{i \ve{Q}\cdot \ve{i}}   m_c. 
\end{eqnarray}
The above Ans\"atze   break  the U($N$) symmetry   down to a  U($N$/2) $ \times$ U($N$/2)   one such that it becomes   convenient to introduce the 
notation: 
\begin{eqnarray}
	 \hat{c}^{\dagger}_{\ve{i},\mu,\sigma}  \equiv \hat{c}^{\dagger}_{\ve{i}, \mu +  \frac{1 + \sigma}{2}  \frac{N}{2}},    \nonumber \\
	  \hat{f}^{\dagger}_{\ve{i},\mu,\sigma}  \equiv \hat{f}^{\dagger}_{\ve{i}, \mu +  \frac{1 + \sigma}{2}  \frac{N}{2}},  
\end{eqnarray}
with $\mu = 1,\ldots, N/2$  and $\sigma = \pm 1 $.   
The mean-field  Hamiltonian is then given by: 
\begin{eqnarray}
\label{SDW_SUN.eq}
	\hat{H}_{MF}   & & =   \sum_{\mu=1}^{N/2} \sum_{\ve{k} \in MBZ, \sigma=\pm 1}  
\begin{pmatrix} 
\hat{c}^{\dagger}_{\ve{k},\mu,\sigma} \\
\hat{c}^{\dagger}_{\ve{k} + \ve{Q},\mu,\sigma} \\
\hat{f}^{\dagger}_{\ve{k},\mu,\sigma} \\
\hat{f}^{\dagger}_{\ve{k} + \ve{Q},\mu,\sigma} 
\end{pmatrix}^{T}  \\
&&  \times
\begin{pmatrix} 
\epsilon(\ve{k})        & \frac{Jm_f \sigma}{2N}  &    JV \frac{N-1}{N}    &   0  \\
 \frac{Jm_f \sigma}{2N} & -\epsilon(\ve{k})       &       0                &  JV \frac{N-1}{N}  \\
 JV \frac{N-1}{N}       &     0                   &       0                & -\frac{Jm_c \sigma}{2N} \\
       0                &  JV \frac{N-1}{N}       & -\frac{Jm_c\sigma}{2N} &   0
\end{pmatrix} 
\nonumber  \\
& & \times \begin{pmatrix} 
\hat{c}^{}_{\ve{k},\mu,\sigma} \\
\hat{c}^{}_{\ve{k} + \ve{Q},\mu,\sigma} \\
\hat{f}^{}_{\ve{k},\mu,\sigma} \\
\hat{f}^{}_{\ve{k} + \ve{Q},\mu,\sigma} 
\end{pmatrix}  
+ J L^2 \left( \frac{ m_c m_f}{4}  + V^2 \frac{N-1}{2} \right).
\nonumber 
\end{eqnarray}
Here, $\epsilon(\ve{k})  = -2t  \left(  \cos(k_x) + \cos(k_y) \right) $ such that  $ \epsilon(\ve{k} + \ve{Q}) = -    \epsilon(\ve{k})  $ and particle-hole symmetry  pins the average  $f$-occupation to $N/2$.    The saddle-point equations then read: 
\begin{equation}
    \frac{\partial F} {\partial m_{c} } =  \frac{\partial F} {\partial m_{f} } =  \frac{\partial F} {\partial V }   = 0,
\end{equation}
with $F = -\frac{1}{\beta} \ln{ \text{Tr}   e^{- \beta  \hat{H}_{MF}}} $. 
Several comments are in order. 
\begin{itemize}
\item  The underlying particle-hole symmetry pins the $f$-occupation to half-filling such that  no  Lagrange multiplier is required  to enforce this constraint on average.
\item Since the $\mu$ index does not appear in the Hamiltonian matrix, the above  has  a U($N$/2) $\times $ U($N$/2)   symmetry that  generalizes the  U(1) $\times$  U(1) symmetry presented in 
Refs.~\cite{MF00,Capponi01}.  One will notice that at  $N=2$ we recover precisely Eq.~(45) of Ref.~\cite{Capponi01}. 
In this case, and assuming $m_f=m_c=0$ but $V\ne 0$  as appropriate for the KI phase, one finds the single-particle  dispersion relation:
\begin{equation}
 E_{\ve{k}}^{\pm}  = \frac{1}{2} \left[\epsilon(\ve{k}) \pm \sqrt{ \epsilon^2(\ve{k})+ (JV)^2} \right], 
\end{equation}	
QP gap $\Delta_{qp} = - E^{-}_{\ve{k}=(\pi,\pi)}$ and residue:
\begin{equation}
Z_{(\pi,\pi)}  =  \frac{1}{2} \left[ 1 - \frac{\epsilon(\ve{k}=(\pi,\pi)) }{ \sqrt{\epsilon^2(\ve{k}=(\pi,\pi))+ (JV)^2 }}\right],
\end{equation}
for a doped hole away from half-filling. Solving self-consistently the saddle-point equation  for the hybridization order parameter $V$ and using the above relations 
for $\Delta_{qp}$ and $Z_{(\pi,\pi)}$  lead us to the large-$N$ results shown in Figs.~\ref{Charge} and \ref{Ak_cmp}(b) in the main text.  
On the other hand, assuming that the AF order is present $m_f\ne 0$ and $m_c\ne 0$ and the spin degrees of freedom are frozen such that $V=0$, the corresponding 
dispersion relation reads: 
\begin{equation}
E^{\pm}_{\ve{k}}  = \pm\sqrt{ \epsilon^2(\ve{k}) + \left(\frac{Jm_f}{2N}\right)^2},
\end{equation}
and the QP gap tracks $J/N$ as does the QMC data in Fig.~\ref{Charge}(a) in the main text.
\item  While the magnetic energy  scales as order $N^0$ the kinetic and hybridization energies scale as order $N$.   This can be seen explicitly in the  last  constant term of Eq.~(\ref{SDW_SUN.eq}) and is consistent with the above
discussion of the Kondo and RKKY energy scales.    As a consequence, we expect the  $J=0$ and $N \rightarrow \infty$ point to be singular.    For the ordering of limits   $\lim_{N \rightarrow \infty }  \lim_{J \rightarrow 0} $  we expect an AF ground state whereas for   $ \lim_{J \rightarrow 0} \lim_{N \rightarrow \infty }  $ a paramagnetic one. 
\item It is very tempting to follow ideas presented in Ref.~\cite{Assaad03}  and to formulate a  U($N$/2) $\times$  U($N$/2) field theory that  possesses the above mean-field Hamiltonian as a saddle point in the large-$N$ limit and that reproduces the U(2) invariant KLM model  at $N=2$.   At $N=2$, and in  the magnetically ordered phase, we observe a remarkable coexistence of Kondo screening and antiferromagnetism that stands at odds with the mean-field results predicting only a very narrow coexistence  region~\cite{MF00,Capponi01}.   As a function of $N$, fluctuations  around the magnetically ordered  
saddle point are reduced  and we expect a stronger mean-field-like competition between   magnetic ordering and Kondo screening. It is very interesting to see that the QMC data supports this line of thought. 
\end{itemize}